\documentclass[aps,prb,superscriptaddress,twocolumn]{revtex4}
\usepackage{amsmath,amssymb,bm}
\usepackage{graphicx}
\usepackage{epstopdf}
\usepackage{latexsym}
\usepackage{subfigure}
\usepackage{color}
\usepackage{amssymb}
\usepackage{wasysym}
\usepackage{verbatim}

\newcommand{\bM}{\mathbf{M}}
\newcommand{\bN}{\mathbf{N}}
\newcommand{\bJ}{\mathbf{J}}
\newcommand{\ep}{\epsilon}
\newcommand{\px}{\partial_x}

\newcommand{\nlsm}{NL$\sigma$M}
\newcommand{\br}{\mathbf{r}}
\newcommand{\bk}{\mathbf{k}}

\begin{document}
\date{\today}
\title{Ground states of spin-$\frac{1}{2}$ triangular antiferromagnets in a magnetic field}

\author{Ru Chen}
\affiliation{Department of Physics, University of California, Santa Barbara, Santa Barbara, CA, 93106}

\author{Hyejin Ju}
\affiliation{Department of Physics, University of California, Santa Barbara, Santa Barbara, CA, 93106}

\author{Hong-Chen Jiang}
\affiliation{Kavli Institute of Theoretical Physics, University of California, Santa Barbara, Santa Barbara, CA, 93106}

\author{Oleg A. Starykh}
\affiliation{Department of Physics and Astronomy, University of Utah, Salt Lake City, UT 84112}

\author{Leon Balents}
\affiliation{Kavli Institute of Theoretical Physics, University of California, Santa Barbara, Santa Barbara, CA, 93106}

\begin{abstract}
  We use a combination of numerical density matrix renormalization group
  (DMRG) calculations and several analytical approaches to
  comprehensively study a simplified model for a spatially anisotropic
  spin-1/2 triangular lattice Heisenberg antiferromagnet: the three-leg
  triangular spin tube (TST).  The model is described by three
  Heisenberg chains, with exchange constant $J$, coupled
  antiferromagnetically with exchange constant $J'$ along the diagonals
  of the ladder system, with periodic boundary conditions in the shorter
  direction.  Here we determine the full phase diagram of this model as
  a function of both spatial anisotropy (between the isotropic and
  decoupled chain limits) and magnetic field.  We find a rich phase
  diagram, which is remarkably dominated by quantum states -- the phase
  corresponding to the classical ground state appears only in an
  exceedingly small region.  Among the dominant phases generated by
  quantum effects are commensurate and incommensurate coplanar
  quasi-ordered states, which appear in the vicinity of the isotropic
  region for most fields, and in the high field region for most
  anisotropies.  The coplanar states, while not classical ground states,
  can at least be understood semiclassically.  Even more strikingly, the
  largest region of phase space is occupied by a spin density wave
  phase, which has incommensurate collinear correlations along the
  field.  This phase has no semiclassical analog, and may be ascribed to
  enhanced one-dimensional fluctuations due to frustration.  Cutting
  across the phase diagram is a magnetization plateau, with a gap to all
  excitations and ``up up down'' spin order, with a quantized
  magnetization equal to 1/3 of the saturation value.  In the TST, this
  plateau extends almost but not quite to the decoupled chains limit.
  Most of the above features are expected to carry over to the two
  dimensional system, which we also discuss.  At low field, a dimerized
  phase appears, which is particular to the one dimensional nature of the TST,
  and which can be understood from quantum Berry phase arguments.
\end{abstract}

\maketitle

\section{Introduction}
\label{sec:intro}

The nearest-neighbor spin-1/2 Heisenberg antiferromagnet on the
triangular lattice is an archetypal model of frustrated quantum
magnetism.  While the isotropic model in zero field is rather well-understood
 and is known to order into a coplanar ``120$^\circ$''
state\cite{PhysRevLett.69.2590}, away from this limit the situation is less clear.  Two
deformations of the Hamiltonian are of particular
physical and experimental importance: the application of an external
magnetic field and the introduction of spatial anisotropy into the
exchange interactions.

\begin{figure}[t]
  \begin{center}
  \scalebox{0.8}{\includegraphics[width=\columnwidth]{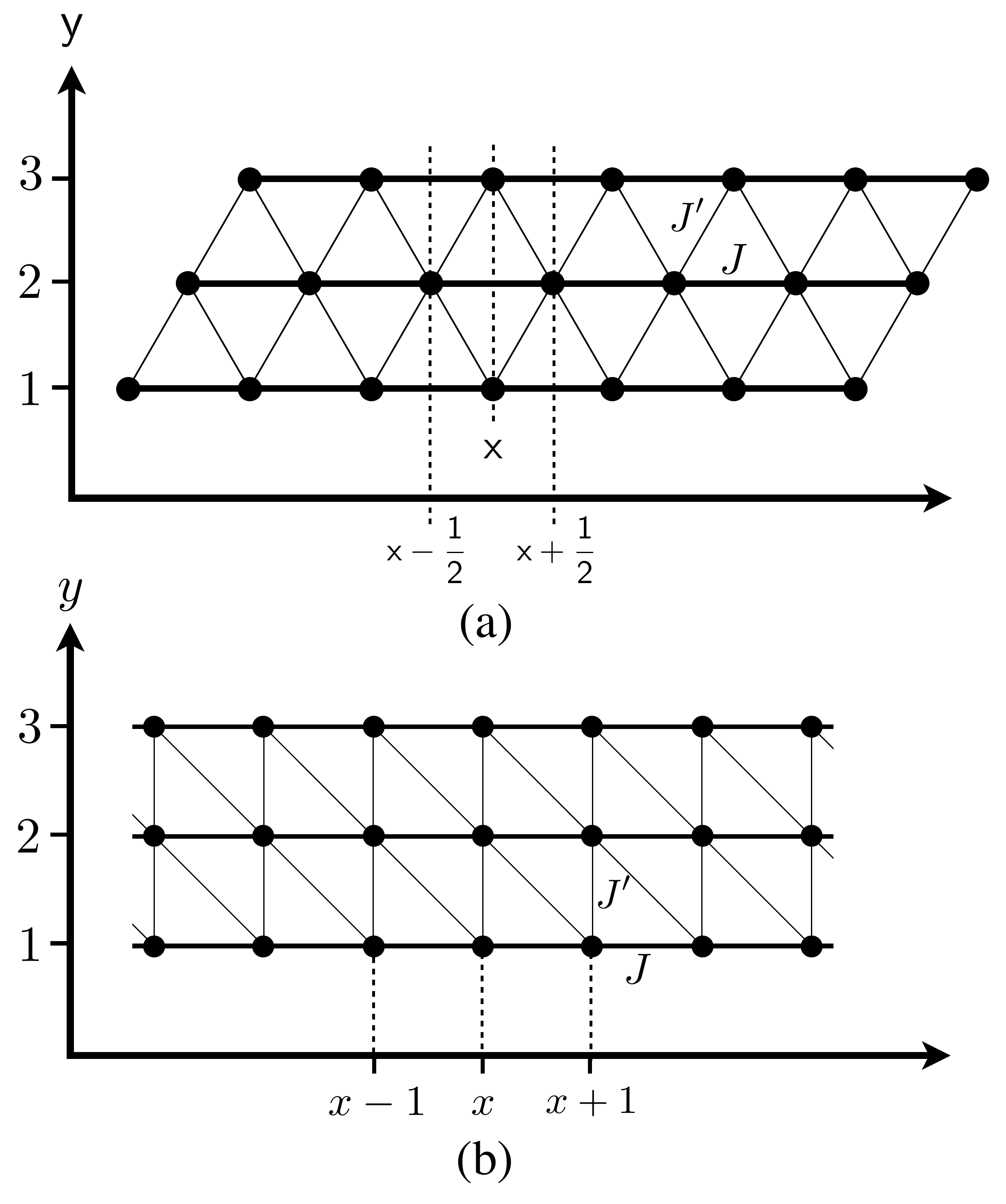}}
  \end{center}
  \caption{TST in (a) Cartesian and (b) sheared coordinates with intrachain interactions $J$ and interchain interactions $J'$.}
        \label{fig:lattice}
\end{figure}

The spatial anisotropy is introduced by decomposing the lattice into
chains with bonds of strength $J$, arranged into a parallel array,
with inter-chain interactions of strength $J'$ (see
Fig.~\ref{fig:lattice}). Here, we define $R\equiv 1-J'/J$ as the degree of
anisotropy, and $h$ measures the applied magnetic field.
There have been many extensive studies that consider these effects separately.
However, a two-dimensional (2d) phase
diagram, taking both effects together, remains to be understood.
This problem is of considerable experimental interest.  The
application of a magnetic field is one of the few general means to
tune quantum magnets {\sl in situ}, and provides very important
information on the quantum dynamics, as well as clues to the
underlying spin Hamiltonian, which is often not well-known.
 Two materials whose behavior in magnetic fields has
been extensively studied are Cs$_2$CuCl$_4$ and Cs$_2$CuBr$_4$, which
are known to be approximately described by the spatially anisotropic
version of the model, with larger anisotropy in the chloride ($R
\approx 0.7$) than the bromide ($R\approx 0.3-0.5$).  Both materials
exhibit a rich structure of multiple phases in applied magnetic
fields, for which a theoretical view of the phase diagram would be
quite helpful.

\begin{figure}[t]
  \begin{center}
  \scalebox{0.7}{\includegraphics[width=\columnwidth]{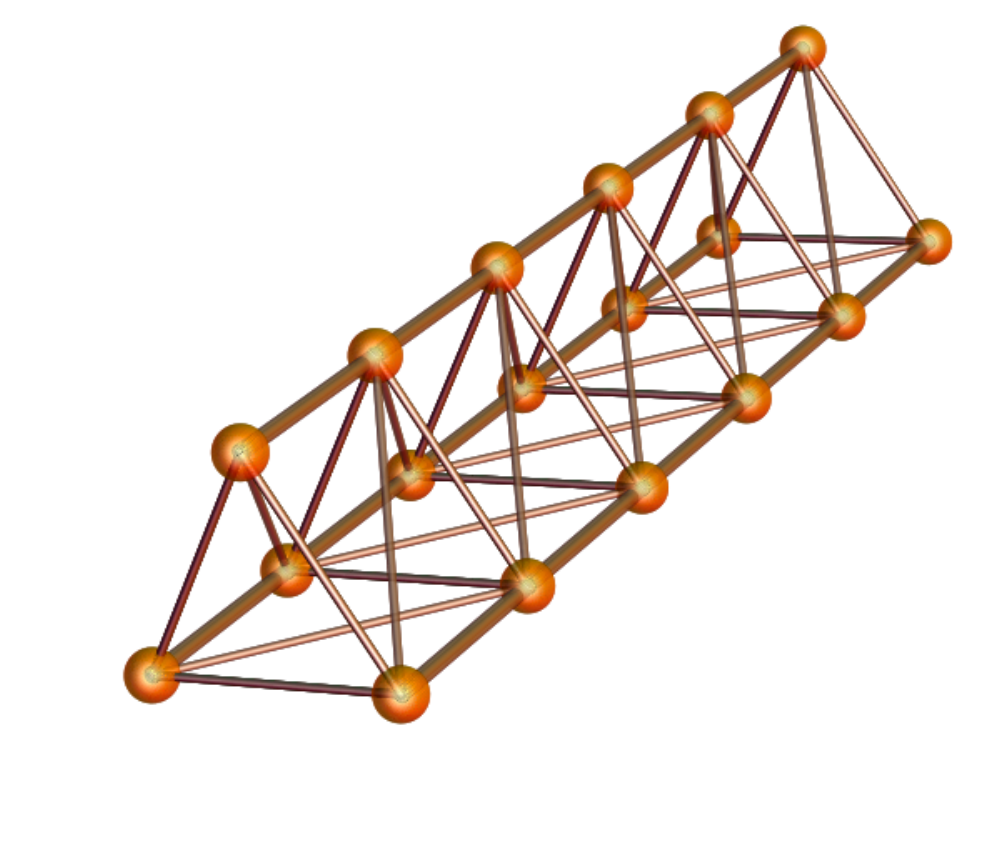}}
  \end{center}
  \caption{TST in sheared coordinates with period of three lattice spacings in the $y$-direction.
        It is {\sl crucial} to note that this geometry allows one to write
        $\sum_{y=1}^3 \sum_x  \mathcal{O}_y \mathcal{O}_{y+1} =  \sum_{y=1}^3 \sum_x  \mathcal{O}_y \mathcal{O}_{y+2}$ for an operator $\mathcal{O}$.}
  \label{fig:lattice2}
\end{figure}

\begin{figure}[t]
  \begin{center}
  \scalebox{1}{\includegraphics[width=\columnwidth]{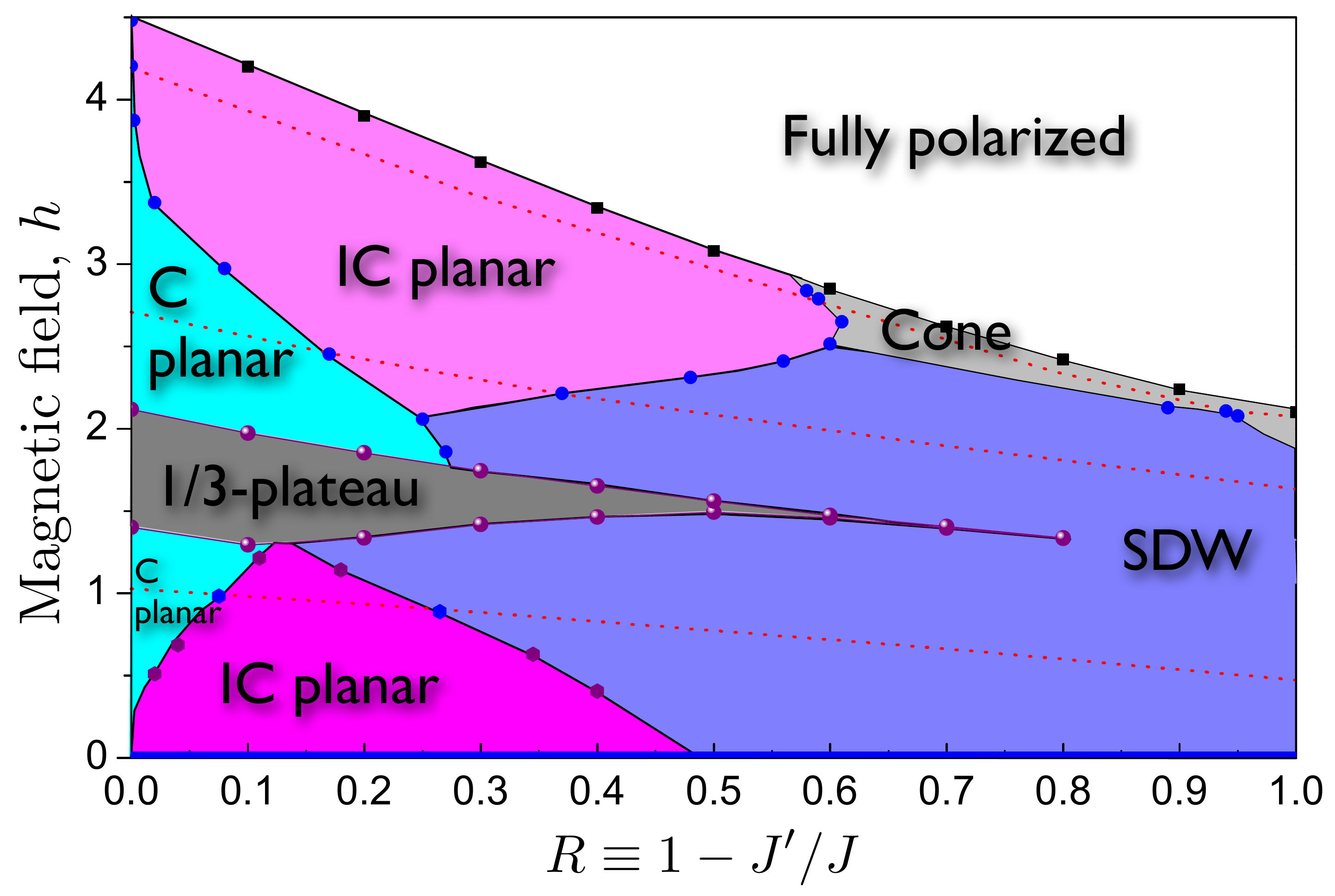}}
  \end{center}
  \caption{(Color online) Phase diagram for the spatially anisotropic spin-1/2 TST in a magnetic field.
        Here, we use the following abbreviations to label the various phases of the diagram:
        C = commensurate; IC = incommensurate; SDW = spin density wave. $R\equiv 1-J'/J$ is the degree of anisotropy.
        The dashed lines indicate constant magnetization lines, where the upper, middle and lower ones are at $M/M_s=5/6,1/2$ and $1/6$, respectively.}
\label{fig:phase}
\end{figure}

The solution of the ground state of a fully two dimensional frustrated
quantum spin model in a two-parameter phase space is quite ambitious.
Here, we consider a somewhat simpler task, by concentrating on the
problem defined by the model confined to a cylinder with a
circumference of three lattice spacings (i.e. making $y$ periodic with
period 3), which we refer to as a {\sl Triangular Spin Tube}, or TST (see Fig.~\ref{fig:lattice2}).
By a combination of analytical approaches and extensive numerical
simulations using the Density Matrix Renormalization Group \cite{DMRG} (DMRG), we
reveal a rich and complex phase diagram for the TST, shown in
Fig.~\ref{fig:phase}.  We argue in the
Discussion (Sec.\ref{sec:discussion}) that much of this diagram translates to the fully
2d model.  Whenever possible, we use a nomenclature for
the ground state phases which translates directly to two dimensions,
though there are, of course, differences due to the absence of
spontaneously broken continuous symmetry in one dimension.

Different parts of this phase diagram will be discussed in detail in
the bulk of the paper, but we will highlight a few aspects here, where
strong quantum features occur.  First, the isotropic line, $R=0$, as
a function of magnetic field has been considered many times in the
two-dimensional limit.  There, semi-classical
methods\cite{chubukov1991quantum,alicea2009quantum}
predict the stabilization of both coplanar spin configurations by quantum
fluctuations, and, most interestingly, a magnetization plateau, at
which the magnetization of the system is fixed (at $T=0$) at 1/3 of
the saturation magnetization over a range of magnetic fields. We will refer to this state as the ``1/3 plateau" throughout this text.  On the
plateau, the spins order into a collinear configuration.
Stabilization of such a plateau is very much a quantum effect and is one
of the more striking quantum features of the TST.  The presence of the
plateau has been confirmed 
for both the one-dimensional \cite{okunishi03,dagotto2007,hikihara2010} and the two-dimensional spin-1/2 Heisenberg
models by exact diagonalization \cite{honecker2004magnetization}, coupled-cluster \cite{farnell2009} and variational \cite{tay2010variational} methods.  
Our DMRG study of the TST is
also consistent with the semi-classical picture along the $R=0$ line.
We directly confirm the two ``coplanar'' phases, and accurately locate
the boundaries of the 1/3 plateau.
%
%

Another regime of strong quantum fluctuations occurs when $R$ is close
to 1, where the system is composed of weakly coupled (strictly) one
dimensional (1d) chains.  There, an approach based on scaling and
bosonization methods is possible, following
Refs.~\onlinecite{starykh2010extreme, starykh2007ordering}.  Those techniques
(explained in this context in Sec.~\ref{sec:weak-coupled}) predict a {\sl spin density
  wave} (SDW) state over a wide range of applied fields.  In this SDW
state, the dominant spin correlations are those of the Ising
component parallel to the field, in sharp contrast to the classical
behavior.  Our DMRG simulations show that the SDW state dominates a
remarkably broad region of the phase diagram, extending far beyond
 the decoupled line, $R=1$.

In two dimensions, the quasi-1D approach of
Refs.~\onlinecite{starykh2010extreme,starykh2007ordering} shows the
existence of a (very narrow) 1/3 plateau arising out
of the SDW phase, leading to the speculation that the plateau persists
for all $R$ in two dimensions.  In the TST, we find that the plateau
is also very robust, and persists almost, but not quite, to the
1D limit.  The suppression relative to two dimensions can
be understood as a result of enhanced fluctuations due to the
one-dimensionality of the TST.  To check this, we have also carried
out some DMRG studies of wider cylinders consisting of 6 and 9 sites
in the periodic direction.  Our results appear consistent with the
existence of a plateau for all $R$ in two dimensions.

The last quantum regime we discuss here is clearly specific to the
periodic boundary conditions imposed around the TST.  This occurs at
zero field, where for all values of $R$, we observe a spontaneously
dimerized ground state.  The dimerization is most clearly observed in
the entanglement entropy, which shows a pronounced oscillatory
behavior along the chain.  We argue that this can be understood as an
effect of one-dimensional quantum fluctuations upon an underlying
short-range spiral magnetically ordered state, somewhat similar to the
formation of a Haldane gap in integer spin chains with collinear
classical states.  The elementary excitations of the dimerized state
are solitons, and we show how the behavior at small magnetization can
be understood in terms of a dilute system of such solitons.

%
%

The remainder of the paper is organized as follows.
In Sec.~\ref{sec:DMRG} we introduce the model and then describe key technical aspects
of our DMRG simulations, including the procedure to
determine the phase boundaries using the second derivative of the
ground state energy and entanglement entropy, and careful finite size
scaling.
In Sec.~\ref{sec:iso}, we review and compare the semi-classical
predictions to the DMRG results in the isotropic limit.  Next, we
discuss the high field region in Sec.~\ref{sec:high field}.  In the
vicinity of the saturation field, the problem can be modeled as a
dilute system of spin-flip bosons.  We compare an analysis of this
limit, built upon an analytic solution of the Bethe-Salpeter equation,
to the DMRG, and find a transition between coplanar and cone phases,
and a commensurate-incommensurate transition.  In
Sec.~\ref{sec:weak-coupled}, we study the regime of weakly coupled
chains, and in particular discuss the spin density wave (SDW) state
and show that the 1/3 plateau terminates in a Kosterlitz-Thouless
transition around $R\sim 0.7\pm0.1$ for the TST. We consider the low
field region in Sec.~\ref{sec:lowfield}, showing the persistent
dimerization, the evidence for solitons at small magnetization, and
the commensurate to incommensurate transition near $R=0$.  DMRG
numerical results will be presented throughout these sections,
presenting the important features used to identify each
phase. Physical quantities, like entanglement entropy, vector
chirality and the spin density profile will be shown for some
representative large system size.
Finally, we conclude in Sec.~\ref{sec:discussion} with a summary and
discuss some generalizations of our results to larger spin and two-dimensional systems.
%
%

\section{Model and DMRG method}
\label{sec:DMRG}


\subsection{Hamiltonian and notation}
\label{sec:hamiltonian-notation}

The explicit Hamiltonian studied in this paper is written as
\begin{eqnarray}
\label{eq:hami}
H & = & \sum_{x,y} \left[ J \, \mathbf{S}_{x,y} \cdot
\mathbf{S}_{x+1,y} + J' \, \mathbf{S}_{x,y} \cdot \left(
  \mathbf{S}_{x,y+1}+ \mathbf{S}_{x-1,y+1} \right) \right] \nonumber\\
&& - h \sum_{x,y} \mathbf{S}_{x,y}^z,
\end{eqnarray}
where $x$ is the direction along the chains, and $y$ is perpendicular
to it, and $h$ is the magnetic field.  Importantly, we choose
coordinates, as shown in Fig.~\ref{fig:lattice}b, where the triangular
lattice is ``sheared'' to embed it in a square one.  This is convenient for the
application of periodic boundary conditions in the TST.

Many previous works on the anisotropic triangular lattice in two
dimensions, including those by some of the
authors\cite{starykh2010extreme,schnyder2008spatially}, use
instead ``cartesian'' coordinates, as shown in in
Fig.~\ref{fig:lattice}a.  Both for convenience in certain calculations
(especially in the quasi-one-dimensional limit), and to clarify the
connection to this prior work, we give the relation between the
sheared and cartesian coordinates here.  In cartesian coordinates, we
take the distance between sites along the chains and the (normal)
distance between chains to unity.  Defining the cartesian coordinates
as ${\sf x}, {\sf y}$, and ${\bf\sf r}=({\sf x}, {\sf y})$, then
\begin{equation}
  \label{eq:9}
  {\sf x} = x + y/2, \qquad {\sf y} = y.
\end{equation}
 From this, we may also obtain the relationship between wavevectors in
 the two coordinate frames.  We require ${\bf q}\cdot {\bf r} =
 {\bf\sf q}\cdot {\bf\sf r}$, which implies
 \begin{equation}
   \label{eq:10sf}
   q_x = {\sf q}_x, \qquad q_y =\tfrac{1}{2} {\sf q}_x +   {\sf q}_y.
 \end{equation}
\begin{figure}[t]
  \begin{center}
  \scalebox{0.95}{\includegraphics[width=\columnwidth]{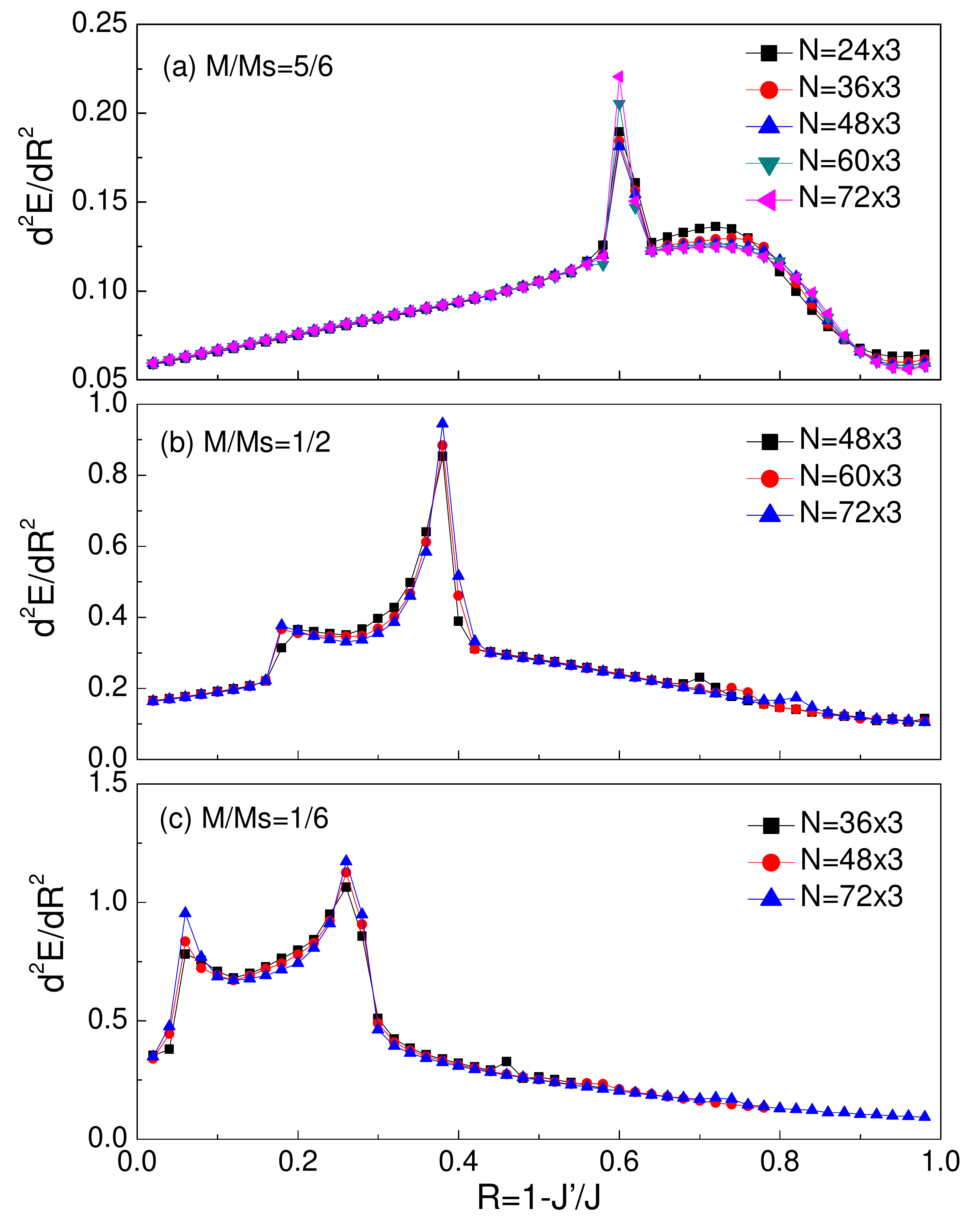}}
  \end{center}
  \caption{(Color online) Second derivative of the ground state energy with respect to
    $R$, for different values of magnetization.
    These plots are used to locate the phase boundaries in Fig.~\ref{fig:phase}.}
  \label{fig:dE2}
\end{figure}

\subsection{DMRG}
\label{sec:dmrg}

Throughout this paper, we rely extensively on DMRG simulations.
For the present study, we kept up to $m=3072$ states in the DMRG
block, performing more than 24 sweeps to obtain fully converged
results. In doing so, we find that our truncation error is of the order $10^{-7}$.
We also take advantage of the cylindrical boundary condition to study large systems and to reduce finite-size effects
for a more reliable extrapolation to the thermodynamic limit.
In particular, in the regions above the 1/3 plateau, we find that observables have much better convergence, with a truncation error of the order $10^{-9}$.
Even in the regions below the 1/3 plateau not close to the dimerized phase, we find reasonable convergence,
with a slightly larger truncation error on the order of $10^{-7}$.
However, when we approach the dimerized phase near zero magnetization, finite-size effects dominate:
system sizes up to $N=180 \times 3$ do not provide a reliable extrapolation to the thermodynamic limit.

The phase boundaries in Fig.~\ref{fig:phase} were determined from
the simulations.  We describe the methodology for doing so here,
leaving the characterization of the phases which occur for
subsequent Sections.  For the case of continuous transitions, it is
common to calculate the second derivative of the ground state
energy, $\frac{\partial^2 E_0}{\partial R^2}$. The calculation follows standard procedure of using three data
points at $R+dR$, $R$, and $R-dR$, according to the formula
$\partial^2E_0/\partial R^2=[E_0(R+dR)+E_0(R-dR)-2E_0(R)]/dR^2$. The derivative
diverges when the infinite-size system undergoes a transition. 
For finite systems, however, one will observe a finite
peak that increases with system size. We then determine the
phase boundaries numerically by looking at the peak position as a
function of tuning parameter R. For example, as shown in Fig.4 (a),
sharp peaks are located at R=0.6. We observe that the
peak value increases significantly with sample size, for all system
sizes studied.   We have not attempted to
carry out detailed finite size scaling analyses of the peaks, as our
focus here is on the phases, not the critical behavior at the transitions between them.  
This transition corresponds to the upper dashed line in
Fig.~\ref{fig:phase}, where there is a transition between an
incommensurate planar and a cone phase.  We use similar procedures to
determine phase boundaries at other magnetizations, e.g. $M/M_s = 1/2,
1/6$ in Figs.~\ref{fig:dE2}(b,c) correspond to the middle and lower
dashed lines in Fig.~\ref{fig:phase}, respectively.

In addition to these divergent peaks, there are some other features
(which are {\sl{not}} phase transitions) due to finite size
effects. For example, in Fig.~\ref{fig:dE2}(a) for $M/M_s = 5/6$, a
broad peak near $R = 0.8$ actually decreases (and eventually goes to
zero) in the thermodynamic limit. Therefore, we can confidently say
that the cone phase dominates in the region $R > 0.6$, and that there
is no transition at $R = 0.8$. Similarly, for
Figs.~\ref{fig:dE2}(b,c), the fluctuations in the plots near $R
\approx 0.7, 0.45$, respectively, are finite size effects and vanish
in the thermodynamic limit.

Finally, we use the structure factor
\begin{equation}
\label{eq:strfac}
S^{\mu \mu}(q)=\frac{1}{N}\sum_{{\bf r},{\bf r}'} e^{-i {\bf q} \cdot ({\bf r}-{\bf r}')} \langle S_{\bf r}^{\mu} S_{{\bf r}'}^{\mu} \rangle .
\end{equation}
to determine the boundaries between the commensurate and incommensurate phases.
For example, for small $R$, the transverse and longitudinal components of the structure factor peak at commensurate momenta
${\bf Q}=(4\pi/3,2\pi/3)$ and $(2\pi/3,4\pi/3)$, respectively.
This defines the ``C planar" regions in Fig.~\ref{fig:phase}.

\section{Semi-classical behavior in the isotropic case}
\label{sec:iso}

\subsection{Two-dimensional model}

The isotropic model, $J'=J$, has been extensively studied in two
dimensions, and it is believed that a semi-classical
description, with weak quantum fluctuations included via spin wave
theory, is qualitatively correct in this case\cite{chubukov1991quantum}.  We find that the
semi-classical analysis largely carries over to the TST, with small
modifications to allow for one-dimensional fluctuations.  Therefore
we review the established semi-classical results first.

In the classical limit, where spins are described as O(3) vectors, the
isotropic problem is known to display an ``accidental" degeneracy in a
non-zero applied magnetic field \cite{kawamura1985}.  This can be seen from the fact that
this model can be rewritten as
\begin{equation}
    H = \frac{J}{2} \sum_{\bigtriangleup}  \left( \mathbf{S}_{\bigtriangleup} - \frac{h}{3J}\hat{\mathbf{z}} \right)^2,
\end{equation}
where $ \mathbf{S}_{\bigtriangleup} = \mathbf{S}_1 + \mathbf{S}_2 + \mathbf{S}_3$ is the sum of the spins on a triangle, and the sum is over all triangles on the lattice.
The ground state configuration is given by the constraint
\begin{equation}
     \mathbf{S}_{\bigtriangleup} - \frac{h}{3J} \hat{\mathbf{z}}= 0.
\end{equation}
At zero magnetization, this constraint is solved by placing all spins in a plane,
with the three spins in each triangle at $120^\circ$ angles to one
another in a three sublattice structure.  A specific ground state
is specified by three angles, e.g. two determining the plane of the
spins and one determining the angle within the plane.  All such states are
related by O(3) spin symmetry; so this is a symmetry-demanded
degeneracy. a previous DMRG study\cite{HCJiang2009} on the 2d model also confirms the three sublattice structure.

\begin{figure}[t]
  \begin{center}
  \scalebox{.8}{\includegraphics[width=\columnwidth]{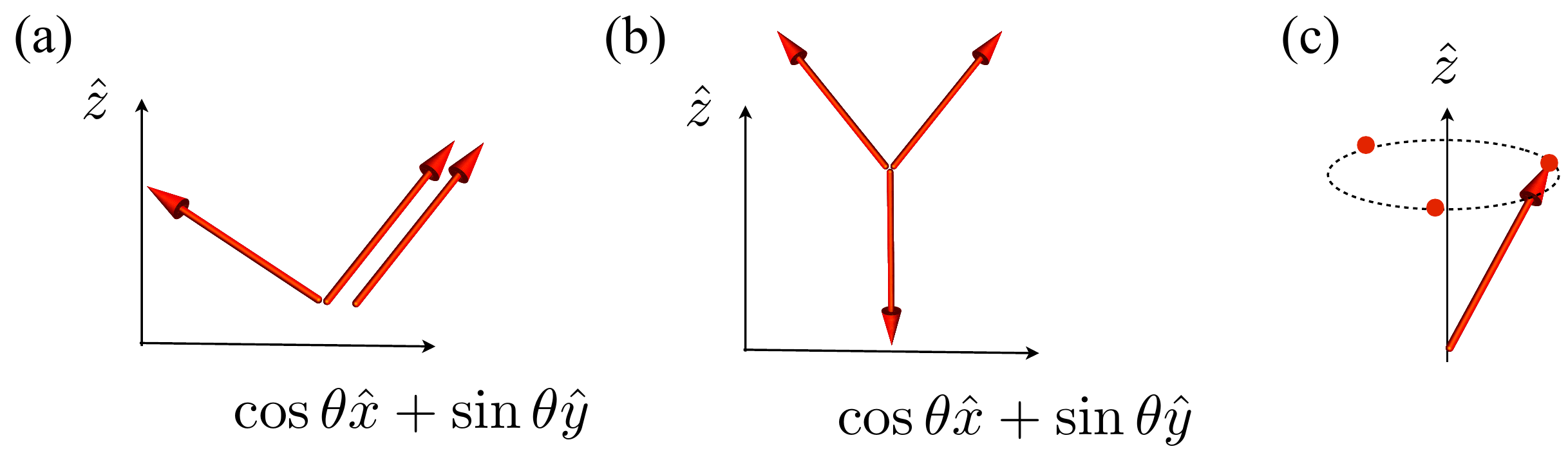}}
  \end{center}
  \caption{Degenerate classical spin configurations in the isotropic limit.
    With the magnetic field taken in the $z$-direction, (a) shows the ``V" configuration above the 1/3 plateau, (b) depicts the ``Y" phase below the 1/3 plateau while
    (c) shows the cone (or umbrella) state.}
  \label{fig:comm-planar}
\end{figure}

In a non-zero field, the ground states retain a three-sublattice
structure, with three arbitrary angles remaining to determine the
specific ground state.  However, the presence of the field reduces the
O(3) symmetry to O(2) (or U(1)), and only one of these angular degrees
of freedom is symmetry demanded.  The remaining two angular degrees of
freedom constitute an {\sl accidental} degeneracy.  Two simple states
within the degenerate manifold are the coplanar and umbrella ones,
shown in Fig.~\ref{fig:comm-planar}.

%
%

As first shown by Chubukov and Golosov~\cite{chubukov1991quantum}, this
accidental degeneracy is lifted by quantum fluctuations.  They showed
by a $1/S$ spin wave expansion that the degeneracy is lifted in favor
of the coplanar states.  Additionally, they demonstrated the existence
of the 1/3 plateau, in which the spins adopt a 3 sublattice ``up up down''
structure.    Away from the plateau, the coplanar state retains a 3
sublattice structure with ordering wavevector ${\bf\sf Q} =
(4\pi/3,0)$, or ${\bf Q} = (4\pi/3, 2\pi/3)$~\cite{chubukov1991quantum,alicea2009quantum,griset2011deformed}.
Below the plateau, the 3 spins form a ``Y'' with one spin antiparallel
to the field and two spins with equal positive projection to the field
but at opposite angles from each other.  This can be viewed as a
deformation of the $120^\circ$ state with spins in a plane containing
the magnetic field.  Here the spin configurations can be parametrized
by
\begin{eqnarray}
\label{eq:comm-planar1}
\langle S_{\mathbf{r}}^+ \rangle & = & a \, e^{i \theta} \sin \left( {\bf Q} \cdot {\bf r}  \right) \nonumber \\
\langle S_\br^z \rangle & = & b - c \cos^2 \left( {\bf Q} \cdot {\bf r} \right),
\end{eqnarray}
where $\theta$ is an arbitrary angle specifying the plane of the spins,
 while $a,b,c$ are constants dependent upon the
field magnitude.  Since $ {\bf Q} \cdot {\bf r} = 2\pi (2x+y)/3$, we see
from Eq.~\eqref{eq:comm-planar1} that when $2x+y$ is a multiple of 3, one of
the spin is aligned with the magnetic field.  Above the plateau one finds instead a ``V'' configuration,
with two spins identical and the third chosen to give zero moment
normal to $z$. In this case, we have
\begin{eqnarray}
\label{eq:comm-planar2}
\langle S_\br^+ \rangle & = & a \, e^{i \theta} \cos \left( {\bf Q} \cdot {\bf r}  \right)   \nonumber \\
\langle S_\br^z \rangle & = & b - c \cos^2 \left( {\bf Q} \cdot {\bf r} \right).
\end{eqnarray}
Note that the cosine in the first line of Eq.~\eqref{eq:comm-planar2}
never vanishes on lattice sites, so that spins are never parallel to the
field in the V state.

\subsection{One dimension}
\label{sec:one-dimension}

We will see that the semi-classical results summarized in the previous
subsection for the two-dimensional case remain qualitatively correct, at least at
short distances, in the TST.  However, we must still account for the
effects of quantum fluctuations on long length scales, since the one
dimensional system {\sl cannot} break the U(1) spin-rotational
symmetry about the field axis.   Since the U(1) symmetry is unbroken
in the plateau state, there are no essential effects of
one-dimensional fluctuations there.  However, they have qualitative
effects in the Y and V phases, since $\langle S_r^+ \rangle=0$ there,
in contrast to Eqs.~(\ref{eq:comm-planar1},\ref{eq:comm-planar2}).
Note that the modulation of $\langle S^z_r \rangle$ is perfectly
consistent with one-dimensionality, and is expected to persist
directly without qualitative modifications.

To incorporate one-dimensional fluctuations, we regard the
semiclassical results in
Eqs.~(\ref{eq:comm-planar1},\ref{eq:comm-planar2}) as defining the
local spin ordering, with a {\sl fluctuating} quantum phase
$\theta(x,\tau)$ ($\tau$ is imaginary time), that is, we make the
replacement
\begin{equation}
  \label{eq:1}
  S_r^+(\tau) \rightarrow a \, e^{i \theta(x,\tau)} \sin \left( {\bf
      Q} \cdot {\bf r}  \right),
\end{equation}
in the Y phase, and
\begin{equation}
  \label{eq:2}
   S_r^+(\tau) \rightarrow a \, e^{i \theta(x,\tau)} \cos \left( {\bf
      Q} \cdot {\bf r}  \right),
\end{equation}
in the V phase.  Note that these formulae are {\sl not} invariant
under translations, reflecting the three-sublattice structure
of the coplanar phases.  This can also be seen from the oscillations in
the $\langle S^z_r\rangle$ expectation values. Even when
one-dimensional fluctuations are taken into account, translational
symmetry is broken.  This is still consistent with the Mermin-Wagner
theorem, since the broken translational symmetry is discrete.
Translating by one or two lattice spacings, one obtains two other
symmetry related but distinct ground states.

In both the Y and V
phases, the field $\theta(x,\tau)$, representing the ``would-be"
Goldstone mode of the spontaneously broken U(1) symmetry, is governed
by the usual massless free relativistic boson action,
\begin{equation}
  \label{eq:3}
  S_\theta = \int \! dx d\tau  \left\{ \frac{v K}{2} (\partial_x
    \theta)^2 + \frac{K}{2v} (\partial_\tau \theta)^2\right\}.
\end{equation}
%

\subsection{Comparison to DMRG}

We now turn to a comparison of the semi-classical predictions,
corrected as in the previous subsection for one-dimensional
fluctuations, to the DMRG.

\subsubsection{Entanglement entropy}

The simplest comparison arises immediately
from Eq.~\eqref{eq:3}: the low energy physics is that of a single
massless scalar field, which is a conformal field theory with central
charge $c=1$.  This central charge can be directly measured using the
entanglement entropy.

\begin{figure}[t]
  \begin{center}
  \scalebox{0.95}{\includegraphics[width=\columnwidth]{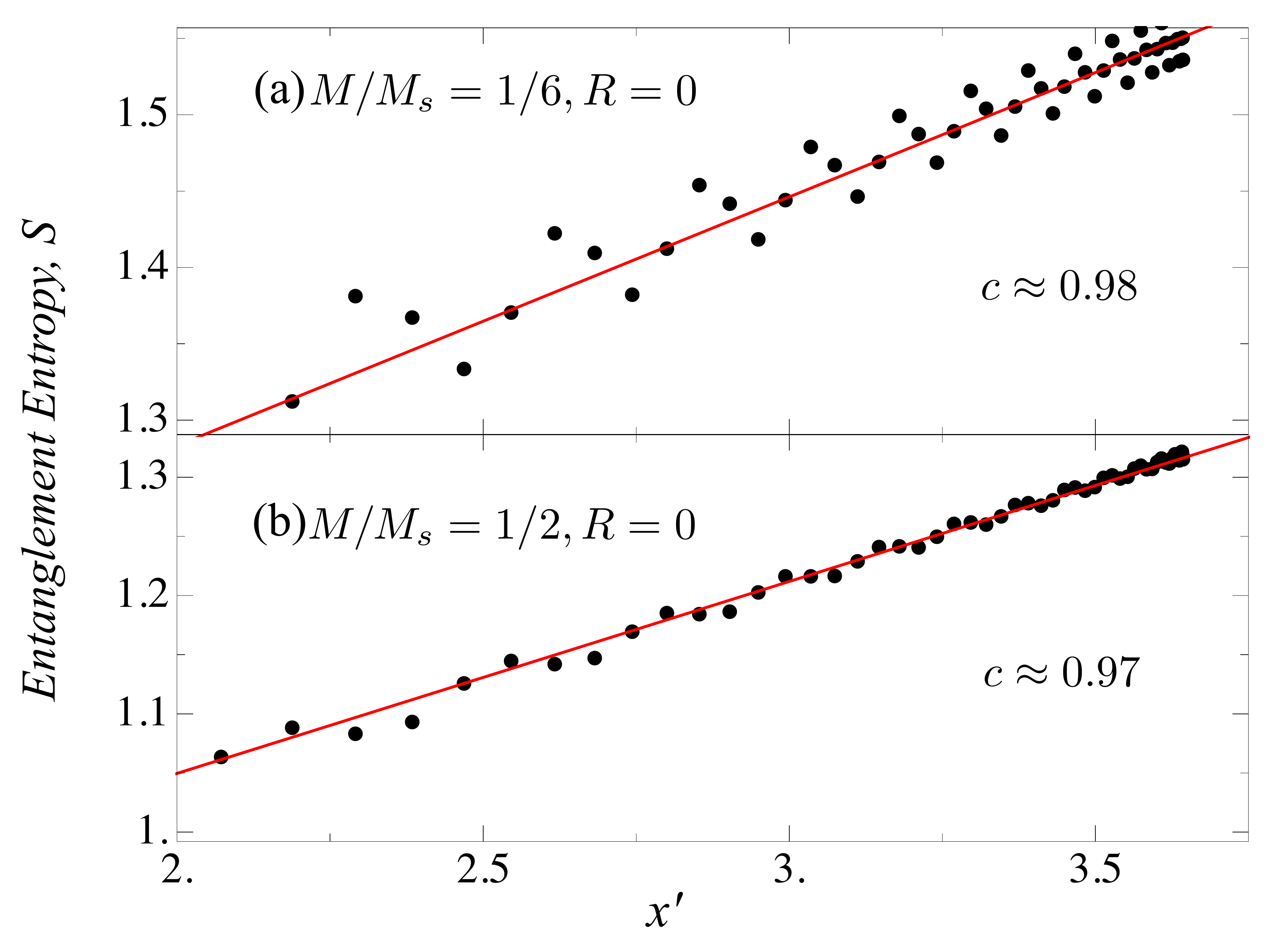}}
  \end{center}
  \caption{Entanglement entropy in the isotropic limit, $R = 0$ for system size $N_x = 120$.
        Note that the reduced coordinate $x' \equiv \ln \left[ \frac{L}{\pi} \sin (\frac{\pi x}{L}) \right]$ is plotted on the $x$-axis.
        We show the von Neumann entanglement entropy for (a) $M/M_s = 1/6$, the commensurate Y phase and (b) $M/M_s = 1/2$, the commensurate V state.
        The solid line is a linear fit, where by Eq.~\eqref{eq:ent}, we can extract the central charge, $c$.}
  \label{fig:EE_isotropic}
\end{figure}

According to conformal field theory\cite{Cardy}, in a one dimensional critical
system with open boundary conditions and total length $L$, the von Neumann entanglement entropy
associated to a region with length $x$ and its complement of length $L-x$ is given by
\begin{equation}
    \label{eq:ent}
    S(x,L) = \frac{c}{6} \ln\left[ \frac{L}{\pi} \sin\left( \frac{\pi x}{L} \right) \right].
\end{equation}
By plotting the entropy $S(x,L)$ versus the reduced coordinate
$x'=\ln[\frac{L}{\pi} \sin\left( \frac{\pi x}{L} \right)]$, we can
directly extract $c$ from the numerics.  As shown in
Fig.~\ref{fig:EE_isotropic}, we can indeed obtain $c=1$ with high
accuracy for both $\rm Y$ and $\rm V$ phases. For example, the
obtained central charge $c=0.98$ at $M/M_s=1/6$ in the $\rm Y$ phase
below the plateau, and $c=0.97$ at $M/M_s=1/2$ in the $\rm V$ phase
above the plateau. Both are consistent with the theoretical prediction.

\subsubsection{$S^z$ profile}
\label{sec:s_z-profile}

%
%

\begin{figure}[t]
  \begin{center}
  \scalebox{0.95}{\includegraphics[width=\columnwidth]{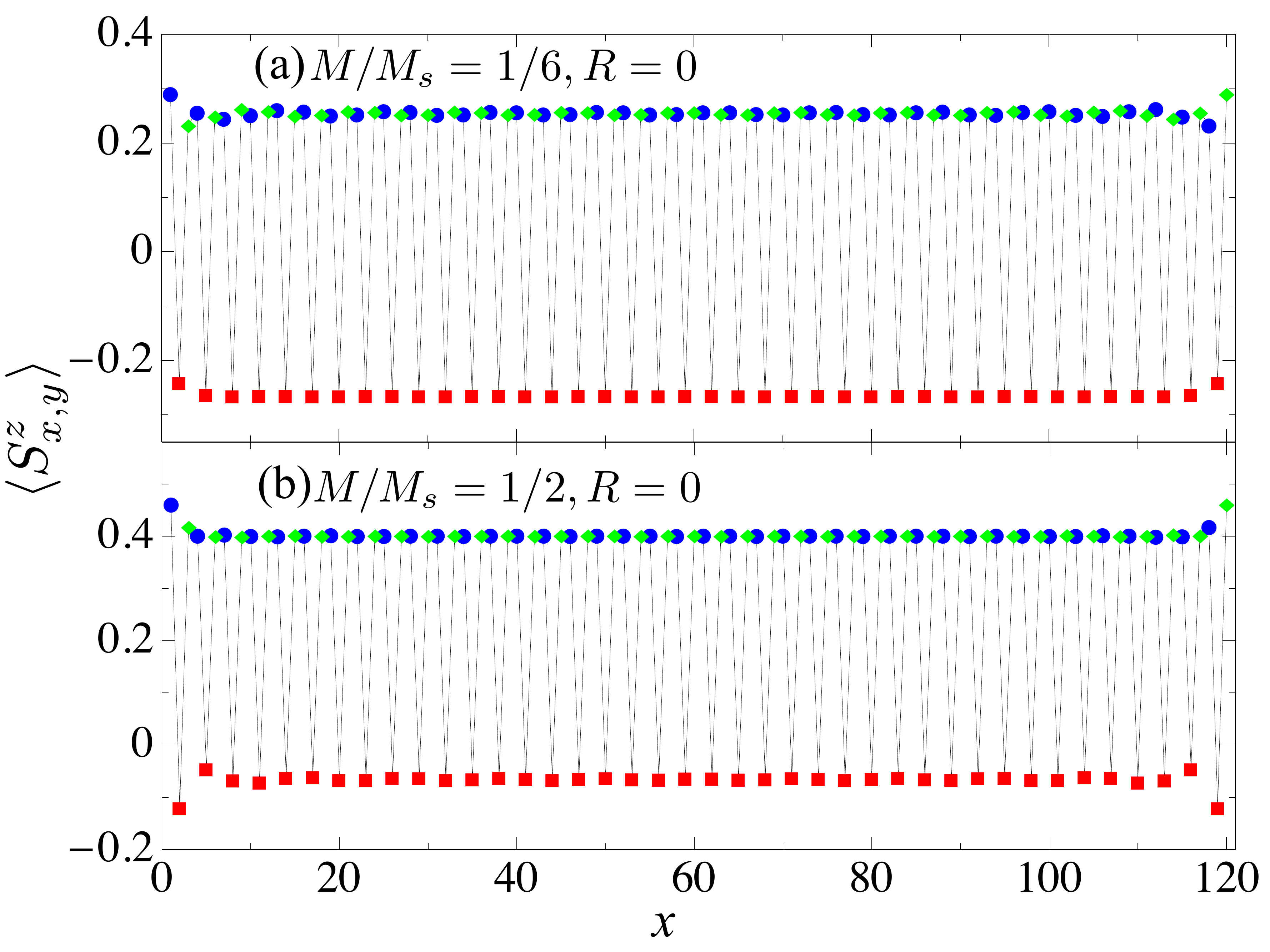}}
  \end{center}
  \caption{(Color online) $S^z$ profile in the isotropic limit, $R=0$ at (a) $M/M_s=1/6$, the commensurate Y phase, and (b) $M/M_s=1/2$, the commensurate V state.
        The square (red), diamond (green) and circle (blue) data points show the 3-sublattice structure of the isotropic case.
        The magnitude of $S^z$ does not decay because the discrete translational symmetry is spontaneously broken in these states.}
  \label{fig:Sz-iso}
\end{figure}

\begin{figure}[t]
  \begin{center}
  \scalebox{0.95}{\includegraphics[width=\columnwidth]{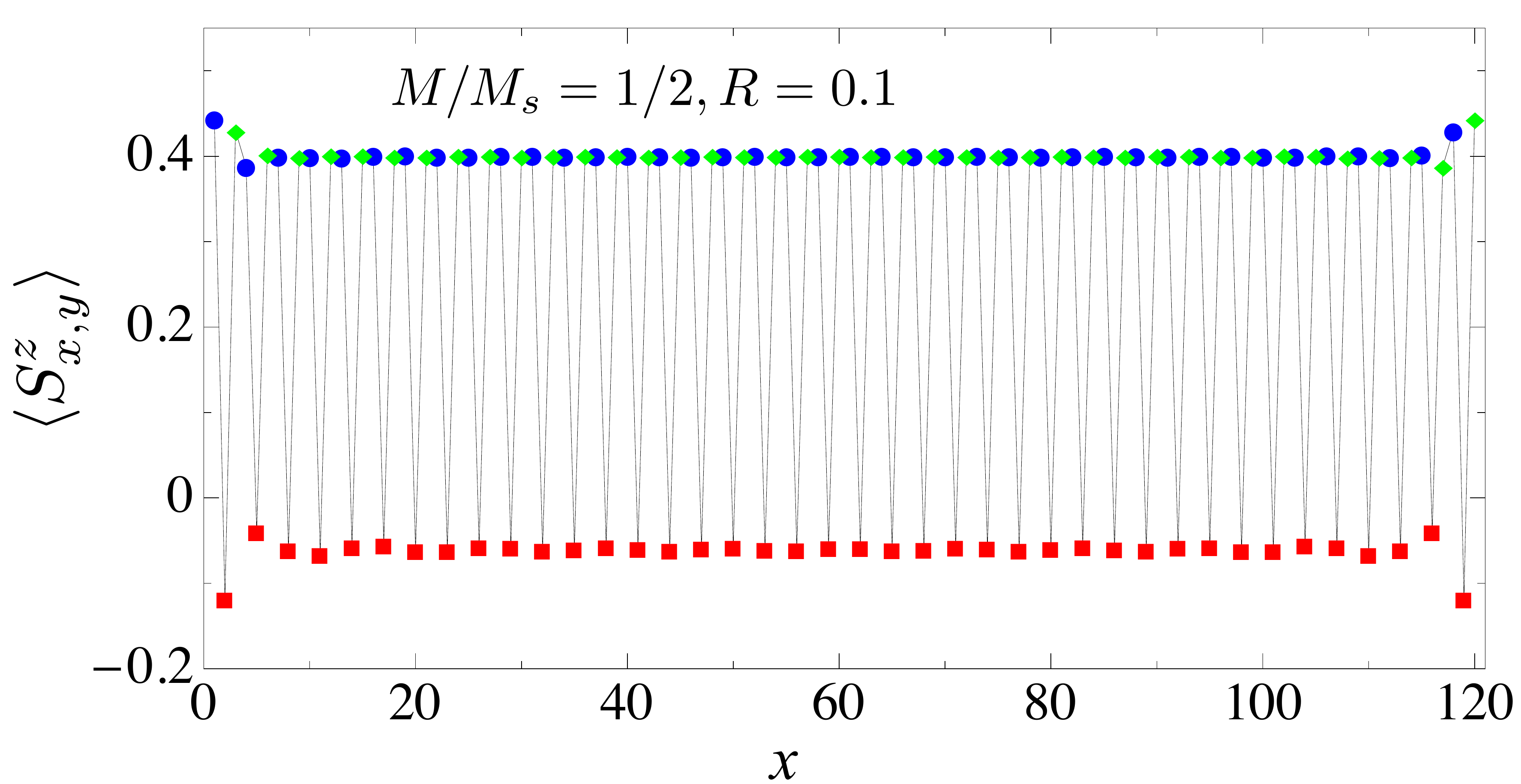}}
  \end{center}
  \caption{(Color online) $S^z$ profile for the commensurate V phase at $M/M_s=1/2$ and $R=0.1$.
        We find that the wave vector remains commensurate, even for a non-zero, but small $R$.}
  \label{fig:Sz-iso2}
\end{figure}

The modulation of $\langle S^z_\br\rangle$ predicted by the
semi-classical theory in
Eqs.~(\ref{eq:comm-planar1},\ref{eq:comm-planar2}) can be directly
compared to the DMRG results.  This is shown in Figs.~\ref{fig:Sz-iso},\ref{fig:Sz-iso2}.  Note that
a particular symmetry broken state is chosen in the simulations,
presumably due to pinning by the boundaries, which explicitly break
translational symmetry.  The origin of the coordinate ${\bf r}$ in
Eqs.~(\ref{eq:comm-planar1},\ref{eq:comm-planar2}) must be
appropriately chosen to match the chosen ground state.

\subsubsection{$S^\pm$ correlations}
\label{sec:spm-correlations}

\begin{figure}[t]
  \begin{center}
  \scalebox{0.95}{\includegraphics[width=\columnwidth]{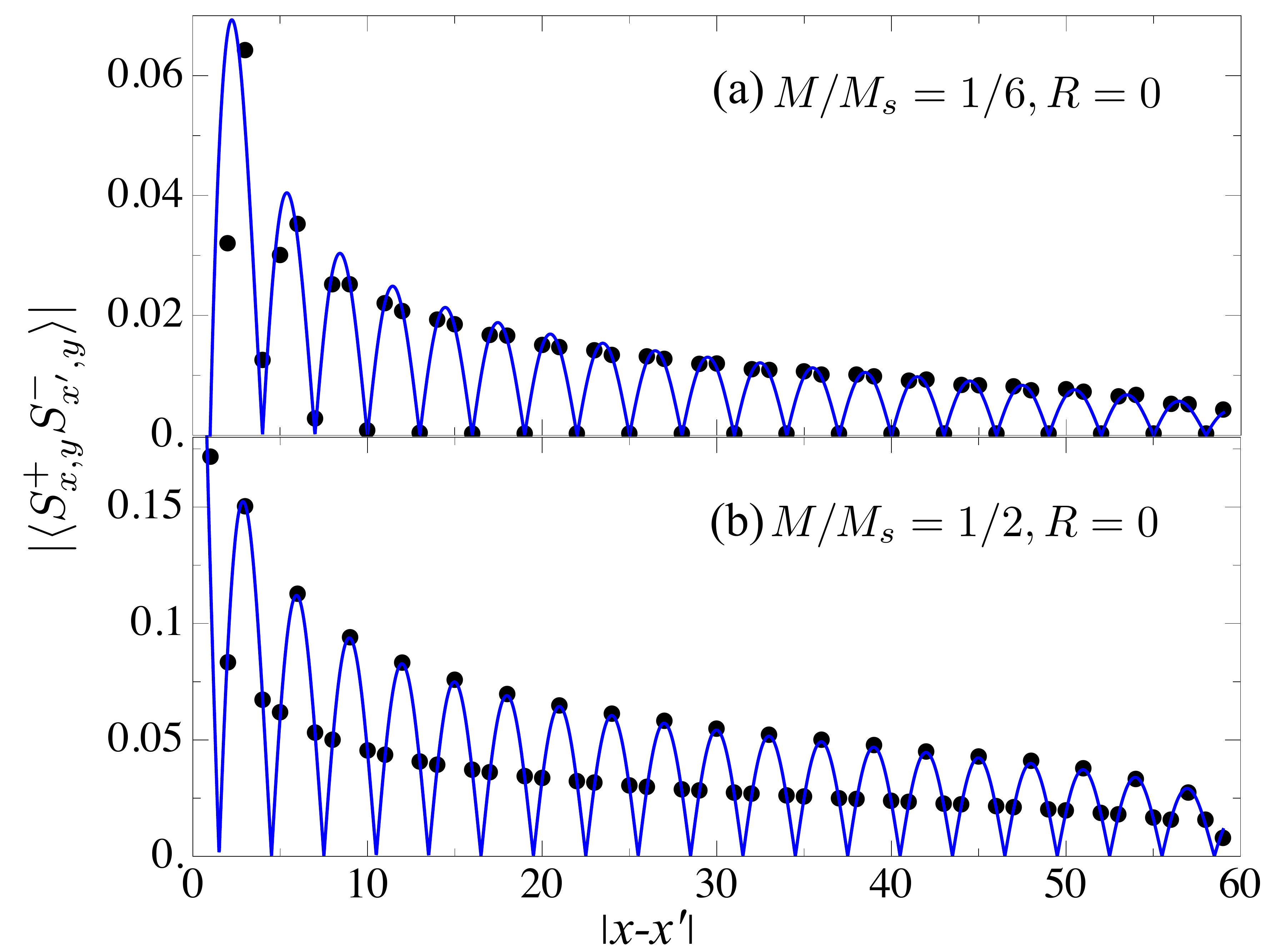}}
  \end{center}
  \caption{(Color online) Transverse spin-spin correlations in the isotropic limit, $R=0$ in the (a) commensurate Y phase and (b) V phase for $N_x = 120$ and $x'=N_x/2$. Data points are shown as (black) circles while the theoretical fit from Eq.~\eqref{eq:ruadd1} is shown as the (blue) line.}
  \label{fig:XY_isotropic}
\end{figure}

Due to quantum fluctuations of the phase $\theta$, the single spin
expectation value $\langle S_r^+\rangle=0$. Therefore, we must instead turn to
correlation functions to detect the Y and V structure of the local
ordering.  Using Eq.~\eqref{eq:1}, we obtain
\begin{equation}
  \label{eq:4}
  \langle S_\br^+ S_{\br'}^- \rangle \sim a^2 \sin  \left( {\bf
      Q} \cdot {\bf r}  \right) \sin  \left( {\bf
      Q} \cdot {\bf r}'  \right) \left\langle e^{i(\theta(x) - \theta(x'))}\right\rangle,
\end{equation}
%
%
%
in the Y phase below the 1/3 plateau. A similar formula, with the sines
replaced by cosines, describes the correlation function of the V phase above the plateau. The correlation function is evaluated with respect
to Eq.~\eqref{eq:3}, where a finite-size form, first derived in Ref. \onlinecite{hikihara2004correlation}, is as follows
\begin{eqnarray}
\label{eq:ruadd1}
\langle e^{i(\theta(x) - \theta(x'))}\rangle&=&C_\eta(x,x'),
\end{eqnarray}
where
\begin{eqnarray}
\label{eq:ruadd1a}
C_\eta(x,x')&=&a_0^\eta\frac{[f(2x)f(2x')]^{\eta/2}}{[f(x-x')f(x+x')]^{\eta}} ,\\
f(x)&=& \left[ \frac{2(L+1)}{\pi}\sin\left(\frac{\pi|x|}{2(L+1)}\right)\right].\nonumber
\end{eqnarray}
Here $a_0$ is a cut-off dependent factor, which we can take to unity,
absorbing the dependence in $a$ in Eq.~\eqref{eq:4}.  The function,
$f(x)$, originates from a quantum average over the normal modes of the bosonic field $\theta$.
One is now able to fit the DMRG measurement of the transverse spin-spin correlation function to Eqs.~(\ref{eq:4},\ref{eq:ruadd1})
to obtain the ordering wave vector and the additional fit parameter, $\eta$. A comparison is plotted in Fig.~\ref{fig:XY_isotropic}, where we show the correlation function along each chain (i.e., $y=1,2,3$) for $R=0$ and $M/M_s = 1/6, 1/2$.
The fitting in Fig.~\ref{fig:XY_isotropic}a yields a commensurate wave vector ${\bf Q}=(4\pi/3,2\pi/3)$ and $\eta=0.65$ for $M/M_s=1/6$, which corresponds to the Y phase below the plateau. Above the plateau, in the $V$ phase shown in Fig.~\ref{fig:XY_isotropic}b, the ordering wave vector still shows commensurability, ${\bf Q}=(4\pi/3,2\pi/3)$ with $\eta=0.43$. One can show that in the thermodynamic limit, the correlation function in Eq.~\eqref{eq:ruadd1} reduces to a simple power-law relation $\propto |x-x'|^{-\eta}$, which is reflected by our data for distances $|x-x'|\ll L/2$.

%
%

\subsection{Behavior for small non-zero $R$}
\label{sec:behavior-small-non}

If we perturb slightly away from the isotropic limit, i.e. $0<R\ll 1$, we expect the
semi-classical picture to still hold.  This has been analyzed in
Refs.~\onlinecite{alicea2009quantum, griset2011deformed}.  Classically,
the minimum energy spin configuration changes immediately when $R>0$
from a commensurate state to an incommensurate one, with an ordering wavevector
${\bf\sf Q} \neq (4\pi/3,0)$ or ${\bf Q} \neq (4\pi/3, 2\pi/3)$.  However, we expect that quantum
fluctuations will stabilize the commensurate
state for a range of anisotropies for a
generic value of the magnetic field.  The reason is that coplanar
phases break discrete translational symmetries of the lattice.  Since
there are three equivalent ground states connected by translations, the
symmetry breaking can be described by a $\mathbb{Z}_3$ order
parameter.  Specifically, the combination
\begin{equation}
  \label{eq:5}
  \zeta_r = S_r^z e^{2\pi i (x+2y)/3},
\end{equation}
defines a $\mathbb{Z}_3$ order parameter with $\langle \zeta_r\rangle = |\zeta|
e^{i\vartheta}$ and $\vartheta=0,2\pi/3, 4\pi/3$ in the three distinct
$\mathbb{Z}_3$ domains.
To restore this discrete symmetry, a phase transition is required.
More specifically, there are topological
excitations of the coplanar state which are domain walls, also called
solitons, connecting different symmetry broken states.  There is a
non-zero energy gap to create a domain wall in any phase with long-range
$\mathbb{Z}_3$ order.  For the $\mathbb{Z}_3$ order to be destroyed,
solitons must proliferate in the ground state.  Small changes of
parameters, such as $R$, cannot instantly lower the gap for the domain
walls to zero, which implies stability of the phase for a range of $R$
values.  This is correct, at least, away from the exceptional points where $h=0$
(where the symmetry breaking becomes continuous) and $h=h_{\rm sat}$
(where the symmetry breaking vanishes).  We will discuss the vicinity of
these exceptional points in subsequent sections.
%
%

In general, with increasing anisotropy, $R$, we will encounter a phase
transition to an incommensurate phase, which corresponds to the
proliferation of solitons and a vanishing of their gap.  Beyond that
point, $\langle \zeta_r\rangle$ becomes zero, and $S^z$ correlations
peak at a wavevector other than $\mathbf{Q}=(4\pi/3,2\pi/3)$.   This
transition is discussed in Sec.~\ref{subsec:incomm-comm}.

A useful test for this phase is the measurement of the central charge via
entanglement entropy.  In the commensurate regions, even for $R>0$,
we expect $c=1$, while incommensurate phases may have $c>1$.
We observe this effect in Fig.~\ref{fig:EE_isotropic}, which shows $c=1$ in the commensurate state,
whereas Fig.~\ref{fig:cone-planar} shows $c=2$ in the incommensurate state.
In addition, we can check for commensurability using structure factor measurements, as discussed in Sec.~\ref{sec:dmrg}.

\subsection{Phenomenological analysis at low field}
\label{sec:CI}

We now address the region slightly away from $R=0$ and at low applied
magnetic field.  We begin the discussion from a 2d
point of view, though it largely applies to the TST as well.
Commensurate coplanar spin order is described by the order parameter
${\bf d} = {\bf n}_1 - i {\bf n}_2$, where ${\bf n}_1$, ${\bf n}_2$
are mutually orthogonal vectors with identical norm spanning the plane
of the spin order.  Then, a spin at coordinate $\br$ can be written as
\begin{equation}
\label{eq:CI1}
{\bf S}_\br = M + {\rm Re}({\bf d} e^{ i {\bf Q}\cdot{\bf r}}) = M + {\bf n}_1 \cos[{\bf Q}\cdot{\bf r} ] + {\bf n}_2 \sin[{\bf Q}\cdot{\bf r}].
\end{equation}
Lattice translations transform ${\bf d} \to {\bf d} e^{-i 2\pi/3}$,
while lattice inversion, ${\bf r} \to -{\bf r}$, results in complex
conjugation, ${\bf d} \to {\bf d}^*$. The effective Ginzburg-Landau
Hamiltonian describing the coplanar state should remain invariant
under these operations (see Ref.~\onlinecite{mzh1996} for a closely
related discussion). Then,
\begin{eqnarray}
H_{\rm comm} &=& - r {\bf d}^* \cdot {\bf d} + a_0 |\partial_x {\bf d}|^2 + a_1 ({\bf d}^* \cdot {\bf d})^2 + a_2 |{\bf d} \cdot {\bf d}|^2 \nonumber\\
&& + \chi_1 h^2 {\bf d}^* \cdot {\bf d} + \chi_2 |{\bf h} \cdot {\bf d}|^2 \nonumber\\
&&+ \frac{1}{2}\chi_3 [ ({\bf h} \cdot {\bf d})^3 + ({\bf h} \cdot {\bf d}^*)^3].
\label{eq:Hcomm}
\end{eqnarray}
Here, at mean-field level, $r>0$ is required to obtain non-zero ${\bf n}_{1,2}$, and
$a_{0,1} >0$, for stability in the ordered phase.  Furthermore,
$a_2 > 0$ energetically imposes the orthogonality condition
${\bf n}_1 \cdot {\bf n_2} =0$ in zero field.  To favor coplanar
(rather than umbrella) spin structures in a finite magnetic field,
requires $\chi_2 < 0$.  We may expect that $\chi_2$ is a
function of the anisotropy, being negative for the isotropic limit
$R=0$ and changing sign to positive values for sufficiently large $R$,
where the order by disorder physics favoring coplanar states gives way
to the classical energetic preference for umbrella states.  Here we
restrict ourselves to the small anisotropy regime, for which we expect
$\chi_2$ to remain negative.  With the preference for coplanar states
set by $\chi_2<0$, for field oriented along ${\hat z}$, the
preferred configurations of ${\bf d}$ may be parametrized as
\begin{equation}
  \label{eq:18}
  {\bf d} = |d| e^{i\tilde\theta} \left[ {\bf \hat z} + i( \cos\theta
    {\bf \hat x} + \sin\theta {\bf \hat y})\right],
\end{equation}
where $\theta$ describes the orientation of the plane of the spins, and
$\tilde\theta$ the angle of the spins within that plane.  With this
form for ${\bf d}$, we obtain the spin operators as
\begin{eqnarray}
  \label{eq:73}
  S^z_{x,y} & \sim & M + |d| \cos ({\bf Q}\cdot {\bf r} +
  \tilde\theta), \nonumber \\
  S^+_{x,y} & \sim &  - |d| e^{-i\theta} \sin({\bf Q}\cdot {\bf r} +
  \tilde\theta).
\end{eqnarray}

The last term in Eq.~\eqref{eq:Hcomm} describes the commensurate
locking of the spin to the lattice by the finite magnetic field.
Using Eq.~\eqref{eq:18}, it may be rewritten as a sine-Gordon term
\begin{equation}
  \label{eq:19}
  H_{sg} = \chi_3 |d|^3 h^3 \cos[3 \tilde\theta].
\end{equation}
The sign, $\chi_3 > 0$, is fixed
by the condition that one of the three spins in a sublattice must be
oriented opposite to the external field in the commensurate state.
Thus, in the commensurate state, $\tilde\theta = \pi$ in \eqref{eq:18}.

Now we move away from the isotropic line to $R>0$.  Here 3-fold
rotational symmetry is broken, which allows the introduction of an
additional term, linear in derivatives, into the effective Hamiltonian:
\begin{eqnarray}
H_{\rm incomm} &=& \frac{i}{2} b_1 ({\bf d}^* \cdot \partial_x {\bf d} - {\bf d} \cdot \partial_x {\bf d}^*) \nonumber\\
&& = - b_1 |d|^2 \partial_x \tilde\theta.
\end{eqnarray}
Since this term must vanish at $R=0$ and be analytic, $b_1 \sim R$.
This term competes with the sine-Gordon term in Eq.~\eqref{eq:19},
with the commensurate state with constant $\tilde\theta$ favored at
small $R$ and destabilized at larger $R$.  Thus the
commensurate-incommensurate transition in two dimensions can be
described by a Hamiltonian of the phase
\begin{equation}
\label{eq:H-CIC}
H_{\rm C-IC} = \int d^2 {\bf r} \{ \tilde{a}_0 (\partial_x
\tilde\theta)^2 - \tilde{b}_1  \partial_x \tilde\theta +\tilde{\chi}_3
 h^3 \cos[3 \tilde\theta]\}.
\end{equation}
Here, the coefficients with tildes, $\tilde{a}_0, \tilde{b}_1,
\tilde{\chi}_3$, are rescaled by unimportant factors, such as the
amplitude $|d|$.

The sine-Gordon model of the form in Eq.~\eqref{eq:H-CIC} appears in
several guises in this paper, and is analyzed in
Appendix~\ref{sec:sine-gordon-model}.  It encodes a
commensurate-incommensurate transition (CIT) with increasing $\tilde{b}_1$.
This transition is mean-field like for $d=2$, and we may apply the
results of Appendix~\ref{sec:dgeq-2:-mean}.  This gives a critical value
for the CIT of $\tilde{b}_{1,{\rm cr}} \sim \sqrt{\tilde{a}_0 \tilde{\chi}_3 h^3}$
for the incommensurate state, which translates to
\begin{equation}
  \label{eq:92}
  h_{\rm C-IC} \sim R^{2/3},
\end{equation}
since $\tilde{b}_1 \sim R$. This is roughly consistent with shape of the boundary in the lower left
corner of Fig.~\ref{fig:phase}.

For the TST, the situation is complicated by one-dimensional
fluctuations.  At zero field, $h=0$, we know that, in fact, the ground
state is {\sl not} a spiral but rather a dimerized phase.  Hence, we
cannot directly apply the above analysis at the lowest fields.  The
dimerized phase is broken fairly rapidly by the field, and so, above some
small critical field, we may expect to be able to use results of this
type.  Even so, we should really use
results for the $d=1$ case, where a non-mean-field analysis applies, as
described in Appendix~\ref{sec:d=1:-quant-fluct}.  Using
Eq.~\eqref{eq:28}, the critical value
$\tilde{b}_{1,{\rm cr}}$ is suppressed by a factor of $(\tilde\chi_3
h^3/\tilde{a}_0)^{\Delta_3/(4-2\Delta_3)}$, so that the net result is
$\tilde{b}_{1,{\rm cr}} \sim h^{\frac{3-\Delta_3}{2-\Delta_3}}$, and
hence
\begin{equation}
  \label{eq:93}
  h_{\rm C-IC} \sim R^{\frac{2-\Delta_3}{3-\Delta_3}}.
\end{equation}
Here, $\Delta_3$ is the scaling dimension of the $\cos 3\tilde\theta$
term.  Assuming the commensurate phase is at all stable for small $R$
implies $\Delta_3 <2$, so that the cosine term is relevant in the
isotropic case, $R=0$.  It is also bounded below by zero, so that the
exponent in Eq.~\eqref{eq:93} varies between $0$ and $2/3$.  Once
again, we caution that the expression must be taken with care, since it
does not in fact apply at the lowest fields.

\section{High Field Region}
\label{sec:high field}
\subsection{Spin flip bosons}
\label{subsec:dilute}

In this section, we study the phase diagram near saturation, i.e. for
applied fields sufficiently large that the magnetization is close to
its maximum of $1/2$ per site.  {\sl At} saturation, the ground state
of the model is the trivial product state with all spins aligned in
the direction selected by the field.  For fields above the saturation
field, this is the exact ground state, and the lowest excited states
consist of single magnons, in which just one spin has been flipped
relative to the saturated state.  These magnons are bosons with
$S^z=1$, and upon reducing the field to the saturation value, the
minimum energy required to create a magnon vanishes.  Below the
saturation field, therefore, we can expect Bose-Einstein condensation
(BEC) of these magnons.  In the one-dimensional TST, strict BEC is not
possible due to phase fluctuations, but these fluctuations are readily
taken into account and a quasi-condensate description remains
appropriate.

To formalize the magnon BEC picture, one may transform the spin model
to a bosonic
one\cite{matsubara1956lattice,batyev1984antiferromagnet,batyev1986,nikuni1995hexagonal,ueda2009magnon,kolezhuk2012},
using the equivalence of the spin $s=1/2$ Hilbert space to that of
hard-core bosons:
\begin{eqnarray}
\label{eq:bec1}
S_\br^+ & = & {\mathcal P}_\br \, b_\br\, {\mathcal P}_\br \\
S_\br^z & = & \frac{1}{2} - n_\br,
\end{eqnarray}
where $n_\br = b_\br^\dagger b_\br^{\vphantom\dagger}$ is the boson
occupation number, and one must project onto the space of no double
boson occupancy, ${\mathcal P}_\br = |n_\br=0\rangle \langle n_\br=0| +
|n_\br=1\rangle\langle n_\br=1|$.  Eq.~\eqref{eq:bec1} is equivalent to
the Holstein-Primakoff bosonization formula, truncated to quadratic
order in boson operators and taking $s=1/2$, provided the no double
occupancy constraint is imposed.  The generalization to $s>1/2$ will
be briefly discussed later in Sec.~\ref{sec:highS}.

It is convenient to implement the no double occupancy constraint by
first relaxing the constraint, adding an on-site interaction $U$ to
the Hamiltonian, and then realizing the projection by taking the
$U\rightarrow \infty$ limit.  In this way we can proceed
simply by rewriting the Heisenberg model using Eq.~\eqref{eq:bec1},
forgetting the projection operators, i.e. taking ${\mathcal
  P}_\br\rightarrow 1$.  We thereby obtain a boson Hamiltonian with
hopping terms ($J$), on-site energies ($J,h$), an on-site ($U$) and
nearest-neighbor ($J,J'$)  interactions.  Fourier transforming to
diagonalize the quadratic terms, we find
\begin{eqnarray}
\label{eq:bec2}
H = &&\sum_\bk \left[ \epsilon( \bk ) - \mu \right] b_\bk^\dagger b_\bk^{\vphantom\dagger} + \nonumber\\
    &&\frac{1}{2N} \sum_{\bk,\bk',\mathbf{q}} V(\mathbf{q}) b_{\bk+\mathbf{q}}^\dagger b_{\bk'-\mathbf{q}}^\dagger b_{\bk'}^{\vphantom\dagger}b_\bk^{\vphantom\dagger},
\end{eqnarray}
where
\begin{eqnarray}
\label{eq:bec3}
\epsilon({\bf k}) & = & J( {\bf k} ) - J_{\text{min}},\\
\mu & = & h_{\text{sat}} - h, \text{ where } h_{\text{sat}} = J(0) - J_{\text{min}},\\
V(\bk) &=& 2\left( \epsilon(\bk) + U \right).
\end{eqnarray}
Here, $J(\bk)$ is the Fourier transform of the exchange interaction,
$\mu$ is the bosonic chemical potential, and $h_{\text{sat}}$ is the
saturation field.
We will use this formalism to derive an effective action for the dilute bosons,
and also to locate (if any) a transition between the planar and cone phases near saturation.

\subsection{Effective field theory for dilute bosons}
\label{sec:effect-field-theory}

For $h>h_{\text{sat}}$, the vacuum is an exact ground state of this
Hamiltonian, i.e. $b_{\bf k} | 0 \rangle = 0$.  Below the saturation
field, a finite density of magnons is introduced into the system, and
a BEC or quasi-BEC is expected.  The phase of the system, and
correspondingly the magnetic order (correlations), is determined by
the structure of this condensate (or quasi-condensate).  To determine
this structure, we construct an effective model.  The lowest energy
magnon excitations in the triangular lattice occur at non-zero
momenta $\pm {\bf Q}$, which minimize the
dispersion\cite{nikuni1995hexagonal,ueda2009magnon}.
In our (sheared) coordinates, the dispersion relation is
\begin{equation}
  \label{eq:11}
  J_{\rm TST}({\bf k}) = J \cos k_x + J' [ \cos k_y + \cos (k_y-k_x)].
\end{equation}
In two dimensions, we can choose arbitrary $k_x$ and $k_y$, and the
minima occur at ${\bf k}=\pm {\bf Q}_{2d}$, with ${\bf
  Q}_{2d}=(Q_{2d},Q_{2d}/2)$, and
\begin{equation}
  \label{eq:10}
  Q_{2d} = 2 \arccos \left[ -\frac{J'}{2J} \right].
\end{equation}
Note that in the conventional cartesian coordinates this wavevector is
${\bf\sf Q}=(Q_{2d},0)$.  For the TST, we must quantize
$k_y=0,2\pi/3,4\pi/3$.  With this restricted choice of $k_y$, the 2d
wavevector ${\bf Q}_{2d}$ cannot generally be achieved.  Instead, we
find that the minimum energy wavevector is  ${\bf k}_{\rm TST} = \pm {\bf
  Q}_{\rm TST} = \pm (Q_{1d},2\pi/3)$, with
\begin{equation}
  \label{eq:12}
  Q_{1d} = \pi + \arctan \left( \frac{\sqrt{3}J'}{2J-J'}\right).
\end{equation}
The two wavevectors coincide when $J=J'$.

In a
low-energy description, the modes away from these two minima may
be integrated out, leaving an effective theory in terms of two
``flavors'' of bosons, $\psi_1$ and $\psi_2$, defined via
\begin{equation}
\label{eq:bec4}
b_{\bf k} = \psi_{1,{\bf Q+k}}  + \psi_{2,{\bf -Q+k}}  + \bar{b}_{\bf k}.
\end{equation}
Here, $\psi_{1,{\bf q}}$ ($\psi_{2,{\bf q}}$) is defined as a boson ``centered'' on the
minimum energy momentum ${\bf Q}$ ($-{\bf Q}$), with weight only for small
$|q|<\Lambda$, where $\Lambda \ll 2\pi$ is a cut-off introduced by
integrating out the modes away from the minima.   The third operator
$\bar{b}_{\bf k}$ represents the high energy modes which remain uncondensed,
and are integrated out.   In two dimensions, Fourier transforming in
$q_x,q_y$ back
to real space leads to slowly varying continuum fields
$\psi_a({\bf r})$, where ${\bf r}$ is a two dimensional spatial
coordinate.  For the TST, we need to keep only the mode with minimum
energy $q_y$, and so, we Fourier transform only in $q_x$, which leads to
a continuum field dependent only on the position along the chain, $x$.

In this continuum limit, the boson fields are governed by an effective
action of the form
\begin{widetext}
\begin{eqnarray}
\label{eq:6}
    {\mathcal S} & = &  \int d^d{\bf r} d\tau \, \Bigg\{ \psi_1^\dagger
          ( \partial_\tau -\frac{1}{2m}\nabla^2 ) \psi_1 +
          \psi_2^\dagger ( \partial_\tau -\frac{1}{2m}\nabla^2  ) \psi_2 - \mu \left( \rho_1 + \rho_2 \right) +
          \frac{1}{2} \Gamma_1 \left( \rho_1^2 + \rho_2^2 \right) +  \Gamma_2 \rho_1 \rho_2 \Bigg\},
\end{eqnarray}
\end{widetext}
where $\rho_\alpha = | \psi_\alpha |^2$. We have written the action,
Eq.~\eqref{eq:6}, in a form which includes both the TST ($d=1$) and two
dimensional ($d=2$) cases.  We expand to fourth order in $|\psi_a|$ and to lowest order in
derivatives, which is justified near saturation due to the diluteness
of the magnons.  The quadratic terms in Eq.~\eqref{eq:6} can be
readily extracted from the exact single-magnon dispersion, which is
given in Eq.~\eqref{eq:bec3} (in general in two dimensions the
quadratic term may have an anisotropic effective mass tensor \cite{ueda2009magnon}, which is not
explicitly shown in Eq.~\eqref{eq:6}).  The quartic interaction terms are more
subtle, because though the magnons may be assumed dilute, the
lattice-scale interactions in Eq.~\eqref{eq:bec2} are not weak.
Therefore the parameters $\Gamma_1, \Gamma_2$ must be obtained from a
more careful analysis, which we return to below.

\subsection{Order parameter structure}
\label{sec:order-param-struct}

Taking for the moment the $\Gamma_a$ as phenomenological parameters,
we discuss the structure of the condensed or quasi-condensed phase.
If $\mu<0$, there are no bosons in the system, and the vacuum is the ground state.
When $\mu>0$, a finite density of bosons is present.  Depending upon
their interactions, different phases may result \cite{nikuni1995hexagonal}.  To discuss the
nature of these phases, a mean field analysis of Eq.~\eqref{eq:6} is
sufficient.  We comment on the modifications to the mean field results
at the end of this subsection.

In mean field theory, we simply minimize ${\mathcal S}$ in Eq.~\eqref{eq:6} for
constant values of $\psi_\alpha$.  When $\mu>0$ and $\Gamma_1 <
\Gamma_2$, then $\rho_1 \neq 0, \rho_2 = 0$ or vice versa, which means
that the magnons condense at one of the two minima: a single-Q
condensate.  Here, in minimizing the energy, one finds that $\rho_1 =
\langle \rho_1\rangle = \mu/\Gamma_1$ and $E/N = -\mu^2/(2\Gamma_1)$.
By taking $\psi_{1,2} = \sqrt{\rho_{1,2}} e^{i\theta_{1,2}}$, one can
write the spin operator as follows
\begin{eqnarray}
\label{eq:bec6}
S_\br^+ & = & \overline{\psi} \,e^{i({\bf Q}\cdot{\bf r}+\theta_1)}\\
S_\br^z & = & \frac{1}{2} - \langle \rho_1\rangle,
\end{eqnarray}
where $\overline{\psi} = \sqrt{ \langle \rho_1\rangle}$ in mean field
theory.  We see that the $z$-component of the spins is non-zero but
constant in space, while the $xy$ components rotate as one moves in
space.  Such a configuration is called a cone or umbrella phase,
because the spins trace out a cone as one proceeds through the lattice,
see Figure~\ref{fig:comm-planar}(c).

When $\Gamma_2 < \Gamma_1$, then $\rho_1 = \rho_2$, which means that
the bosons condense at both $+{\bf Q}$ and $-{\bf Q}$.  This is a double-Q
condensate with density $\langle \rho\rangle = \langle \rho_1\rangle +
\langle \rho_2\rangle = \mu/(\Gamma_1 + \Gamma_2)$ in mean field
theory.  Here, the energy $E/N = \mu^2/(\Gamma_1+\Gamma_2)$.  Again,
by letting $\psi_{1,2} = \sqrt{\rho_{1,2}} e^{i\theta_{1,2}}$ and
$\theta_{1,2} = \theta \pm \tilde\theta$,
\begin{eqnarray}
\label{eq:bec7}
S_\br^+ & = & 2 \overline{\psi} \, e^{i \theta} \cos \left( {\bf Q}\cdot{\bf r} + \tilde\theta \right)\\
S_\br^z & = & \frac{1}{2} - 4 \langle \rho\rangle \, \cos^2 \left( {\bf Q}\cdot{\bf r}+ \tilde\theta \right),
\end{eqnarray}
where $\overline{\psi}= \sqrt{\langle\rho\rangle}$ in mean field
theory.  In this phase, the $z$-component of the spins is not constant,
but the phase of $S_r^+$ is constant.  This implies that the spins
remain in a plane, i.e. this is a coplanar phase. Instead of a cone, the spins in
this phase sweep out a ``fan'' -- so this is sometimes called a fan
state.

How much of this survives beyond mean field theory?  In general, the
dependence of the density on chemical potential is affected by
fluctuations.  Note that in the original spin problem, this dependence
gives the behavior of the magnetization versus field in the vicinity
of saturation, as is seen from Eq.~\eqref{eq:bec1}.  As is
well-known\cite{Subirbook}, the BEC transition at $\mu=0$ is a very
simple example of a quantum critical point, whose upper critical
dimension is $d=2$.  Thus in two dimensions, the deviations from mean
field theory are minimal and consist just of logarithmic corrections.
However, in $d=1$ the corrections are much more significant, and the
dependence of the density on chemical potential is quite different.

In mean field theory, we see that there is a first order transition
between the cone and fan states upon varying $\Gamma_1-\Gamma_2$
through zero.  In fact, the location of this transition at
$\Gamma_1=\Gamma_2$ is correct and moreover, exact, beyond mean field theory.
To see this, note that when $\Gamma_1=\Gamma_2=\Gamma$, the
interaction terms may be rewritten as
$\frac{\Gamma}{2}(\rho_1+\rho_2)^2$, which implies that the action has
an enlarged {\em SU(2) symmetry} under rotations $\psi_\alpha
\rightarrow \sum_\beta U_{\alpha\beta} \psi_\beta$, where $U$ is an
arbitrary $SU(2)$ matrix.  This guarantees the degeneracy of the cone
and fan states at this point, since one can be rotated into the other
by such an $SU(2)$ rotation, and therefore, fixes the location of the
cone to coplanar transition.

When $\Gamma_1 \neq \Gamma_2$, the $SU(2)$ symmetry of Eq.~\eqref{eq:6} is
reduced to U(1)$\times$U(1), corresponding to independent phase
rotations of $\psi_1$ and $\psi_2$.  As a consequence, there will be one
gapless mode in the theory described by Eq.~\eqref{eq:6} for each bose
field with non-zero amplitude, i.e. one in the cone state and two in the
fan.  The fluctuations of these gapless modes lead, in the one
dimensional TST, to power-law
correlations of the spin components transverse to the magnetic field, rather
than the long range order (broken symmetry states) obtained in mean
field.

Physically, the overall U(1) symmetry under simultaneous and equal
rotations of both fields reflects conservation of $S^z$, and is
microscopically mandated by the Heisenberg model.  The ``orthogonal''
symmetry under the rotation of the two boson fields by opposite phases
is {\sl emergent}, however.  It is a consequence of the {\sl discrete}
translational symmetry of the lattice, and the (generically)
incommensurate nature of the wavevector $Q$.  In general, this symmetry
is broken by terms (which should be added to $\mathcal{S}$ in
Eq.~\eqref{eq:6}) of the form
\begin{equation}
\label{eq:bec8}
{\mathcal S}' = -\sum_n w_n \int d^d{\bf x} d\tau \, \left( \psi_1^\dagger
  \psi_2 \right)^n \, e^{-i n {\bf q}_n\cdot {\bf r}} + h.c.,
\end{equation}
where na\"ively ${\bf q}_n=2{\bf Q}$, but in fact we can take ${\bf
  q}_n=2{\bf Q}- {\bf K}/n$, where ${\bf K}$
is any reciprocal lattice (RL) vector, since ${\bf r}$ is a lattice
coordinate.  So henceforth we work with
\begin{equation}
  \label{eq:8}
  {\bf q}_n = {\rm min}_{{\bf K} \in {\rm  RL}} [ 2{\bf Q}- {\bf K}/n],
\end{equation}
i.e. we choose ${\bf K}$ to minimize the magnitude of ${\bf q}_n$.
When the wavevector ${\bf Q}$ is incommensurate and the magnitude of these
terms are small, their oscillations average to zero over short distances,
and they can thereby be neglected.  However, if $2n{\bf Q}$ is close to a
reciprocal lattice vector, then ${\bf q}_n$ is small and the corresponding $w_n$ term
becomes slowly varying, and it can have effects that persist into the
continuum theory.  This occurs only if $2n{\bf Q}$ is close to a reciprocal
lattice vector {\sl and} the amplitude of both $\psi_1$ and $\psi_2$ is
non-zero, i.e. within the coplanar or fan state.  This leads to
commensurate-incommensurate transitions, discussed in
Sec.~\ref{subsec:incomm-comm}.

In the cone state, such effects are not important.  In this case we
expect one gapless ``Goldstone'' mode ($\theta_1$) and power-law
transverse spin correlations.  But actually there is some hidden long
range order.  Note that in Eq.~\eqref{eq:bec6} we have
(arbitrarily) chosen the minimum with $\rho_1\neq 0$ and $\rho_2=0$,
instead of the one with $\rho_1=0$, $\rho_2\neq0$.  In doing so, the
system spontaneously breaks discrete symmetries.  In particular, for the
TST, this choice breaks both inversion symmetry and a ``charge
conjugation'' symmetry, the latter being the anti-unitary symmetry of
the Scr\"odinger equation under complex conjugation of the wavefunction.
Although the fluctuations of the phase $\theta_1$ above will reduce the
mean field magnetic order to quasi-long-range order in the TST, the discrete symmetry breaking is robust to one
dimensional fluctuations.  This symmetry breaking can be most directly
sensed by the vector chirality \cite{kolezhuk05,hikihara2010},
\begin{equation}
\label{eq:chirality}
V_{x,y}=\hat{z}\cdot \langle{\bf S}_{x,y} \times {\bf S}_{x+1,y}\rangle.
\end{equation}
Replacing $S_r^+$ in Eq.~\eqref{eq:chirality} by the ansatz in
Eq.\eqref{eq:bec6}, we find $V = \overline{\psi}^2 \sin Q$, i.e. a
non-zero and constant value in the cone state.  The opposite sign would
be obtained for the solution with $\rho_1=0$, $\rho_2\neq0$, so this
serves as an Ising-type order parameter for the cone state.

\begin{figure}[t]
  \begin{center}
  \scalebox{0.85}{\includegraphics[width=\columnwidth]{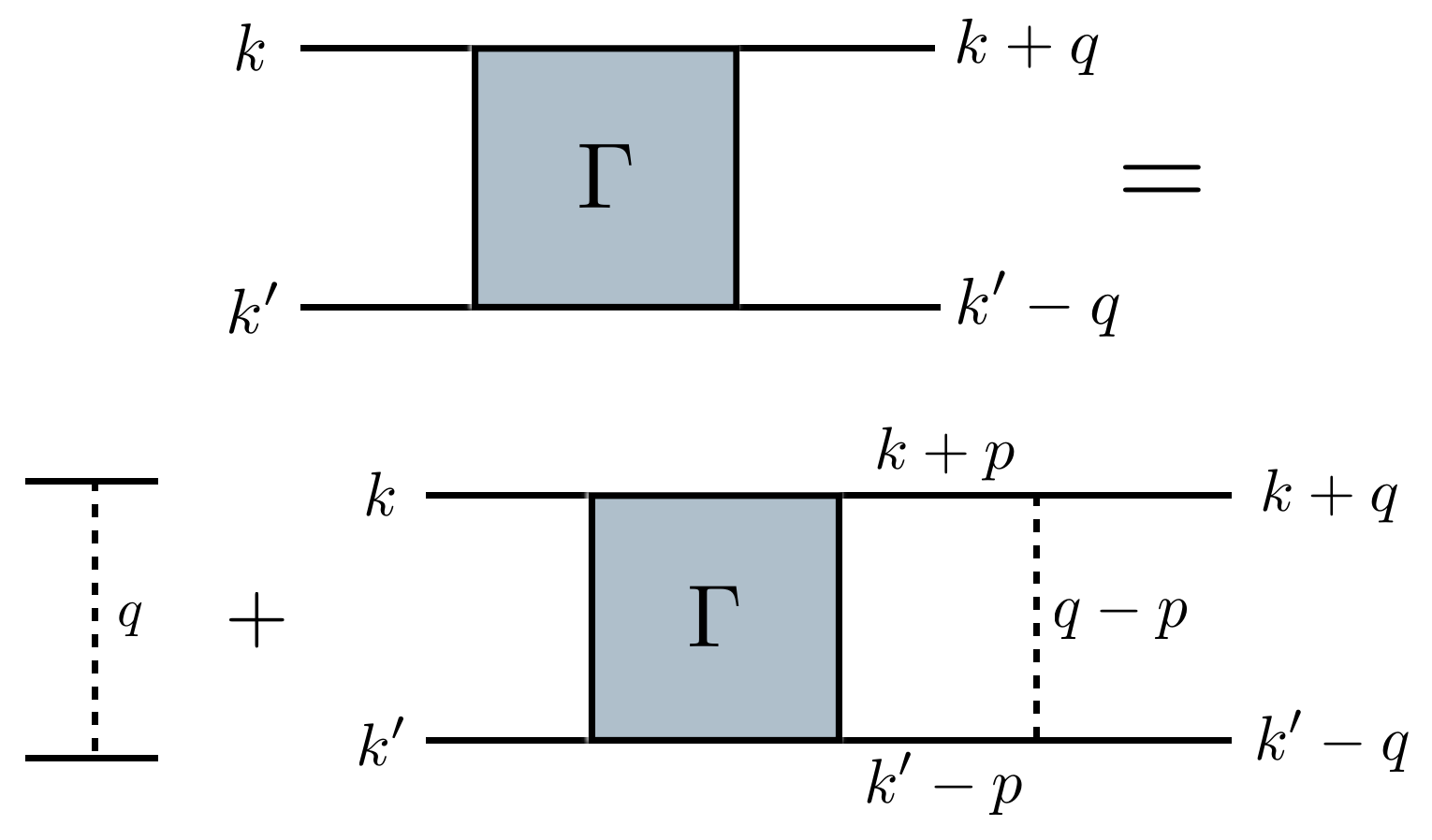}}
  \end{center}
  \caption{Ladder approximation of Eq.~\eqref{eq:bec9}.
      Here, $k, k'$ are incoming momenta while $k+q$, $k'-q$ are outgoing momenta.}
  \label{fig:ladder}
\end{figure}
%

\subsection{Incommensurate planar to cone state transition at the saturation}
\label{subsec:planar to cone}

\subsubsection{Bethe-Salpeter equation}
\label{subsubsec:bs}

Now that we have described the phases of Eq.~\eqref{eq:6}, we will
briefly outline the methods to compute $\Gamma_1, \Gamma_2$.  When the
external field is sufficiently close to the saturation field, then the
density of magnons, or spin flips, is dilute.  In this case, we can
safely use the ladder
approximation\cite{abrikosov1975methods,beliaev1958application,beliaev1958energy}
to renormalize the interaction vertex in a controlled manner.  In
fact, we strictly speaking analyze the interactions for fields {\sl
  above} the saturation field, where there are no bosons present in
the ground state, and we consider just two bosons interacting pairwise
above the vacuum.  We require the behavior in the limit in which the
saturation field is approached, i.e. in which the energy of the two
interacting bosons approaches zero.  This limit should be familiar
from ultra-cold atomic systems, in which the complicated interactions
between atoms can be replaced by one or a few scattering lengths,
which represent the effective interactions in the dilute limit.  Here
we obtain the effective interactions from the Bethe-Salpeter (BS)
equation, which reads
\begin{equation}
\label{eq:bec9}
\Gamma( k, k'; q ) = V(q) -\int_p \frac{ V(q-p) \Gamma(k,k';p) }{
  \epsilon(k+p) + \epsilon(k'-p)+ \Omega}.
\end{equation}
Here
$\Gamma(k,k';q)$ is the irreducible four-point interaction vertex
taken with all external frequencies equal to zero, and $\Omega = 2(h - h_{\text{sat}}) = -2\mu$.
The $k, k'$ are the incoming momenta and $k+q, k'-q$ are the
outgoing momenta, as shown in Fig.~\ref{fig:ladder}.   From this, one
obtains that $\Gamma_1 = \Gamma( Q, Q, 0 )$ and $\Gamma_2 = \Gamma(Q,
-Q, 0 ) + \Gamma( Q, -Q, -2Q )$.  In Eq.~\eqref{eq:bec3}, we
introduced a factor of $U$ into the definition of $V(q)$ to enforce
the spin-$1/2$ constraint, which is equivalent to taking the limit $U
\to \infty$. This limit in the BS language, Eq.~\eqref{eq:bec9},
provides us with an additional constraint which reads \cite{batyev1984antiferromagnet,nikuni1995hexagonal}
\begin{equation}
\label{eq:bec10}
\int_p \frac{\Gamma(k,k';p)}{\epsilon(k+p) + \epsilon(k'-p)+\Omega} = 1.
\end{equation}
Both Eq.~\eqref{eq:bec9} and Eq.~\eqref{eq:bec10} can be applied
either in two or three dimensions, or for the one dimensional TST; in the
latter case, the integral over $p$ should be regarded as an integral
over $p_x$ and a {\sl sum} over the discrete $p_y= 0, 2\pi/3, 4\pi/3$.
Notice that in two or fewer dimensions, since $\epsilon(k) \sim k^2,
V(k) \sim 1$ near $k=0$, the integral is at least logarithmically
divergent when $\Omega$ approaches zero.  This reflects the fact that
weak interactions are marginally relevant at the zero density fixed
point in $d=2$, and relevant for $d<2$.  We use this to our advantage,
since we are interested precisely in this limit: the singular parts
dominate the vertex function as $\Omega \rightarrow 0^+$, and we
extract these dominant singular terms analytically to obtain the
asymptotic behavior.  For $d > 2$, the integrals become non-singular,
and one can directly take the $\Omega=0$ limit.

\subsubsection{Calculation of $\Gamma_1$ and $\Gamma_2$ in 2d}
\label{sec:BS-2d-lattice}
We first give a brief summary of our calculations for the 2d case. The dispersion minima occurs at ${\bf k},{\bf k'} = \pm {\bf Q}_{2d} =
\pm (Q_{2d},Q_{2d}/2)$, where $Q_2d$ is given in Eq.~\eqref{eq:10}. To solve the BS equation, we use the following ansatz:
\begin{widetext}
\begin{equation}
\label{eq:ansatz}
\Gamma(k,k';q;\Omega) = A_0 + A_1 \cos q_x + A_2 \sin q_x + A_3 \cos q_y+ A_4 \sin q_y+ A_5 \cos (q_y-q_x)+A_6 \sin(q_y-q_x),
\end{equation}
\end{widetext}
where $A_i$ are coefficients dependent on $k, k', J, J'$ and
$\Omega$. With Eqs.~(\ref{eq:bec9}, \ref{eq:bec10}, \ref{eq:ansatz}), one can solve a set of linear equations for the
coefficients $A_i$, which gives an explicit form of $\Gamma(q)$ for a
given set of $k,k',J,J'$ and $\Omega$.  Details of the 2d case are given in
Appendix~\ref{app:2d}. From the solution, we simply obtain
\begin{equation}
\label{eq:2dgamma}
\Gamma_1 > \Gamma_2, \qquad  \textrm{for } 0<R< 1,
\end{equation}
which implies that {\it for all range of anisotropies, $0\leq R \leq 1$, the ground state near saturation field is always an
incommensurate planar (or fan) state}.

To see how the incommensurate planar state dominates over the cone state in the weakly coupled chains region, we expand the expression of $\Gamma$'s in the leading order of both $1/\ln\Omega$ and $j \equiv J'/J$
\begin{eqnarray}
\label{eq:gamma2d-decoup}
\Gamma_1/J&=&[-4\pi j+\frac{\pi}{2}j^3+O(j^5)]\frac{1}{\ln\Omega} \nonumber \\
&&+[-8j\pi \ln(4j)+\alpha+O(j^3)]\frac{1}{(\ln\Omega)^2}+...,  \nonumber \\
\Gamma_2/J&=&[-4\pi j+\frac{\pi}{2}j^3+O(j^5)]\frac{1}{\ln\Omega} \nonumber \\
&&+[-8j\pi \ln(4j)+O(j^3)]\frac{1}{(\ln\Omega)^2}+..., \nonumber \\
\alpha &=& \frac{8j\pi(24-16\ln2-3\pi\ln2)}{16+3\pi} > 0.
\end{eqnarray}
Since the extra factor $\alpha$ is always larger than zero, the ground state always prefers the fan state in the limit of decoupled chains.

One can analytically check this result in the same limit, $J' \ll J$.
We discuss this extension in Appendix \ref{sec:BS-1d}.

\subsubsection{Calculation of $\Gamma_1$ and $\Gamma_2$ in the TST}
\label{sec:BS-TST}

We now present a brief overview of our calculations on the TST.  We
consider an infinitely long system, where $q_x$ is continuous and
$q_y=0,2\pi/3, 4\pi/3$ is discretized by periodic boundary conditions.
The dispersion minima occur at ${\bf k},{\bf k'} = \pm {\bf Q}_{1d} =
\pm (Q_{1d},2\pi/3)$, given in Eq.~\eqref{eq:12}.  We are now in a
position to solve the BS equation, where we follow similar procedures as the two-dimensional case.
We use the same
ansatz, Eq.~\eqref{eq:ansatz}, to solve for the coefficients $A_i$.
From these coefficients, we can obtain the explicit forms of $\Gamma(q)$,
for which we provide details in Appendix~\ref{app:tst}.
Our results are as follows
\begin{equation}
\label{eq:2d4}
\begin{array}{cc}
\Gamma_1 > \Gamma_2, \qquad & \textrm{for } 0<R< 0.48, \\
\Gamma_1 < \Gamma_2, \qquad& \textrm{for } 0.48<R<1.
\end{array}
\end{equation}
This tells us that for $R < R_c = 0.48$, the incommensurate (fan) state is favored,
while for $R > R_c$, the cone (umbrella) state is favored.
This result is in agreement with the analytical result, in Appendix \ref{sec:BS-1d},
where it was shown that spins order into a cone state in the decoupled chains limit.

\subsection{Commensurate-Incommensurate Transitions (CIT)}
\label{subsec:incomm-comm}

In the previous subsection, we found that near saturation,
the ground state of the two-dimensional model for all $R$ and of the TST for $R > 0.48$
is coplanar, with modulation of the $z$-component of the spin at wavevector
$2Q$. As mentioned in Section~\ref{sec:order-param-struct}, this implies spontaneous
breaking of the discrete translational symmetry, which is sensitive to
commensurability effects via the terms in
Eq.~\eqref{eq:bec8}.  In particular, we expect that the wavevector $Q$ will {\sl lock}
to commensurate values, where $2Qn$ is a reciprocal lattice vector, over a finite range of field and anisotropy, $R$.
We now turn to a description of these commensurate-incommensurate
transitions (CITs), both in the 2d case and for the TST.

To study the CITs, we must now consider the full action,
Eqs.~(\ref{eq:6},\ref{eq:bec8}), for $h<h_{\rm sat}$, i.e. for
$\mu>0$, where the bosons are at non-zero density.  In two-dimensions,
we can regard them as condensed, while in the TST, true condensation is
impossible but the system can be viewed as a quasi-condensate or a
Luttinger liquid.  In either case, amplitude fluctuations of the
$\psi_\alpha$ fields are small, and we can write the effective action
in terms of the phases $\theta_\alpha$, where $\psi_\alpha \sim \psi_0
e^{-i\theta_\alpha}$ in the coplanar/fan region.

Conceptually, the effective action for the phase fields is obtained by
first following the renormalization of the system away from the zero
density fixed point, $\mu=0$, where amplitude fluctuations are still
important.  Once the energy scale set by $\mu$ is reached, these
fluctuations are quenched, and it is sufficient to consider only small
fluctuations in the amplitudes.  To achieve this, we simply make the
assumption of small amplitude fluctuations in
Eqs.~(\ref{eq:6},\ref{eq:bec8}), but with the bare couplings replaced by
{\sl fully renormalized ones, at the scale $\mu$}.  We believe this
procedure properly captures the scaling for small $\mu$, though it is
not quantitatively reliable.

Because the low energy dispersion of the single magnon states is
exactly known and described by the quadratic terms in
Eq.~\eqref{eq:6}, the corresponding couplings are unrenormalized.  The
interactions $\Gamma_1$ and $\Gamma_2$, however, are renormalized by
multiple scatterings, which is exactly what is captured by the BS
equation discussed in Sec.~\ref{subsec:planar to cone}.  From this
analysis, we simply take as our renormalized couplings
$\Gamma_a(\Omega =2\mu)$.  Note that this would be exactly correct if
we replaced $\mu$ by $|\mu|$ for the case $\mu<0$, but on scaling
grounds it should give the correct dependence even for $\mu>0$.

The renormalized interactions can be approximately represented for
small $\mu$ as
\begin{equation}
  \label{eq:21}
  \Gamma_\alpha(\mu) \sim \frac{u_\alpha}{1+ m u_\alpha/\zeta(m\mu)},
\end{equation}
where
\begin{equation}
  \label{eq:31}
  \zeta(m\mu) = \left\{ \begin{array}{cc}
    (m\mu)^{1/2} &\qquad d=1 \\
    1/|\ln(m\mu)| & \qquad d=2\end{array}\right. ,
\end{equation}
and $u_\alpha$ are constants related to the ``bare'' values of
$\Gamma_\alpha$.  We can in principle use the renormalize
$\Gamma_\alpha(\mu)$ for the original lattice spin model, which have
the same leading and first sub-leading terms for small $\mu$ (up to
second order in $\zeta \ll 1$) as in Eq.~\eqref{eq:21}, but with
considerably more complicated coefficients. Beyond second
order in $\zeta$, the lattice $\Gamma_\alpha$ differ somewhat, and the
expression is unwieldy.  The above form is sufficient for our
purposes, and is exact for a continuum model.

Once the $\Gamma_\alpha(\mu)$ are known, the analysis is
straightforward \cite{ueda2009magnon}.  We write $\psi_\alpha = \left[\overline{\rho} +
\sigma_\alpha\right]^{1/2} e^{-i\theta_\alpha}$, and assume small fluctuations in
$\sigma_\alpha$ around the saddle point value for
\begin{equation}
  \label{eq:40}
  \overline{\rho}=\frac{\mu}{(\Gamma_1(\mu)+\Gamma_2(\mu))}.
\end{equation}
(Here we assume $\Gamma_1(\mu)>\Gamma_2(\mu)$).  Eq.~\eqref{eq:40}
properly captures, through the dependence of $\Gamma_\alpha$ on $\mu$,
the non-mean-field dependence of the boson density on chemical
potential.  In particular, it yields $\overline{\rho} \sim \mu^{1/2}$
in 1+1 dimensions, consistent with the fact that repulsively interacting
bosons behave with an effective hard core at low density, and
consequently have an equation of state similar to free fermions.

Expanding the action to quadratic order in $\sigma_\alpha$ and
neglecting irrelevant terms involving derivatives of $\sigma_\alpha$
and their couplings to higher derivatives of $\theta_\alpha$, we
obtain (neglecting constant terms)
\begin{widetext}
  \begin{eqnarray}
    \label{eq:32}
    \mathcal{S} & = & \int d^d{\bf r} d\tau\, \Bigg\{ i
    (\sigma_1 \partial_\tau \theta_1 + \sigma_2 \partial_\tau
    \theta_2) + \frac{\overline{\rho}}{2m}( |\nabla\theta_1|^2 +
    |\nabla\theta_2|^2) + \frac{\Gamma_1}{2} (\sigma_1^2+\sigma_2^2) +
    \Gamma_2 \sigma_1 \sigma_2 \Bigg\}.
  \end{eqnarray}
\end{widetext}
Next, we integrate out the $\sigma_\alpha$ fields, and express the
resulting action in terms of  new linear combinations,
\begin{equation}
  \label{eq:30}
  \theta=\theta_1+\theta_2, \qquad \tilde\theta=\theta_1-\theta_2.
\end{equation}
The result is
\begin{eqnarray}
  \label{eq:33}
  \mathcal{S} & = & \mathcal{S}_\theta + \mathcal{S}_{\tilde\theta},
\end{eqnarray}
where
\begin{eqnarray}
  \label{eq:34}
  \mathcal{S}_\theta &=& \int d^d{\bf r}\, d\tau\, \left\{
  \frac{\kappa_c}{2}(\partial_\tau\theta)^2 +  \frac{\rho_c}{2}
  (\nabla  \theta)^2  \right\},
\end{eqnarray}
with
\begin{equation}
  \label{eq:36}
  \kappa_c = \frac{1}{2(\Gamma_1(\mu)+\Gamma_2(\mu))}, \qquad \rho_c = \frac{\overline{\rho}}{2m},
\end{equation}
and
\begin{eqnarray}
  \label{eq:7}
  \mathcal{S}_{\tilde\theta} & =& \int d^d{\bf r}\, d\tau\, \Big\{
    \frac{\kappa}{2}(\partial_\tau\tilde\theta)^2 +  \frac{\rho}{2} (\nabla
    \tilde\theta)^2  \nonumber \\
    && -
    \sum_n \lambda_n \cos [n (\tilde\theta - {\bf q}_n \cdot {\bf r})]\Big\},
\end{eqnarray}
with
\begin{eqnarray}
  \label{eq:37}
  \kappa & = & \frac{1}{2(\Gamma_1(\mu)-\Gamma_2(\mu))}, \qquad \rho =
  \frac{\overline{\rho}}{2m}, \nonumber \\
  \lambda_n & = & 2 w_n \overline{\rho}^n.
\end{eqnarray}
Here we have restored the term resulting from $\mathcal{S}'$ in
Eq.~\eqref{eq:bec8}.  Note that the ``charge'' field $\theta$
describes the Goldstone mode of the broken (or quasi-broken in 1d)
U(1) symmetry, and thus remains exactly massless.  It completely
decouples from the $\tilde\theta$ field, and can be neglected in the
analysis of the CIT.

We are now in a position to analyze the CIT
 using Eqs.~(\ref{eq:7},\ref{eq:37}) and the results of
Appendix~\ref{sec:sine-gordon-model}.  This is strongly dimension
dependent, so we treat the cases of two dimensions and one dimension
separately.

\subsubsection{Two dimensions}
\label{sec:two-dimensions}

In two dimensions, we begin by presuming that {\sl one} of the cosines
in Eq.~\eqref{eq:7} is almost non-oscillating, i.e when one of the $q_n$ is close to
zero.  Generically, this will happen for one specific minimal $n$,
when
\begin{equation}
  \label{eq:17}
  Q_{2d} = \frac{\pi m}{n} + \delta Q,
\end{equation}
for some specific $m,n$, with $|\delta Q|\ll 1$.  The other rapidly
oscillating cosines can be neglected, and we retain only the weakly
oscillatory one.  Then, in the ${\sf x},{\sf y}$ coordinates, the action takes the form given in
Eq.~\eqref{eq:84}, with $\lambda_n=\lambda$, and $q=q_n = 2 \delta Q$.

We can now directly apply the results of
Appendix~\ref{sec:dgeq-2:-mean}.  Using $\delta=\rho q = 2\rho \delta
Q$, and Eq.~\eqref{eq:87}, we obtain that the commensurate
state is stable for $|\delta Q|< \delta Q_c$, which defines the location
$\delta Q_c$ of the CIT as
\begin{eqnarray}
  \delta Q_c & \sim & \sqrt{\lambda_n/\rho} \sim \sqrt{m w_n} \,
  \overline{\rho}^{(n-1)/2}, \nonumber \\
  \label{eq:ic8}
  & \sim & \sqrt{m w_n} (\Upsilon(\mu)\mu)^{(n-1)/2} ,
\end{eqnarray}
where we used Eq.~(\ref{eq:37}) for $d=2$, and, of course, we assume
$\mu>0$.  Here
\begin{equation}
  \label{eq:35}
  \Upsilon(\mu) = \frac{1}{\Gamma_1(\mu)+\Gamma_2(\mu)} \sim \frac{2|\ln (m\mu)|}{m}  ~\text{for}~ \mu \ll 1,
\end{equation}
is a weak logarithmic function of $\mu$.

For the commensurate state centered around $R=0$ ($J'=J$), we have
$n=3$, and the phase boundary for the C-IC transition is linear in
$\mu$, up to logarithmic corrections. However, as $n$ increases, the
widths of the commensurate phases decrease.

\subsubsection{One dimension}
\label{sec:one-dimension-1}

In the TST, to derive the 1d theory we must sum over
discrete $y$.  This restricts the $\lambda_n$ terms in
Eq.~\eqref{eq:7} to $n$ which are multiples of $3$, so that the $y$
component of ${\bf q}_n$ ($=2nQ_y$) is a multiple of $2\pi$.

Following the discussion for two dimensions, we again consider
wavevectors
\begin{equation}
  \label{eq:94}
  Q_{1d} = \frac{\pi m}{n} + \delta Q,
\end{equation}
with appropriate $m,n$ such that $|\delta Q|\ll
1$, and keep only the dominant cosine term of order $n$, which then
matches the sine-Gordon form in Eq.~\eqref{eq:84} with $q= 2\delta Q$.
Then we take over results from Appendix~\ref{sec:d=1:-quant-fluct}.

According to that discussion, a commensurate phase is stabilized
whenever the scaling dimension of the cosine term, $\Delta_n$, is less
than two.  Using the result in Eq.~\eqref{eq:24} and also, Eq.~\eqref{eq:37}, we obtain
\begin{equation}
  \label{eq:38}
  \Delta_n =  \frac{n^2}{\sqrt{2}\pi} \left( \frac{\mu}{m}\right)^{1/4}
  \sqrt{\frac{u_1-u_2}{u_1 u_2}},
\end{equation}
so that $\Delta_n \ll 1$ for $\mu \ll 1$.  This shows that
$\Delta_n<2$, and the commensurate phase is indeed realized.
Note that if we approximate
$\Delta_n=0$, then this becomes the same classical estimate as in the
previous section, except that $\Gamma_a(\mu)$ has a different
dependence in one dimension.  While this is in principle appropriate
for very small $\mu$, the $1/4$ exponent in Eq.~\eqref{eq:38}
indicates that $\Delta_n$ can be substantial nonetheless, so we will
proceed with the estimate taking $\Delta_n \neq 0$.

\begin{figure}[t]
  \begin{center}
  \scalebox{0.95}{\includegraphics[width=\columnwidth]{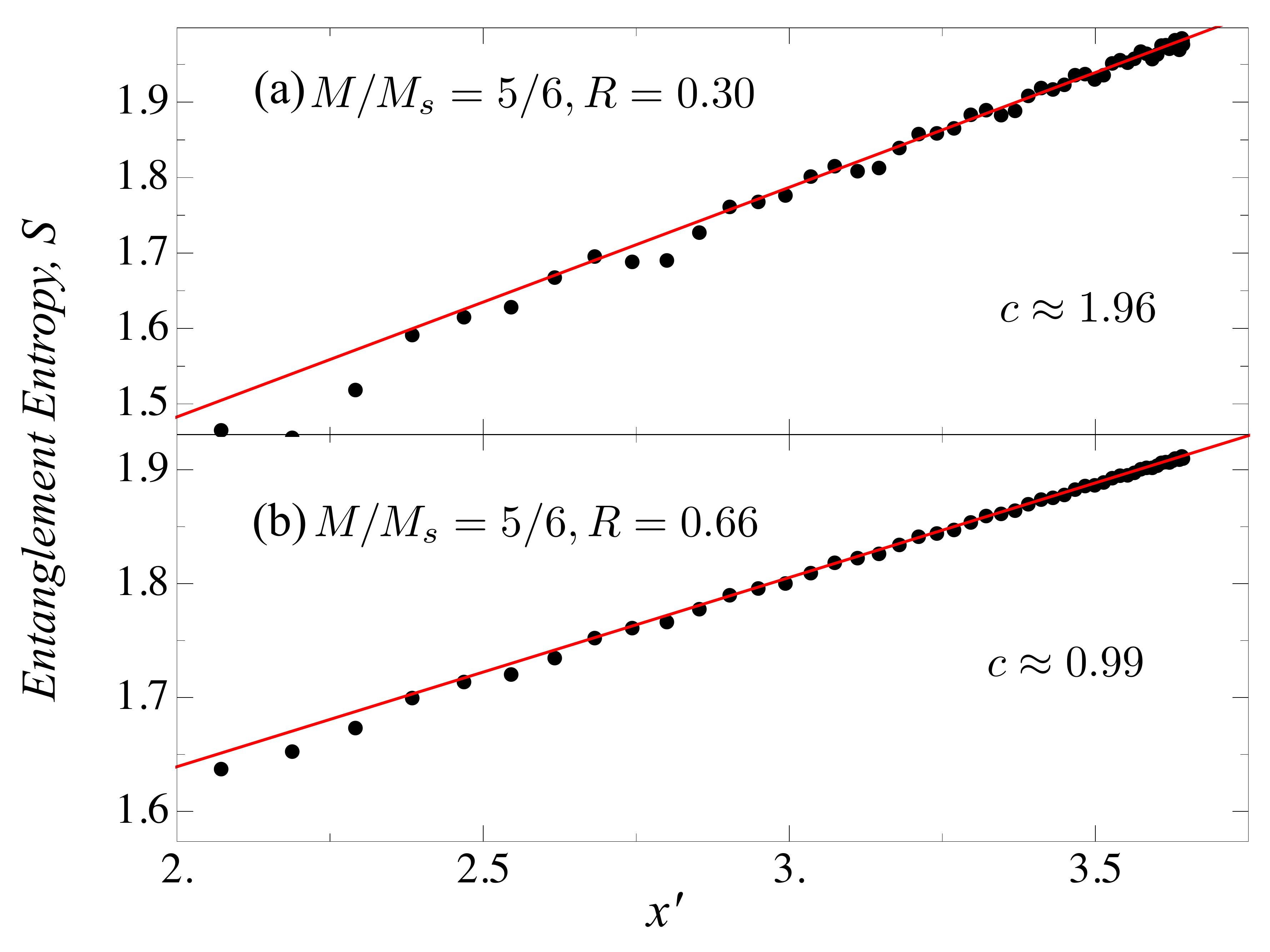}}
  \end{center}
  \caption{Entanglement entropy at (a) $M/M_s=5/6$, $R=0.3$, the incommensurate coplanar phase,
  and (b) $M/M_s=5/6$, $R=0.66$, the cone (or umbrella) phase.
  We take a system size of $N_x=120$.}
  \label{fig:cone-planar}
\end{figure}

Using $\delta= 2\rho \delta Q$ and the estimate for the critical
$\delta_c$ in Eq.~\eqref{eq:28}, and applying Eqs.~\eqref{eq:40} and
\eqref{eq:37}, we find the location of the 1d CIT as
\begin{equation}
  \label{eq:39}
  \delta Q_c \sim \left(w_n m^{\frac{n+1}{2}} n^{\Delta_n} \mu^{\frac{n-1}{2}}\right)^{\frac{1}{2-\Delta_n}}.
\end{equation}
For $n=3$ and assuming $\Delta_n \to 0$, this predicts $\delta Q_c \sim \mu^{1/2}$, which does not
agree with $\mu \sim R$ scaling of the C-IC boundary in the upper left corner of the phase diagram in Fig.~\ref{fig:phase}.
However the range of $\mu$ there is not particularly small, $h$ changes from $4.5$ to approximately $3$
as $R$ changes from $0$ to $0.1$. This observation calls for a more careful analysis of behavior
predicted by Eqs. (\ref{eq:38},\ref{eq:39}) for $\mu \sim O(1)$. We find that numerical coefficients in \eqref{eq:38}
make $\Delta_{n=3}$ to vary in the interval $0.5 - 1$ for $\mu$ relevant to the C-IC boundary in Fig.~\ref{fig:phase},
resulting in an almost linear dependence $\delta Q_c \sim \mu$ away from the strict $\mu\to 0$ limit
and in qualitative agreement between our analysis here and the numerical data in Fig.~\ref{fig:phase}.

\begin{figure}[t]
  \begin{center}
  \scalebox{0.95}{\includegraphics[width=\columnwidth]{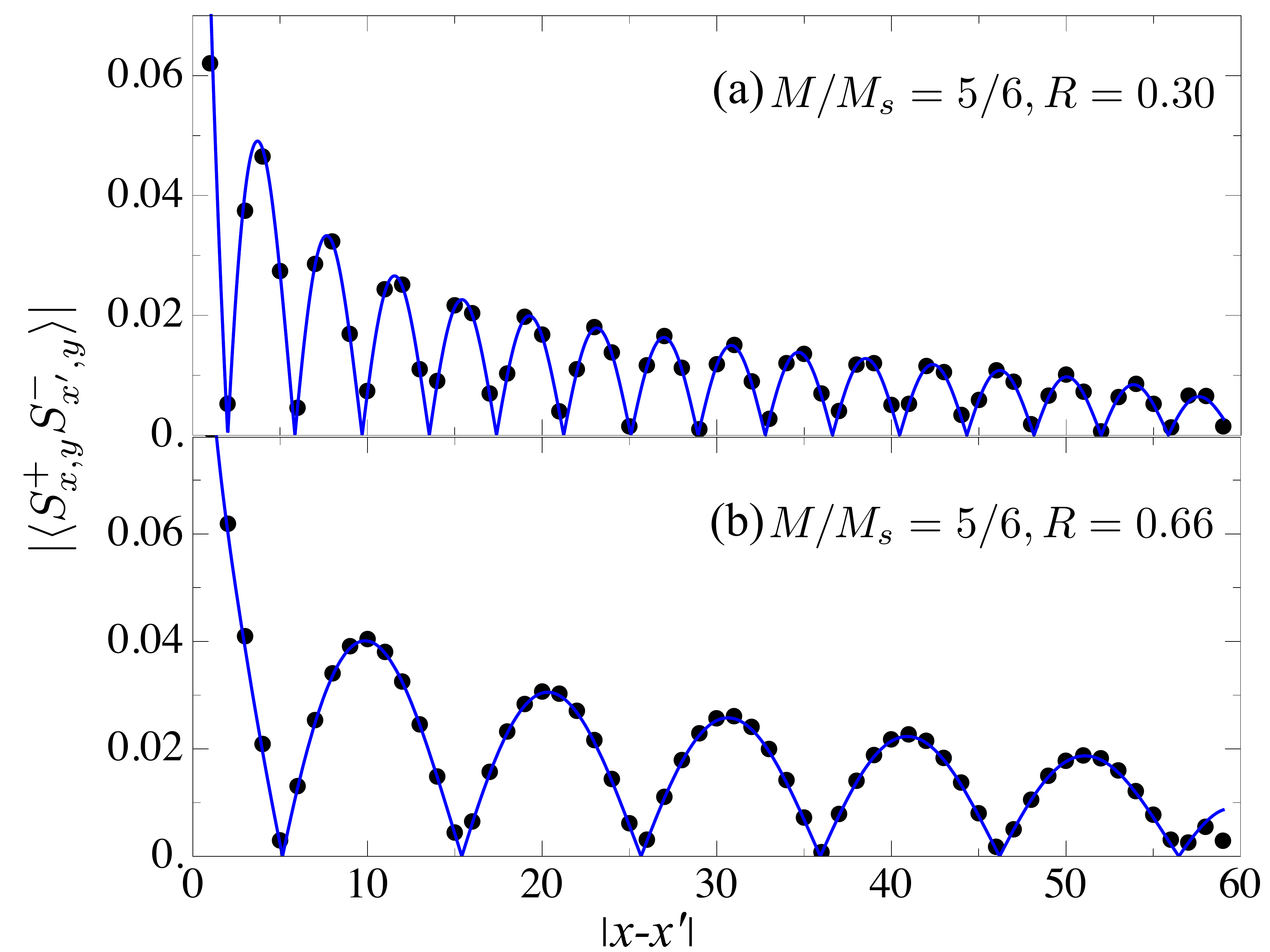}}
  \end{center}
  \caption{(Color online) Transverse spin-spin correlation function for $M/M_s = 5/6$, $N_x = 120$ and $x'=N_x/2$ at (a)$R=0.30$ in the incommensurate coplanar state and (b)$R=0.66$ in the cone state.
  Our DMRG data points are plotted in (black) circles, while the theoretical fit, Eq.~\eqref{eq:ruadd2}, is shown as a solid (blue) line.}
  \label{fig:corr-cone}
\end{figure}
%

%
\begin{figure}[t]
  \begin{center}
  \scalebox{1}{\includegraphics[width=\columnwidth]{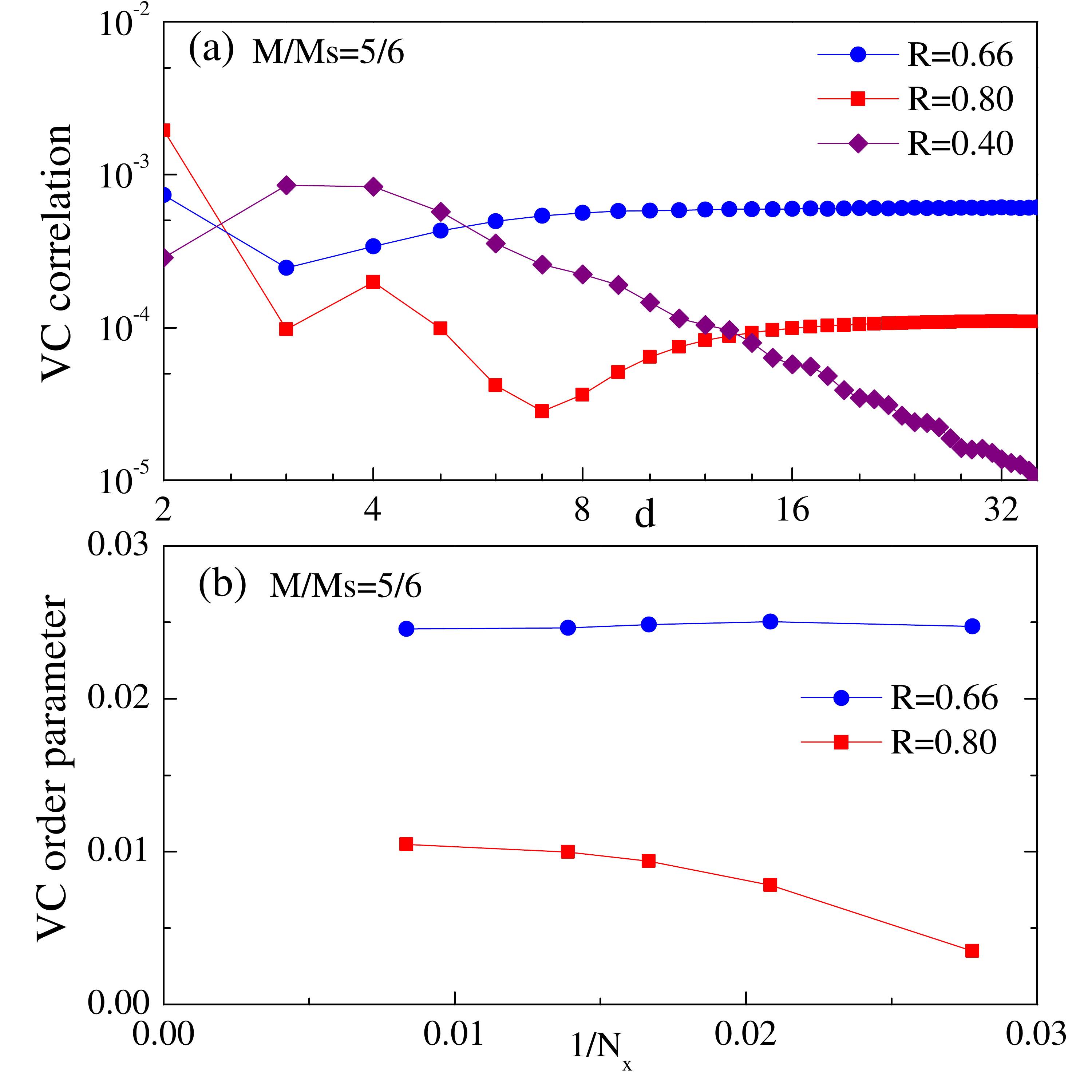}}
  \end{center}
  \caption{(Color online) Spin vector chirality (VC) correlation function, as defined in Eq.~\eqref{eq:chirality}, at $M/M_s = 5/6$ with system size $N_x = 120$ and $x'=N_x/2$.
        In (a), we show $R = 0.4$ (purple diamond) where the system orders into an incommensurate coplanar phase.
        Furthermore, $R = 0.66, 0.80$, where the system is in the cone phase, is shown on the same plot.
        We can see that the VC approaches a constant in the cone phase while decaying in the coplanar phase.
        In (b), we show the finite-size scaling of the VC order parameter in the cone phase.}
  \label{fig:spin-chiral}
\end{figure}
%

\subsection{DMRG results}

In Sec.~\ref{sec:order-param-struct}, we show that the cone state corresponds
to a single-Q condensate bosonic field, while the incommensurate planar state corresponds
to double-Q condensate. This is verified by the central charge measurement, where we find  $c= 2$ to describe the coplanar phase as shown in Fig.~\ref{fig:cone-planar}a,  as opposed to $c=1$ for the cone in Fig.~\ref{fig:cone-planar}b .

The transverse spin-spin correlation function for the cone state can be written as
\begin{eqnarray}
\label{eq:ruadd2}
  \langle S_\br^+ S_{\br'}^- \rangle & \sim & \overline{\psi}^2
  \cos\left({\bf Q} \cdot \left({\bf r}-{\bf r}'\right)\right)
  \left\langle e^{i(\theta(\br) - \theta(\br'))}\right\rangle,
  \nonumber \\
  & \sim & \overline{\psi}^2
  \cos\left({\bf Q} \cdot \left({\bf r}-{\bf r}'\right)\right) C_\eta(x,x')
\end{eqnarray}
With $C_\eta(x,x')$ given in Eq. ~\eqref{eq:ruadd1a}.   We fit the
DMRG results to this formula in Fig.~\ref{fig:corr-cone}b.   The
transverse correlation shows a clear sinusoidal pattern with
incommensurate wavevector ${\bf Q}=(1.10\pi,2\pi/3)$ and $\eta=0.37$
at $M/M_s=5/6$, $R=0.66$. Fig.~\ref{fig:corr-cone}b shows an
excellent fit which yields the exponent $\eta=0.37$.

The whole procedure is repeated for the incommensurate planar state,
 \begin{eqnarray}
 \label{eq:ruadd3}
   \langle S_\br^+ S_{\br'}^- \rangle & \sim & 4\overline{\psi}^2 \left\langle \cos\left({\bf Q} \cdot {\bf r}+{\tilde \theta(x)}\right) \cos\left({\bf Q} \cdot {\bf r}'+{\tilde \theta}(x')\right)\right\rangle  \nonumber \\
   &&\left\langle e^{i(\theta(x) - \theta(x'))}\right\rangle. \\
    &=&  \frac{\overline{\psi}^2}{2} \cos({\bf Q} \cdot (\br-\br')) C_{\eta+\tilde\eta}(x,x')\nonumber
 \end{eqnarray}
The exponent $\eta$ and ${\tilde \eta }$ come from averaging the $\theta$ and $\tilde{\theta}$ fields, respectively. The fitting estimates ${\bf Q}=(1.26\pi,2\pi/3)$ and $\eta+{\tilde \eta }=0.54$ at $M/M_s=5/6$, $R=0.3$, shown in Fig.~\ref{fig:corr-cone}a.

Next we consider the vector chirality (VC), which is defined
as $V_{x,y}=\hat{z}\cdot \langle S_{x,y}\times S_{x+1,y}\rangle$ in
Eq.~\eqref{eq:chirality}. As discussed in Sec.~\ref{sec:order-param-struct}, since the cone state favors XY order,
the VC should be a nonzero and constant value. Indeed, as shown in
Fig.~\ref{fig:spin-chiral}, the VC correlation function does not
decay with distance in the cone state, i.e., $R=0.66$ and $0.80$,
and the finite-size scaling (Fig.~\ref{fig:spin-chiral}(b)) shows
that the corresponding VC order parameter remains finite in the
thermodynamic limit.
Instead, for planar states, the spins are confined to one plane, so
the VC correlation decays exponentially (see $R=0.4$ data in Fig.~\ref{fig:spin-chiral}).

\section{Weakly Coupled Chains}
\label{sec:weak-coupled}

\subsection{Bosonization of a Heisenberg chain}
\label{subsec:bosonization}

In this section, we give a brief overview of applying Abelian bosonization to a single spin-1/2 Heisenberg chain in a magnetic field.
The Hamiltonian of interest is as follows
\begin{equation}
\label{eq:chain}
H_{ch}=J\sum\limits_{x=1}^{L} \mathbf{S}(x) \cdot \mathbf{S}(x+1)- h \sum_{x=1}^L S^z(x),
\end{equation}
where the magnetic field is chosen along the $z$-direction, and the lattice spacing has been set to 1.
Here, the magnetization, $M \equiv \sum_x \frac{1}{L}S^z(x)$, is conserved, and hence, the magnetic field, $h$, can be treated as a chemical potential to relate the properties at $h \neq 0$ to those at $h=0$.
For any magnetizations less than saturation, i.e. $M < M_{\rm sat} = 1/2$, the low energy theory can be described by a canonical set of a massless scalar field, $\theta$, and its dual field $\phi$
\begin{equation}
\label{eq:Hchain}
H_0=\int dx \frac{v}{2}((\partial_x\phi)^2+(\partial_x\theta)^2).
\end{equation}
These two fields satisfy the familiar commutation relations
\begin{equation}
\label{eq:comm}
[\theta(x),\phi(x')]=-i\Theta(x-x')
\end{equation}
where $\Theta$ is the Heaviside step function.
The spin velocity, $v$, in Eq.~\eqref{eq:Hchain}, is a function of the magnetization, $M$.
When $M = 0$, $v/J = \pi/2$, and the $SU(2)$ symmetry is restored.
For the case when $M >0$, $v$ decreases continuously and is numerically determined by the Bethe ansatz integral equations (see Fig. 9 of Ref.~\onlinecite{spinvelocityAffleck}).

At a fixed magnetization, both the longitudinal (along the field direction) and transverse (perpendicular to the field axis) spin fluctuations have gapless excitations.
The longitudinal modes occur at commensurate wave vector $k_x=0$ and incommensurate ones $k_x = \pi \pm 2\delta$, where $\delta = \pi M$,
while the transverse modes are at commensurate wave vector $k_x = \pi$ and incommensurate vectors $k_x = \pm  2\delta$.
Then, one can expand the spin operator around these low energy gapless modes, i.e.
\begin{eqnarray}
\label{eq:spinops}
S^z(x) &=& M+\mathcal{S}_0^z(x)+ e^{i(\pi-2\delta)x} \mathcal{S}_{\pi-2\delta}^z(x) \nonumber \\
    &&+  e^{-i(\pi-2\delta)x} \mathcal{S}_{\pi+2\delta}^z(x), \nonumber \\
S^+(x) &=& e^{-i2\delta x} \mathcal{S}_{-2\delta}^+(x)+  e^{i2\delta x} \mathcal{S}_{2\delta}^+(x)\nonumber\\
    &&+(-1)^x \mathcal{S}_{\pi}^+(x),
\end{eqnarray}
where ${S}_0^z$, $\mathcal{S}_{\pi\pm 2\delta}^z(x)$, $\mathcal{S}_{\pm 2\delta}^+(x)$ and $\mathcal{S}_{\pi}$ are operators whose scaling dimensions depend on $M$.
One can rewrite these operators in terms of the bosonic fields, $\phi$ and $\theta$,
\begin{eqnarray}
\label{eq:bosontoS}
\mathcal{S}_0^z(x) &=& \beta^{-1} \partial_x\phi , \nonumber \\
\mathcal{S}_{\pi-2\delta}^z(x) &=& -\frac{i}{2} A_1 e^{-2\pi i \phi/\beta} ,\nonumber \\
\mathcal{S}_{\pm 2\delta}^+(x) &=&\pm \frac{i}{2} A_2 e^{i\beta \theta} e^{\pm i 2\pi \phi/\beta} ,\nonumber \\
\mathcal{S}_{\pi}^+(x)&=& A_3 e^{i \beta \theta}.
\end{eqnarray}
Here, the parameter $\beta\equiv 2\pi \mathcal{R}$ is related to the compatification radius $\mathcal{R}$ and can be calculated by solving the integral equations, which can be found in Refs.~\onlinecite{bogoliubov1986critical, betaAffleck1997, cabraspinladder}.
The compactification radius takes on a simple form, $2\pi \mathcal{R}^2 = 1$ at zero magnetization, and approaches $2 \pi\mathcal{R}^2 = 1/2$ as $M \to M_{\rm sat} = 1/2$.
The constants, $A_1$, $A_2$ and $A_3$, are determined numerically\cite{hikihara2004correlation}.
Furthermore, at $M=0$, the scaling dimension of ${S}_0^z$ and $\mathcal{S}_{\pm 2\delta}^+(x)$ is $1$, and these operators can be written in its $SU(2)$ symmetric form ${\bf M}={\bf J}_R+{\bf J}_L$.
The scaling dimension of $\mathcal{S}_{\pi\pm 2\delta}^z(x)$ and $\mathcal{S}_{\pi}$, however, is $1/2$ at zero magnetization and is
related to the staggered N\'{e}el order, ${\bf N}$, and dimerization ${\bf \epsilon}$.
Further details for the $M=0$ case are provided in Appendix~\ref{sec:zero-field-analysis}.

Now, in order to compare our DMRG results to this analysis, we must enforce open boundary conditions (BC) along the chain direction to mimic DMRG's BC.
This can be achieved by introducing two additional ``phantom sites" at $x=0$ and $x=L+1$ \cite{OBCAffleck}.
At these positions, we enforce boundary conditions on the bosonic field, $\phi$, where $\phi(x=0)=0$ and $\phi(x=L+1)=0$.
The sum in Eq.~\eqref{eq:chain} now runs from site index 0 to L, and we effectively obtain a periodicity of $L+1$ using these phantom sites.
We can now substitute Eq.~\eqref{eq:bosontoS} into Eq.~\eqref{eq:spinops}, and enforce the open boundary conditions.
The spin operators can now be written as (for brevity, we suppress chain index $y$)
\begin{eqnarray}
\label{eq:spinopscont}
S^z(x) &=& \tilde{M}+\frac{1}{\beta}\frac{d\phi}{dx} - A_1 \sin(\frac{2\pi}{\beta}\phi(x)-(\pi-2\tilde{\delta})x), \nonumber \\
\label{eq:s+}
S^+(x) &=& e^{i\beta \theta(x)}[A_3(-1)^x\nonumber\\
    &&+A_2 \sin(\frac{2\pi}{\beta}\phi(x)+2\tilde{\delta} x)],
\end{eqnarray}
where $\tilde{M}=M L/(L+1)$ and $\tilde{\delta}=\pi \tilde{M}$.
The bosonic field, $\phi$, can also be expanded in terms of its lattice modes as
\begin{equation}
\label{eq:phi}
\phi(x)=\sum\limits_{n=1}^{\infty}\frac{\sin(q_n x)}{\sqrt{\pi n}}(a_n+a_n^+),
\end{equation}
where $q_n=\pi n/(L+1)$.
Here, $a_n$ and $a_n^+$ are the annihilation and creation operators and satisfy the commutation relation $[a_{n},a_{n'}^+]=\delta_{n,n'}$.

\subsection{Triangular spin tube}

We now extend our previous discussion to study the behavior of the TST, described by Eq.~\eqref{eq:hami}, in the limit of weak coupling, $J' \ll J$.
Using the low energy expansions of the spin operators in Eq.~\eqref{eq:spinops}, we can express the low energy Hamiltonian as $H = H_0 + H_1$, where
$H_0$ is described by a sum over the free bosonic modes in Eq.~\eqref{eq:Hchain} on each chain.
Here, $H_1$ describes interchain interactions and is as follows
\begin{eqnarray}
  \label{eq:perturbH1}
  H_1&=& J'\sum\limits_{y=1}^{3} \int\limits_{x=0}^{L}dx
  \{2 \tilde{M}^2+2\mathcal{S}_{y;0}^z
  \mathcal{S}_{y+1;0}^z \\
  &+&\sum_{\sigma=\pm} (1-e^{2i\sigma\tilde{\delta}})\mathcal{S}_{y;\pi+2\sigma\tilde{\delta}}^z \mathcal{S}_{y+1;\pi-2\sigma\tilde{\delta}}^z \nonumber \\
  &+&\frac{1}{2}[\mathcal{S}_{y;\pi}^+ \partial_x \mathcal{S}_{y+1;\pi}^- + {\rm h.c.}] \nonumber\\
  &+& \sum_{\sigma=\pm} \left[
    \left(\frac{1+e^{2i\sigma\tilde{\delta}}}{2}\right) \mathcal{S}_{y;2\sigma\tilde{\delta}}^+
  \mathcal{S}_{y+1;2\sigma \tilde{\delta}}^-  +{\rm h.c.}\right]\}, \nonumber
\end{eqnarray}
where again, $\tilde{M}=M L/(L+1)$.

The first term, $2\tilde{M}^2$, with scaling dimension 0, is the most relevant, but is trivially a constant.
The second term is marginal with scaling dimension 2, and
renormalizes the Luttinger parameters and the velocities of the bosonic fields, $\phi, \theta$, in Eq.~\eqref{eq:Hchain}.
The third term is relevant at $\tilde{M} = 0$ with scaling dimension 1, and becomes marginal as magnetization increases,
approaching a scaling dimension 2 as $\tilde{M} \to M_{\rm sat}$.
This term is responsible for the SDW phase that arises when relevant.
The fourth term, which involves a derivative, is marginal at $\tilde{M} = 0$ with scaling dimension 2 and
becomes increasingly relevant with increasing magnetization, saturating to a scaling dimension of 3/2 as $\tilde{M} \to M_{\rm sat}$.
This is a ``twist" term that favors the cone or XY phase that orders perpendicular to the magnetic field.
The last term is always irrelevant, with scaling dimension $\ge 2$ and can be neglected in the analysis of this theory.

Apart from the trivial constant term, the SDW and the ``twist" terms are the most relevant ones and have competing scaling dimensions as magnetization varies from $0$ to saturation.
With the exception of some subtleties that arise from the TST boundaries (we discuss this in later subsections),
standard scaling arguments can be made about these two operators.
For small $M$, the SDW term dominates, and the system orders into a collinear SDW in which the ordering momentum, $\pi - 2 \tilde{\delta}$ scales linearly with
magnetization.
The twist interaction dominates over the SDW at a larger magnetization, and the system orders into a cone-like state.
Since there is no spontaneous breaking of continuous symmetry in one dimension, the SDW and cone order are not really ordered states, but are Luttlnger liquids with one gapless mode.
This competition between cone and SDW phase was discussed for 2d triangular lattice in Ref.~\onlinecite{starykh2010extreme}, where critical
magnetization, $M_{\rm crit}$, at which the quantum phase transition from the SDW to the cone phase takes place, was
evaluated.  The TST has the same critical $M_{\rm crit} = 0.64 M_{\rm sat}$ as the 2d case, except that the cone state obtained in this
quasi-1d regime is smoothly connected to the cone phase obtained in the high field region in Sec.~\ref{sec:high field}.

Eq.~\eqref{eq:perturbH1} is not complete as it does not account for
several less-obvious relevant terms which are allowed by the lattice
symmetry of the problem. This will be considered in more detail
later.  Within the SDW phase, it is possible to lock the SDW momentum
to a commensurate value by accounting for high-order umklapp
processes. The first of these leads to a commensurate SDW, which
is in fact identical to the 1/3 plateau with the ``up up
down'' structure.  This is discussed
extensively later in Sec.~\ref{sec:plateau}.

Other more relevant intra-chain interaction terms may appear due to
fluctuations that are not accounted for in the na\"ive bosonization in
Eq.~\eqref{eq:spinopscont}.    We will discuss these effects in
Appendix~\ref{sec:cone}.

\subsection{SDW}
\label{subsec:sdw}

In the region of low to intermediate magnetization and small $J'$, we
can neglect all terms in $H_1$ except the marginal one and the SDW
interaction.  Using bosonization, Eq.~\eqref{eq:spinopscont}, the
Hamiltonian can be re-written as follows
\begin{eqnarray}
\label{eq:hamisdw}
H_{sdw} &=& \sum\limits_{y=1}^{3}\int dx \frac{v}{2} \left[(\partial_x\phi_y)^2 + (\partial_x\theta_y)^2\right] +\frac{2J'}{\beta^2} \partial_x \phi_y \partial_x \phi_{y+1}
\nonumber\\
&+&  \gamma_{\rm sdw}\cos[\frac{2\pi}{\beta}(\phi_y-\phi_{y+1})-\frac{\pi-2\tilde{\delta}}{2}] .
\end{eqnarray}
where the bare SDW coupling is given by $\gamma_{\rm
  sdw}=J'A_1^2\sin(\tilde{\delta}) > 0$.

\subsubsection{Scaling considerations}
\label{sec:scal-cons}

Renormalization group arguments give considerable insight into
the physics of Eq.~\eqref{eq:hamisdw}.  All but the last term in
$H_{sdw}$ are scale invariant, and can be considered a fixed point
Hamiltonian.  The remaining SDW term, proportional to $\gamma_{\rm sdw}$,
is not, and renormalizes under the scale transformation $x \rightarrow
b x$, according to the usual linearized relation
\begin{equation}
  \label{eq:29}
  \gamma_{\rm sdw}(b) = b^{2-\Delta_{\rm sdw}} \gamma_{\rm sdw},
\end{equation}
where $b>1$ is an arbitrary scale factor.   As discussed in the
previous subsection, $\Delta_{\rm sdw}<2$, so that the SDW interact is
{\sl relevant}, and grows in strength under rescaling.  Eq.~\eqref{eq:29} is valid
for small dimensionless $\gamma_{\rm sdw}(b)$, and therefore the weak
coupling regime is limited by the condition $\gamma_{\rm sdw}(b)<
v$.  This defines an ``SDW correlation length'' $\xi_{\rm sdw}$ such
that $\gamma_{\rm sdw}(b)=v$:
\begin{equation}
  \label{eq:41}
  \xi_{\rm sdw} \sim (v/\gamma_{\rm sdw})^{1/(2-\Delta_{\rm sdw})}.
\end{equation}
In the weakly coupled chain regime, $\gamma_{\rm sdw}$ is small and so
$\xi_{\rm sdw}$ is large.  On scales large compared to this correlation length, we expect that
the bosonic modes appearing inside the SDW term become ``pinned'' to
values which minimize this interaction.  This pinning corresponds to
the creation of well-established SDW order.

Due to the divergence of $\xi_{\rm sdw}$, however, the establishment
of SDW order can be prevented by finite size effects, even for
reasonably large systems accessible by DMRG.  For a finite system of
length $L$, we must compare the SDW correlation length to $L$, and it
is expected that physical quantities will be functions of the
dimensionless ratio $\Xi_{\rm sdw}\equiv \xi_{\rm sdw}/L$.  For
$\Xi_{\rm sdw} \ll 1$, SDW-like behavior is expected, but when
$\Xi_{\rm sdw} \gtrsim 1$, there may be a non-trivial crossover.  This
occurs particularly in the case of the TST, for which an analysis,
detailed below, shows that the crossover is {\sl discontinuous}.

\subsubsection{L = $\infty$}
\label{sec:l-=-infty}

For an infinitely {\sl long} system, $\Xi=0$, we can understand the nature of the
SDW state by simply minimizing the $\gamma_{\rm sdw}$ term in
Eq.~\eqref{eq:hamisdw}.  When the {\sl width} is also infinite,
i.e. in two dimensions, one can simultaneously minimize each cosine
term (for each $y$) independently.  This occurs by taking
\begin{equation}
\label{eq:phicond}
\left.\frac{2\pi}{\beta}\phi_y\right|_{\rm d=2}=\varphi+\frac{\pi-2\tilde{\delta}}{2}y ,
\end{equation}
where $\varphi$ is an arbitrary constant ($x$- and $y$-independent)
phase. Allowing for small gradients of $\varphi$, which might be
present due to fluctuations or perturbations and by substituting
Eq.(79) into Eq.(73), we see that the spin operator can then be
represented as
\begin{equation}
  \label{eq:51}
 \left. S_y^z(x) \right|_{\rm d=2} \sim \tilde{M} + \frac{\partial_x \varphi}{2\pi} - A_2 \sin
  \big[ \varphi(x) - \tfrac{\pi-2\tilde{\delta}}{2} (2x-y)\big],
\end{equation}
which indeed is the classic form for a spin density wave with
wavevector $\frac{\pi-2\tilde{\delta}}{2}(-2,1)$.  This corresponds to an ideal two dimensional
SDW state, and $\varphi$ gives the ``sliding'' or ``phason''\cite{chaikin2000principles} mode
of the SDW.  For generic irrational $\tilde\delta/\pi$, $\varphi$
remains a gapless pseudo-Goldstone mode associated with translational
symmetry breaking.  In two dimensions, the zero point fluctuations of
this mode do not, however, destroy long-range SDW order.

Now consider the case of the TST ladder, where $y=1,2,3$ and periodic
boundary conditions are applied.  In this case it is generically impossible to
simultaneously minimize each cosine term separately.  Instead, the
minimum occurs when
\begin{equation}
  \label{eq:42}
  \left.\frac{2\pi}{\beta} \phi_y\right|_{L=\infty, {\rm TST}} = \varphi +
  \frac{2\pi}{3} y,
\end{equation}
where again $\varphi$ is an arbitrary constant, reflecting the
invariance of Eq.~\eqref{eq:hamisdw} under uniform translations of all
the $\phi_y$.  Again, one can express the spin operator here using
this form
\begin{eqnarray}
  \label{eq:52}
  \left. S_y^z(x) \right|_{L=\infty,{\rm TST}} &  \sim & \tilde{M} +
  \frac{\partial_x \varphi}{2\pi} \\
  & & - A_2 \sin
  \big[ \varphi(x) - (\pi-2\tilde{\delta})x +\tfrac{2\pi}{3}
  y\big]. \nonumber
\end{eqnarray}
In contrast with Eq.~\eqref{eq:phicond}, the minimum configuration in
the TST, Eq.~\eqref{eq:42} is {\sl independent} of $\tilde\delta$,
manifesting in Eq.~\eqref{eq:52} as a difference dependence on $y$
from Eq.~\eqref{eq:51}.  The difference is due to the frustration of
the intrinsic 2d SDW order by periodic boundary conditions, which tend
to lock the SDW order to a commensurate form in the $y$ direction.
Interestingly, the two results coincide when $\tilde\delta = \pi/6$,
which corresponds to the case $M=M_{\rm sat}/3$.  At this point, the
periodicity of the TST and the SDW order are compatible.

As in the 2d case, at the level of Eq.~\eqref{eq:hamisdw} applied to
the TST, the uniform translation mode $\varphi$ remains gapless.
Unlike the 2d case, however, in one dimension, the zero point fluctuations
of this mode are sufficient to disrupt long range SDW order, which
instead manifests as power law correlations.  Nevertheless, the short
distance physics is still that of an SDW, and moreover the 1d
fluctuations are easily accounted for theoretically.  This is
accomplished simply by treating $\varphi$ as a free massless
boson, as we discuss below in Sec.~\ref{sec:finite-length-linfty}.

\subsubsection{Finite length $L<\infty$}
\label{sec:finite-length-linfty}

As we have discussed in Sec.~\ref{subsec:bosonization}, for a finite
length chain, we must impose the boundary conditions $\phi_y(x=0) =
\phi_y(x=L)=0$.  These conditions are {\sl incompatible} with the
values, in Eq.~\eqref{eq:42}, which minimize the SDW term in the
infinitely long case.  This means that end effects strongly affect,
and tend to suppress SDW ordering.  What do we expect?  For short
systems, where $\Xi \gg 1$, the end effects will dominate, and the
effects of the SDW interaction become negligible.  In other words, all
components $\phi_y$ will be largely not affected by the SDW term, and the
system should behave similarly to three decoupled chains of finite
length.  For long systems, $\Xi \ll 1$, the SDW pinning should be
effective far from the boundaries, and only the pseudo-Goldstone mode
$\tilde\Phi_0$ will behave like a massless field (pinned at the
boundaries).

Let us now address the crossover.  It is convenient to first make a change of
basis \cite{cabra1998magnetization} from the $\phi_1,\phi_2,\phi_3$ to new fields $\Phi_0,\Phi_1,\Phi_2$:
\begin{eqnarray}
\label{eq:sdw11}
\left( \begin{array}{c} \phi_1 \\ \phi_2 \\ \phi_3 \end{array} \right) = \begin{pmatrix} 1/\sqrt{3} & 1/\sqrt{2} & 1/\sqrt{6} \\
1/\sqrt{3} & 0 & -2/\sqrt{6} \\ 1/\sqrt{3} & -1/\sqrt{2} & 1/\sqrt{6} \end{pmatrix} \left( \begin{array}{c} \Phi_0 \\ \Phi_1 \\ \Phi_2 \end{array} \right) .
\end{eqnarray}
The dual fields $\theta_y$ transform similarly. Note that the center
of mass field is just proportional to the SDW phase introduced
earlier: $\Phi_0 = \frac{\sqrt{3}\beta}{(2\pi)} \varphi$. The boundary
conditions $\phi_y=0$ at the ends translate to $\Phi_i=0$ at the ends.
The SDW Hamiltonian now reads $H_{\rm sdw} = H_{\rm sdw}^{(0)} +
H_{\rm sdw}^{(1)}$, where the harmonic part
\begin{equation}
\label{eq:sdw12}
 H_{\rm sdw}^{(0)} = \sum\limits_{n=1}^{3} \int dx \left[ \frac{\tilde{v}_n}{2\kappa_n} (\partial_x\Phi_n)^2 + \frac{\tilde{v}_n\kappa_n}{2}(\partial_x\Theta_n)^2\right]
\end{equation}
is expressed in terms of renormalized stiffnesses $\kappa_0^{-2} = 1 + 4 J'/(\beta^2 v)$ and $\kappa_{1,2}^{-2} = 1 - 2J'/(\beta^2 v)$
and velocities $\tilde{v}_n = v/\kappa_n$.
Its interacting part (the analog of the second line in Eq.~\eqref{eq:hamisdw} written in the new basis) reads
\begin{eqnarray}
\label{eq:sdw13}
H_{\rm sdw}^{(1)} &=& \gamma_{\rm sdw} \int dx  ~2 \cos[\frac{2\pi}{\sqrt{2}\beta} \Phi_1 - \frac{\pi-2\tilde{\delta}}{2}] \cos[\frac{2\pi}{\beta}\sqrt{\frac{3}{2}}\Phi_2]
\nonumber\\
&+& \cos[\frac{2\pi}{\beta} \sqrt{2}\Phi_1 + \frac{\pi-2\tilde{\delta}}{2}].
\end{eqnarray}
Note that the center-of-mass mode $\Phi_0 \propto \varphi$ does not enter in
Eq.~\eqref{eq:sdw13}.  Thus it behaves as a free massless boson,
independent of the strength of the SDW coupling. The distinction between $\delta$ and $\tilde\delta$ in the SDW Hamiltonian is not important when analyzing the crossover, and will be dropped in this subsection from now on.

To analyze the crossover, we first carry out the renormalization group
procedure by integrating out fluctuations of the fields due to modes
with wavelength less than the system size $L$.  In doing so, we
replace $\gamma_{\rm sdw}$ by its renormalized value at this scale,
\begin{eqnarray}
  \label{eq:43}
 \gamma_{\rm sdw} & \rightarrow & \gamma_{\rm sdw}(L) =
  L^{-\Delta_{\rm sdw}}  \gamma_{\rm sdw} .
\end{eqnarray}
Note that we have done the coarse-graining step of the RG of
integrating out modes, but we have not rescaled any fields or
coordinates, so as to keep the original units unchanged for clarity.
Under this coarse-graining transformation, the quadratic terms in
the Hamiltonian remain unmodified.

In this renormalized Hamiltonian, it is appropriate to carry out a
classical saddle point approximation for $\Phi_1$ and $\Phi_2$, which are the
fields pinned by the SDW coupling.  The SDW potential in
Eq.~\eqref{eq:sdw13} is minimized by $\Phi_2=0$, which is compatible
with the boundary condition, and so, we can impose this condition.  Then
only $\Phi_1$ enters the saddle
point condition in a non-trivial way.  For simplicity we specialize to
the case $\delta=\pi/6$, or $M=M_{\rm sat}/3$.  Then we may
define $\Psi = \frac{2\pi}{\sqrt{2}\beta} \Phi_1 + \frac{2\pi}{3}$, for which
the saddle point Hamiltonian, neglecting the decoupled $\Phi_0$ term becomes
\begin{equation}
\label{eq:sdw8}
H_{\rm class} = \int_0^L dx\, \Big\{ K (\partial_x \Psi)^2 - \gamma_{\rm
  sdw}(L) (\cos[2 \Psi] + 2 \cos[\Psi]) \Big\},
\end{equation}
with $K=\beta^2 \tilde{v}_1/4\pi^2 \kappa_1$.

The $\gamma_{\rm sdw}$ term is clearly minimized by $\Psi=0$, while the open
boundaries require $\Psi(0)=\Psi(L)=2\pi/3$, causing the strong
suppression of SDW order by the ends.  There can be a non-trivial
configuration, $\Psi(x)$, which minimizes the functional $H_{\rm
  class}$.  To bring out the crossover physics, we transform to
dimensionless coordinates, letting
\begin{eqnarray}
  \label{eq:44}
  x & = & \sqrt{K/\gamma_{\rm sdw}(L)} z,
\end{eqnarray}
which gives
\begin{equation}
  \label{eq:45}
  H_{\rm class} =  \epsilon_0\int_0^{\tilde{L}}
  dz\, \Big\{  (\partial_z \Psi)^2 - (\cos[2 \Psi] + 2 \cos[\Psi])
  \Big\},
\end{equation}
with
\begin{eqnarray}
  \label{eq:46}
  \epsilon_0 & = & \sqrt{K \gamma_{\rm sdw}(L) }, \\
  \tilde{L} & = & (L/\xi)^{1-\Delta_{\rm sdw}/2} = \Xi_{\rm
    sdw}^{\Delta_{\rm sdw}/2-1}, \\
  \xi_{\rm sdw} & = & (K/\gamma_{\rm sdw})^{1/(2-\Delta_{\rm sdw})}.\label{eq:47}
\end{eqnarray}
Note that Eq.~\eqref{eq:47} agrees, at the level of scaling, with
Eq.~\eqref{eq:41} obtained earlier from general arguments.  For the
purpose of minimization, the overall prefactor $\epsilon_0$ is
irrelevant, so it is clear already from Eq.~\eqref{eq:45} that the
properties are a function of the scaling variable $\Xi_{\rm sdw}$
only, as expected.

We are now prepared for the saddle point approximation, which consists
in minimizing Eq.~\eqref{eq:45}.   Starting from
the Euler-Lagrange equation, which has the usual ``energy'' integral of
motion, one obtains
\begin{equation}
  \left( \frac{d\Psi}{d z}\right)^2 = C - (2 \cos[\Psi] + \cos[2 \Psi]),
\label{eq:sdw9}
\end{equation}
where the integration constant (``energy'') $C$ is fixed by the condition
$d\Psi(z=\tilde{L}/2)/dz = 0$ as $C = (2 \cos[\Psi_{1/2}] +
\cos[2 \Psi_{1/2}])$, where we denote $\Psi_{1/2} \equiv
\Psi(z=\tilde{L}/2)$.  As a result the mid-ladder value of $\Psi$ is
implicitly given by the following integral
\begin{eqnarray}
\label{eq:sdw10}
&&\int_{\Psi_{1/2}}^{2\pi/3} \frac{d \varphi}{\sqrt{ 2 \cos[\Psi_{1/2}] + \cos[2 \Psi_{1/2}] - 2 \cos[\varphi] - \cos[2 \varphi]}} \nonumber\\
&&= \frac{\tilde{L}}{2}.
\end{eqnarray}

\begin{figure}[t]
  \begin{center}
  \scalebox{0.9}{\includegraphics[width=\columnwidth]{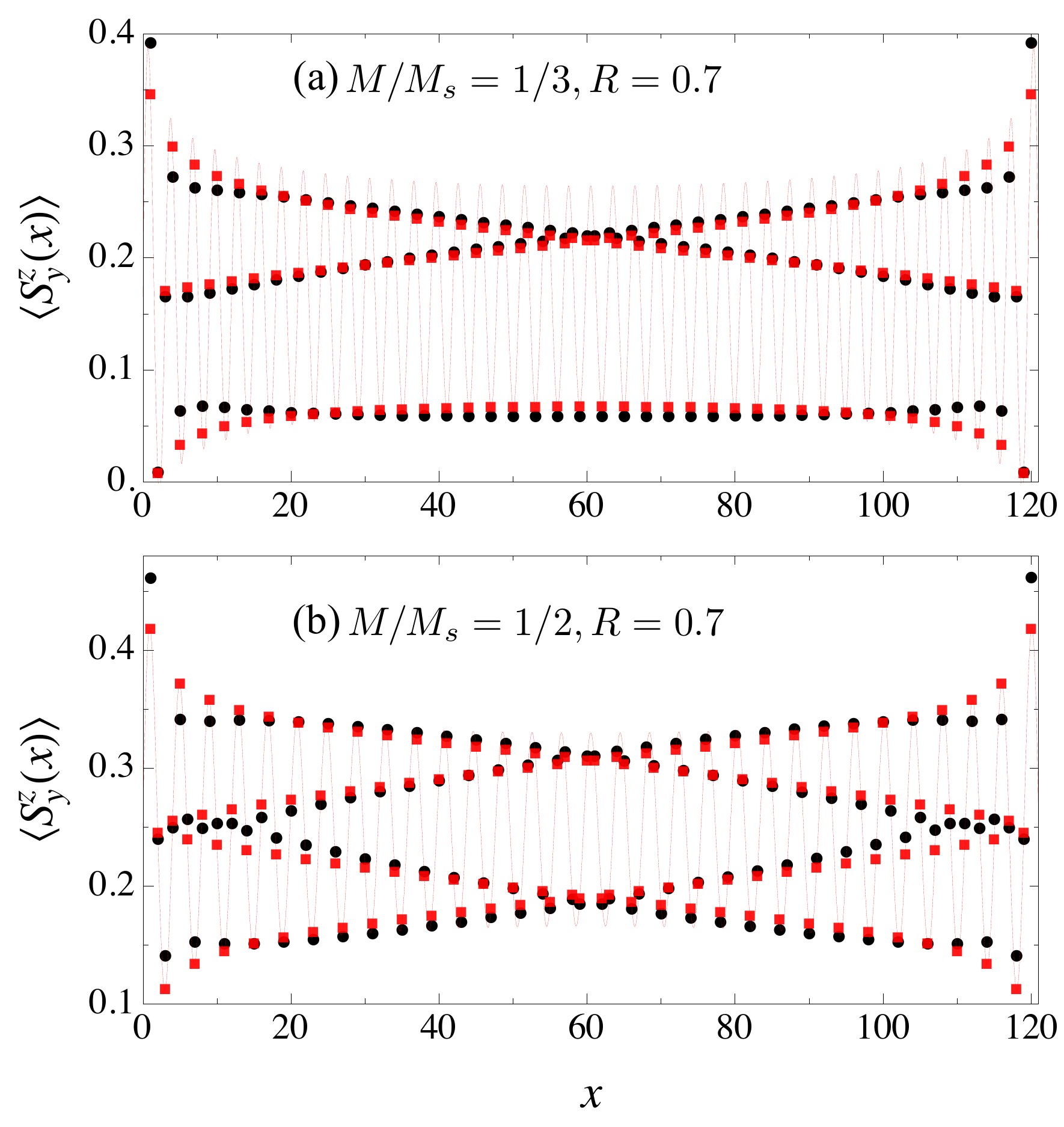}}
  \end{center}
  \caption{(Color online) $S_z$ profile for $R = 0.7$ at (a) $M/M_s=1/3$ plateau and (b)$M/M_s=1/2$
  in the SDW state for the non-frustrated chain (see text).
  We show the DMRG results by (black) circles and the theoretical prediction, Eq.~\eqref{eq:sdw5}, by (red) solid line/square.
  The theoretical line captures all the DMRG data points,
  which appear to form three different curves. The (red) squares show
  the $S^z(x)$ values at discrete lattice site positions $x$, as obtained from \eqref{eq:sdw5}.
  }
  \label{fig:sz2M}
\end{figure}

The full crossover (in this saddle point approximation) is obtained
from Eq.~\eqref{eq:sdw10}.  First, we observe that in the limit
$\Psi_{1/2} \to 0$, the above integral diverges logarithmically,
implying that indeed, $\Psi(\tilde{L}/2) =0$ in the infinite-size
limit.  The short system size limit is less obvious.  For small
$\tilde{L}$, we must choose $\Psi_{1/2}$ to minimize the integral.
However, if we make the obvious choice to let $\Psi_{1/2} = 2\pi/3 -
\epsilon$, with $\epsilon \to 0^+$, one finds that the integral in fact
does not vanish but approaches the {\sl constant} value
$\pi/\sqrt{6}$.  In fact, the integral as a
function of $\Psi_{1/2}$ has a
non-monotonic dependence, and the minimum value of the integral is
$\approx 1.1436 < \pi/\sqrt{6}=1.2826$, which is achieved for
$\Psi_{1/2}\approx 1.3178 < 2\pi/3 =2.0944$.  Regardless, the lower
bound on the integral implies that there is a minimum dimensionless
length, $\tilde{L}_{\rm min} \geq 2.28$, such that for $\tilde{L}<
\tilde{L}_{\rm min}$, the minimum action solution is simply
$\Psi_{1/2}=2\pi/3$, i.e. $\Psi(z)=1/2$ for {\em all} $z$.  For such
short systems, the boundary conditions {\sl completely} disrupt the
SDW order, and the system behaves as though it were just decoupled
chains.  The transition from $\tilde{L}< \tilde{L}_{\rm min}$ to
$\tilde{L}> \tilde{L}_{\rm min}$ is evidently discontinuous, since
$\Psi_{1/2}$ must jump from a value $\Psi_{1/2} \leq 1.3178$ at
$\tilde{L}=\tilde{L}_{\rm min}+\epsilon$ to $\Psi_{1/2}=2\pi/3$ for
shorter systems. To precisely determine the value
of $\tilde{L}_{\rm min}$ requires a comparison of the action of the
non-trivial and trivial solutions to see where they cross.

What are the consequences of this transition?  In numerics, the
transition can be probed by varying $L$ {\sl or} varying $J'/J$ at
fixed $L$.  In either case, on crossing the transition, one expects a
sharp change from SDW-like behavior for $\tilde{L}>\tilde{L}_{\rm
  min}$ to decoupled chain-like behavior for $\tilde{L}<\tilde{L}_{\rm
  min}$.  In the SDW-like regime, the two modes $\Phi_1,\Phi_2$
may be considered to have developed a gap, and consequently, the
entanglement entropy of a bipartite cut of the sample is reduced
compared to the decoupled chain-like regime.  Specifically, in the
SDW-like regime a logarithmic growth with $L$ is expected and consistent
with central charge $c=1$, while in the decoupled chain regime, the
behavior should be closer to $c=3$.  {\sl At} the transition, a sharp
{\sl drop} with increasing $L$ of the entanglement entropy is expected.
More detailed predictions can be made for the spin density profile,
$\langle S^z_y(x)\rangle$.  We make such a comparison in the following
subsection.

\begin{figure}[t]
  \begin{center}
  \scalebox{1}{\includegraphics[width=\columnwidth]{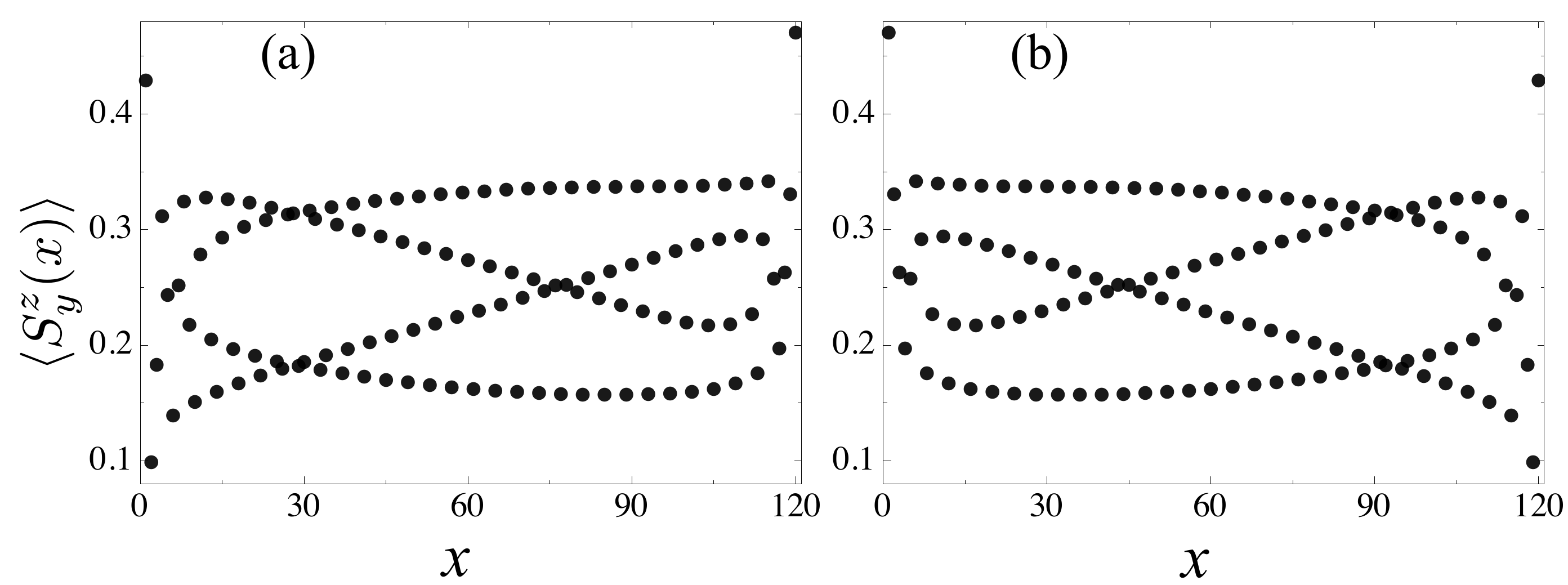}}
  \end{center}
  \caption{$S^z$ profile of DMRG result for SDW state at $M/M_s=1/2$,
    $R=0.7$ for frustrated chains, (a) $y=1$ and (b) $y=3$.  We can
    see from these plots that the translational symmetry is
    spontaneously broken, and that the SDW is strongly affected by
    boundaries.  The DMRG data are seen to obey the symmetry $\langle
    S_1^z(x)\rangle= \langle S_3^z(L+1-x)\rangle$, which follows from
    Eq.~(\ref{eq:sdw5}).}
  \label{fig:sz-frustrated}
\end{figure}

\subsection{DMRG results for SDW}
\label{sec:dmrg-results-sdw}

A number of measurements in the DMRG give evidence of the SDW state.
As discussed in the previous subsection, the SDW regime of long TSTs
can be described by pinning the fields $\Phi_1=\Phi_2=0$, and allowing
for gapless fluctuations of the free massless boson field $\Phi_0$.
In the semiclassical approximation discussed in
Sec.~\ref{sec:finite-length-linfty}, one can do somewhat better by
using the $\Phi_0$ fluctuations {\sl and} replacing $\Phi_2 \rightarrow
0$ and $\Phi_1(x) \rightarrow
\frac{\sqrt{2}\beta}{2\pi}(\Psi(x)-\frac{2\pi}{3})$, with $\Psi(x)$
given by the solution of Eq.~\eqref{eq:sdw9}.  In this way, one
obtains from Eq.~\eqref{eq:spinopscont}
\begin{eqnarray}
\label{eq:sdw5}
\langle S_y^z(x)\rangle &=& \tilde{M} + \frac{2-y}{2\pi} \partial_x \Psi(x) \\
&-& \frac{A_1}{X^{\eta_{\rm sdw}}} \sin[(2-y)(\Psi(x)-\frac{2\pi}{3}) - (\pi - 2 \tilde\delta)x]. \nonumber
\end{eqnarray}
Here the quantity
\begin{equation}
  \label{eq:sdw6}
  X = [\frac{2(L+1)}{\pi}\sin(\frac{\pi|x|}{L+1})],
\end{equation}
arises from the quantum average over the free boson field
$\Phi_0$, which is evaluated along the lines of Ref.~\onlinecite{hikihara2004correlation}, with
the result that the exponent
\begin{equation}
  \label{eq:48}
  \eta_{\rm sdw} = \frac{\pi\kappa_0}{3\beta^2} = \frac{\kappa_0}{6}\frac{1}{2\pi{\cal R}^2}.
\end{equation}
For $M=M_{\rm sat}/3$ and small $J'$, we estimate $\kappa_0 \approx 1$
and $2\pi{\cal R}^2 \approx 1 -1/(2\ln[6 \sqrt{8/(\pi e)}]) = 0.72$ (see Appendix A of Ref.~\onlinecite{starykh2010extreme}),
which leads to $\eta_{\rm sdw} \approx 0.23$, so the spin density profile decays quite slowly with distance
from the boundary in the SDW regime.  Note that the $y=2$ chain does
not depend on $\Psi$, so one can directly compare the numerically
obtained magnetization profile for the `non-frustrated' chain with
Eq.~\eqref{eq:sdw5}, see Fig.~\ref{fig:sz2M} below.

\begin{figure}[t]
  \begin{center}
  \scalebox{0.95}{\includegraphics[width=\columnwidth]{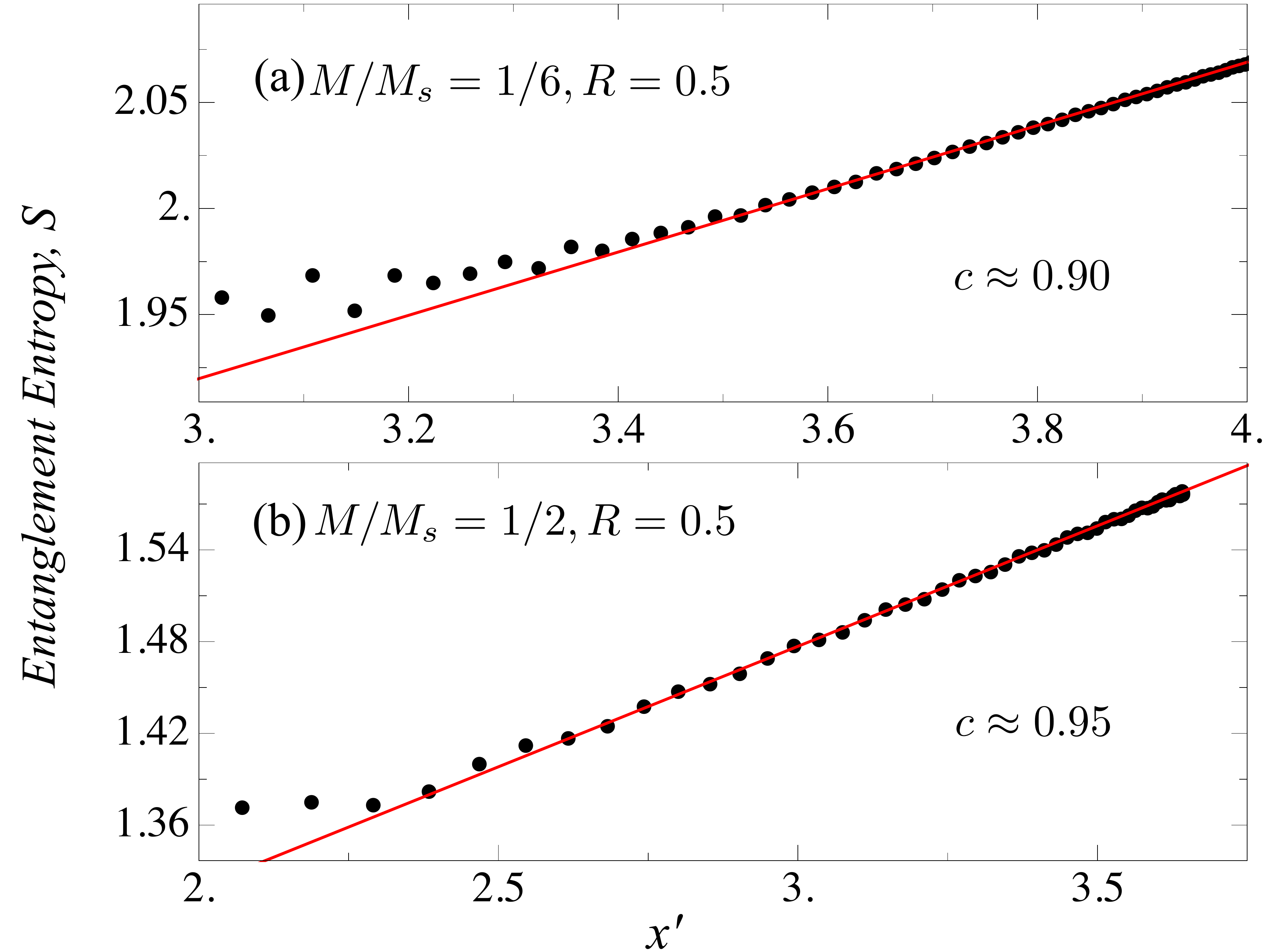}}
  \end{center}
  \caption{Entanglement entropy for SDW phase for $R = 0.5$ at (a)$M/M_s=1/6$ below the 1/3 plateau and (b)$M/M_s=1/2$ above the 1/3 plateau.
  Due to large finite-size effects of this measurement, we chose to run our simulations on a larger system size, $N_x=180$, in (a), to compare to a smaller size, $N_x=120$ in (b). }
  \label{fig:cc-sdw}
\end{figure}

One may wonder about the
selection of the $y=2$ chain.  For the geometry of our simulations,
the model has full translational symmetry, $y\rightarrow y+1$ in the
$y$ direction.  This symmetry is broken by our {\sl combined} choice
of saddle point $\Psi=\Phi_0=0$ in the bulk {\sl and} the boundary
condition $\Phi_0=0$ at the edges.  Examination of the interaction
term in Eq.~\eqref{eq:sdw13} shows that there are apparently two other
minimum solutions, $\Psi=\pi$ and $\Phi_2 = \pm \beta/\sqrt{6}$.  In
the infinite system, these are {\sl equivalent} to the one we have
chosen, insofar as they give identical results for all operators if we
make a suitable translation of $\Phi_0$.  However, the choice of
boundary condition for $\Phi_0$ prevents this translation and results
in a broken symmetry state.  By a different choice of the otherwise
equivalent saddle points, we can obtain formulae analogous to
Eq.~\eqref{eq:sdw5} but with the $y=1$ or $y=3$ chains independent of
$\Psi$.  In principle, for a finite system even the discrete
translational symmetry should be unbroken, but the restoration of this
symmetry is probably only at extremely low energies at which tunneling
occurs between these minima, and indeed we find the symmetry to be
spontaneously broken in our DMRG simulations.

In the decoupled regime, $\tilde{L}<\tilde{L}_{\rm min}$, it is more
appropriate to just calculate the spin expectation value using the
free theory, Eq.~\eqref{eq:sdw12}, for all three fields
$\Phi_0,\Phi_1,\Phi_2$.  Then we obtain, instead of Eq.~\eqref{eq:sdw5},
the result that
\begin{equation}
  \label{eq:49}
  \langle S_y^z(x)\rangle = \tilde{M} + \frac{A_1}{X^{\eta_{\rm dc}}}
  \sin[(\pi - 2 \tilde\delta)x] ,
\end{equation}
where the ``decoupled chains'' exponent is
\begin{equation}
  \label{eq:50}
  \eta_{\rm dc} = \frac{\pi(\kappa_0+\kappa_1+\kappa_2)}{3\beta^2}.
\end{equation}
In the same small $J'$ approximation, this gives $\eta_{\rm dc}
\approx 3 \eta_{\rm sdw}$, so that $\eta_{\rm dc} \approx 0.610$.  Note
that there is a much more rapid decay of the spin density profile from the
boundary in this regime.

We compare the spin density profile in Eq.~\eqref{eq:sdw5} with our
DMRG data and find reasonable agreement.  Fig.~\ref{fig:sz2M} shows a comparison of numerical data with
magnetization profile of the non-frustrated chain, i.e. the $y=2$ result
of Eq.~\eqref{eq:sdw5}, while Fig.~\ref{fig:sz-frustrated} shows that of frustrated chains, $y = 1,3$.

\begin{figure}[t]
  \begin{center}
  \scalebox{0.95}{\includegraphics[width=\columnwidth]{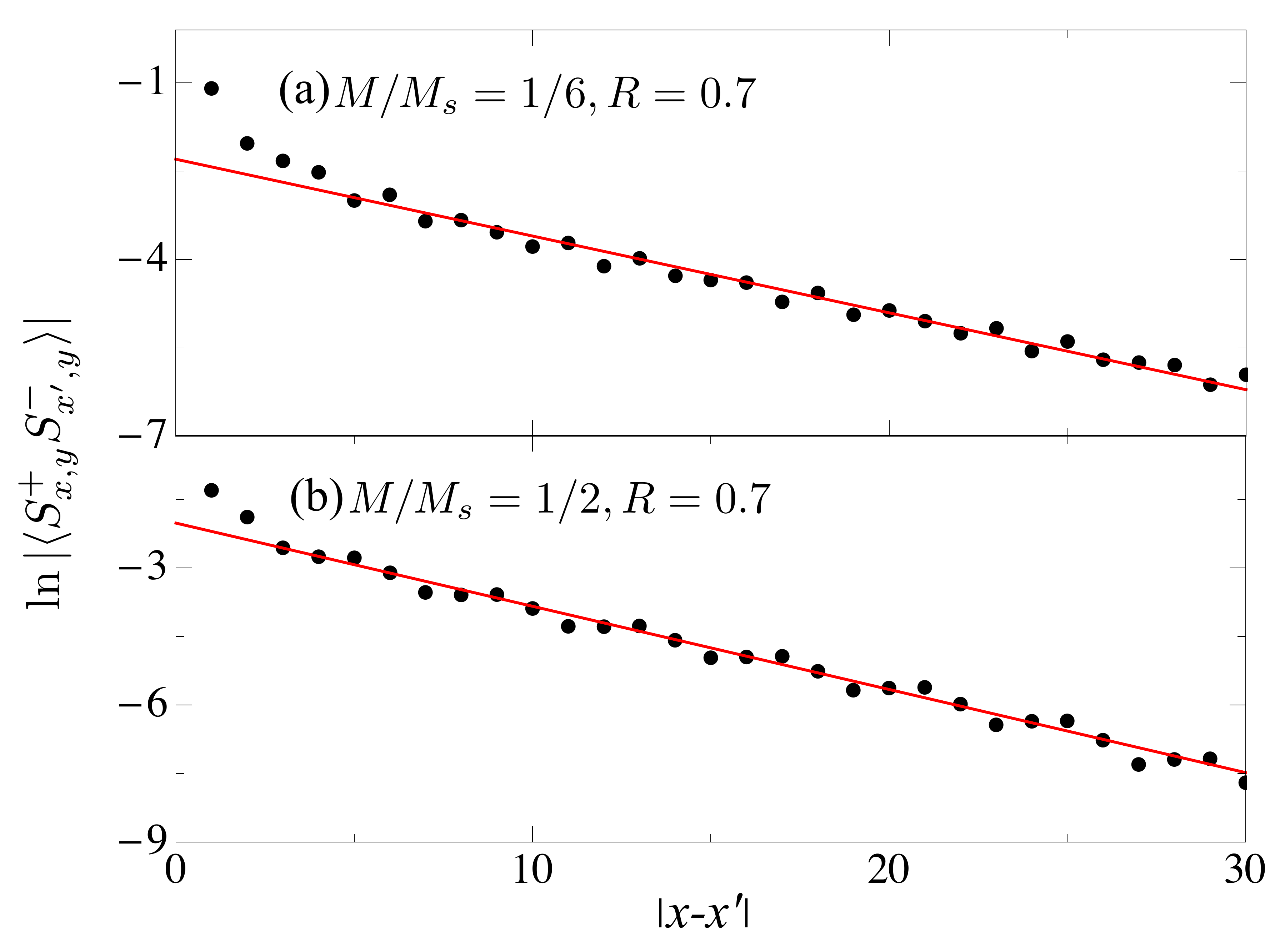}}
  \end{center}
   \caption{(Color online) The transverse spin-spin correlation function on a log-linear scale for $R = 0.7$ and at magnetizations (a)$M/M_s=1/6$ and (b)$M/M_s=1/2$, both in the SDW state for system size $N_x=120$ and $x'=N_x/2$. Data points are shown as (black) circles while the (red) line is a fit to a pure exponential function.}
  \label{fig:corrxy-sdw}
\end{figure}

We can also measure in DMRG the central charge via entanglement entropy, which yields $c=1$ for the SDW phase as opposed to $c=3$ for decoupled chains.
This is shown in Fig.~\ref{fig:cc-sdw}, where the plots show that at magnetizations $M/M_s = 1/6, 1/2$ for $R = 0.5$, the central charges obtained from numerics are $c = 0.9, 0.95$, respectively.
These values are very close to the predicted $c=1$, which gives evidence for the SDW.

\begin{figure}[t]
  \begin{center}
  \scalebox{0.99}{\includegraphics[width=\columnwidth]{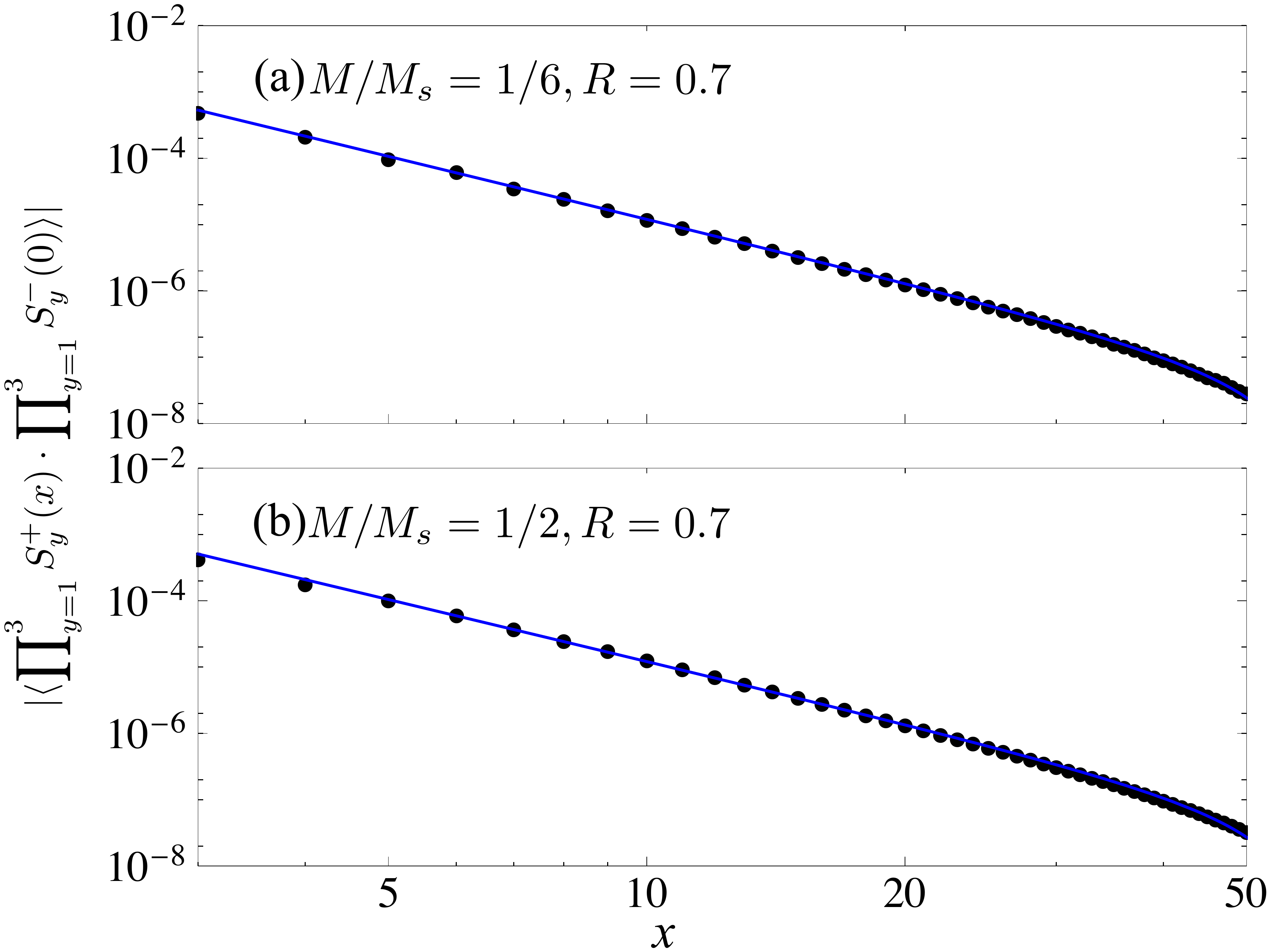}}
  \end{center}
   \caption{(Color online) The ``octupolar" correlation function, $\langle (\Pi_{y=1}^3 S_y^+(x))
(\Pi_{y=1}^3 S_y^-(x'))\rangle$ with $x'=N_x/2$, shown on a log-log scale. We show for $R= 0.7$ at magnetizations (a) $M/M_s=1/6$ and (b)$M/M_s=1/2$ in the SDW state.  Our DMRG data points are plotted in (black) circles, while the theoretical fit to Eq. ~(\ref{eq:16}) is shown as (blue) line.}
  \label{fig:Sd3CorSdw}
\end{figure}

Another measurement we can perform is the transverse spin-spin correlation function, which should decay exponentially to support the SDW state.
We observe exactly this behavior from our simulations, as shown in Fig.~\ref{fig:corrxy-sdw}.
Finally,  power-law behavior is expected for the ``octupolar''
correlation function \cite{hikihara2010},
\begin{equation}
  \label{eq:16}
  \langle (\Pi_{y=1}^3 S_y^+(x)) (\Pi_{y=1}^3 S_y^-(x'))\rangle \sim C_{\eta_3}(x,x').
\end{equation}
The operator $\Pi_{y=1}^3 S_y^+(x)$ may be though of as inserting a
soliton -- an extra period -- into the SDW.  This correlation function
decays in the thermodynamic limit with the power-law exponent
\begin{equation}
\eta_3=\frac{3\beta^2}{2\pi \kappa_0} = \frac{1}{2\eta_{\rm sdw}}.
\end{equation}
We indeed observe such power law behavior in the DMRG, as shown in
Fig.~\ref{fig:Sd3CorSdw}.  Fitting this data (for $M/M_s=1/2$, $R=0.7$)
gives $\eta_3 = 3.1\pm 0.2$, while the $S^z$ profile in
Fig.~\ref{fig:sz2M} for the same parameters is fit to
$\eta_{\rm sdw} = 0.2\pm 0.1$, yielding the product $\eta_3 \eta_{\rm sdw} =
0.62 \pm 0.31$. The uncertainties for each exponent is crudely estimated by tracing out the boundary values when the fitting starts to mismatch the DMRG result. The slow decay of the
$S^z$ profile and strong boundary effects as seen in Fig.~\ref{fig:sz2M}
induce significant uncertainties in the estimate for $\eta_{\rm sdw}$, so we consider the degree of agreement to the expected value $\eta_3 \eta_{\rm sdw} =1/2$ satisfactory.

\section{$M=M_{\rm sat}/3$ plateau}
\label{sec:plateau}

Magnetization plateaux are observed frequently in models of frustrated
magnetism, and in a number of experiments on such materials.
Theoretically, we define a magnetization plateau as a ground state of
a spin system in a magnetic field $h$, such that for a range of
fields, $h_1<h<h_2$, the magnetization (along the field) $M(h) =
M_{\rm p}$ is constant.  This implies that the magnetization is a
good quantum number, and, since by assumption the only term in the
Hamiltonian coupling to the applied field is $h M$, that the ground
state wavefunction itself is independent of the field in this range.
Moreover, since the magnetization $M$ is just the total spin $S^z_{\rm
  tot}$ along the
field direction, the symmetry under rotations generated by $S^z_{\rm
  tot}$ is unbroken.  Thus, there can be no spin expectation values
normal to the field.  Furthermore, no other nearby states must cross
the ground state (in energy) in this field range, since it remains the
ground state, and thus, since states with different magnetization must
have energy depending linearly on the field, there must be a {\sl spin
  gap} to excitations which carry non-zero spin $S^z$ relative to the
plateau state.

There are restrictions on such gapped states, following from the Lieb-Schultz-Mattis theorem
and related arguments\cite{oshikawa1997magnetization}.  One way to understand them is to map the
spins to hard-core bosons, where the boson number $n_i = S_i^z + 1/2$.
A gapped, insulating ground state of bosons in one dimension must have
an {\sl integer} number of bosons {\sl per unit cell}.  This implies
that the total spin $\sum_{i\in {\rm u.c.}} \langle S_i^z\rangle$ per unit cell must
be an {\sl integer} if the unit cell contains an even number of sites,
and must instead be a {\sl half integer} if the unit cell contains an
odd number of sites. Often, such gapped plateau states may be
considered as ordered states with spins arranged in some pattern
parallel and antiparallel to the field within a unit cell.

A prominent feature in the phase diagram we obtain is a
magnetization plateau at one third of the saturation magnetization,
$M=M_{\rm sat}/3$.  This has been extensively studied in the
literature for the isotropic model \cite{richter2009,tay2010variational}, $R=0$, where it is usually
regarded as a result of quantum ``order by disorder''.  The structure
of the plateau state in that case is indeed in agreement with a
semi-classical approach  \cite{chubukov1991quantum}, and has a unit cell consisting of two up and
one down spin, forming a three-sublattice enlargement of the primitive
triangular lattice unit cell.  Based on a combination of our DMRG
studies and an analytic analysis of the quasi-1d limit, $J'/J \ll 1$
(below), we show that, in the 2d system, the plateau state persists in the full range of
anisotropies $0<R\leq 1$ and forms a single phase throughout.
For the one-dimensional TST, however, we find
that the plateau, while present in the isotropic regime, {\sl
  terminates} before reaching the decoupled chains limit.  Both these
results can be understood from the relation between the plateau state
and the SDW phase, as will be explained in the next subsection.

\subsection{Plateau states from SDW}
\label{subsec:1/3fromsdw}

The collinear SDW state shares many of the expected elements of the
plateau phase.  It has an unbroken U(1) symmetry, even in the 2d limit,
and exponentially decaying transverse correlations in the TST.  It has
rather long-range oscillating correlations of the component of the
spin parallel to the field, and consequently a markedly modulated
$\langle S^z_y(x)\rangle$ profile in finite systems.  The distinction
between the SDW and the plateau phase is that the former is
generically incommensurate and gapless.

Both these differences may be removed due to further interactions
neglected up to now, which {\sl pin} the gapless phason mode $\varphi$
at specific discrete values.  This has been discussed at length
already in Ref.~\onlinecite{starykh2010extreme} for the two dimensional case.
There, it was argued that an infinite sequence of plateaux occur at
$T=0$ within the SDW phase, the strongest of these being the 1/3 plateau, and that all these plateaux exist at arbitrarily
small $J'/J$.  In the two-dimensional system, the plateau width (in
a magnetic field) can be estimated to scale as $J (J'/J)^{9/2}$, see
Ref.~\onlinecite{starykh2010extreme}.  Here, we will restrict the discussion to
the TST, and find that one-dimensional fluctuations
suppress most of these plateaux, including the
1/3 plateau for sufficiently small $J'/J$.

The plateau formation is due to additional interactions neglected in
the sine-Gordon Hamiltonian presented so far in
Eqs.~(\ref{eq:sdw12},\ref{eq:sdw13}), which involve {\sl higher
  harmonics} of the phason mode $\varphi$.  The allowed terms are
obtained directly from a symmetry analysis.  The action of the
symmetries of the problem on $\varphi$ may be understood directly from
the expression for the spin operator in the SDW phase of the TST in
Eq.~\eqref{eq:52}.  Under each symmetry, which is a lattice space
group operation, $\varphi$ must be chosen to transform appropriately
{\sl so that $S_y^z(x)$ is a scalar}.  This dictates the following
transformation rules
\begin{enumerate}
\item translation along $x$, $x\rightarrow x+1$: $\varphi \rightarrow
  \varphi + \pi-2\delta$.
\item translation along $y$, $y \to y+1$: $\varphi \rightarrow \varphi -
  2\pi/3$.
\item 2D inversion, $x \to -x$, $y\to 2-y $: $\varphi
  \rightarrow -\pi/3 - \varphi$.
\end{enumerate}
In addition, there is a ``gauge invariance'' arising because of the
ambiguity of $\varphi$ due its definition as a phase variable, which
forces the invariance of the Hamiltonian under {\sl local} shifts of
$\varphi$ by $2\pi$.  Note that in this section, we always consider
the infinite $L$ limit, and neglect the difference between
$\tilde\delta$ and $\delta$.

Using the local gauge invariance, we seek terms of the form
\begin{equation}
  \label{eq:53}
  H_{\rm pin} = \sum_n \int \! dx\, t_n \sin (n \varphi + \alpha_n),
\end{equation}
where $t_n$ and $\alpha_n$ are arbitrary parameters.  (In general we
can also allow $\alpha_n$ to be a slowly varying linear function of
$x$, which is important for a full analysis of commensurate to
incommensurate transitions, but we do not require this here for the
more limited purpose of just identifying the relevant plateau states).
Using the translational symmetry along $y$, we immediately obtain the
constraint that $t_n=0$ unless $n$ is a multiple of $3$, and so we set $n=3 k$. The
inversion symmetry then forces $\alpha_n=0$ (mod $2\pi$), so finally,
we find
\begin{equation}
  \label{eq:54}
  H_{\rm pin} = \sum_{k \in \mathbb{Z}} \int \! dx\, t_k \sin 3k\varphi,
\end{equation}
where we have redefined the $t_k$ appropriately.  Now it remains to
apply translation symmetry along $x$.  This simply gives the condition
that $3k( \pi - 2\delta)$ is an integer multiple of $2\pi$.  Writing $\delta = \pi
M = (\pi/2) M/M_{\rm sat}$, we have
\begin{equation}
  \label{eq:55}
  \frac{M}{M_{\rm sat}} = \frac{3k - 2p}{3k},
\end{equation}
with $k,p$ integers.  This gives a rational family of potential
magnetization plateaux, whose strength decreases with increasing $k$.

\begin{figure}[t]
  \begin{center}
  \scalebox{0.95}{\includegraphics[width=\columnwidth]{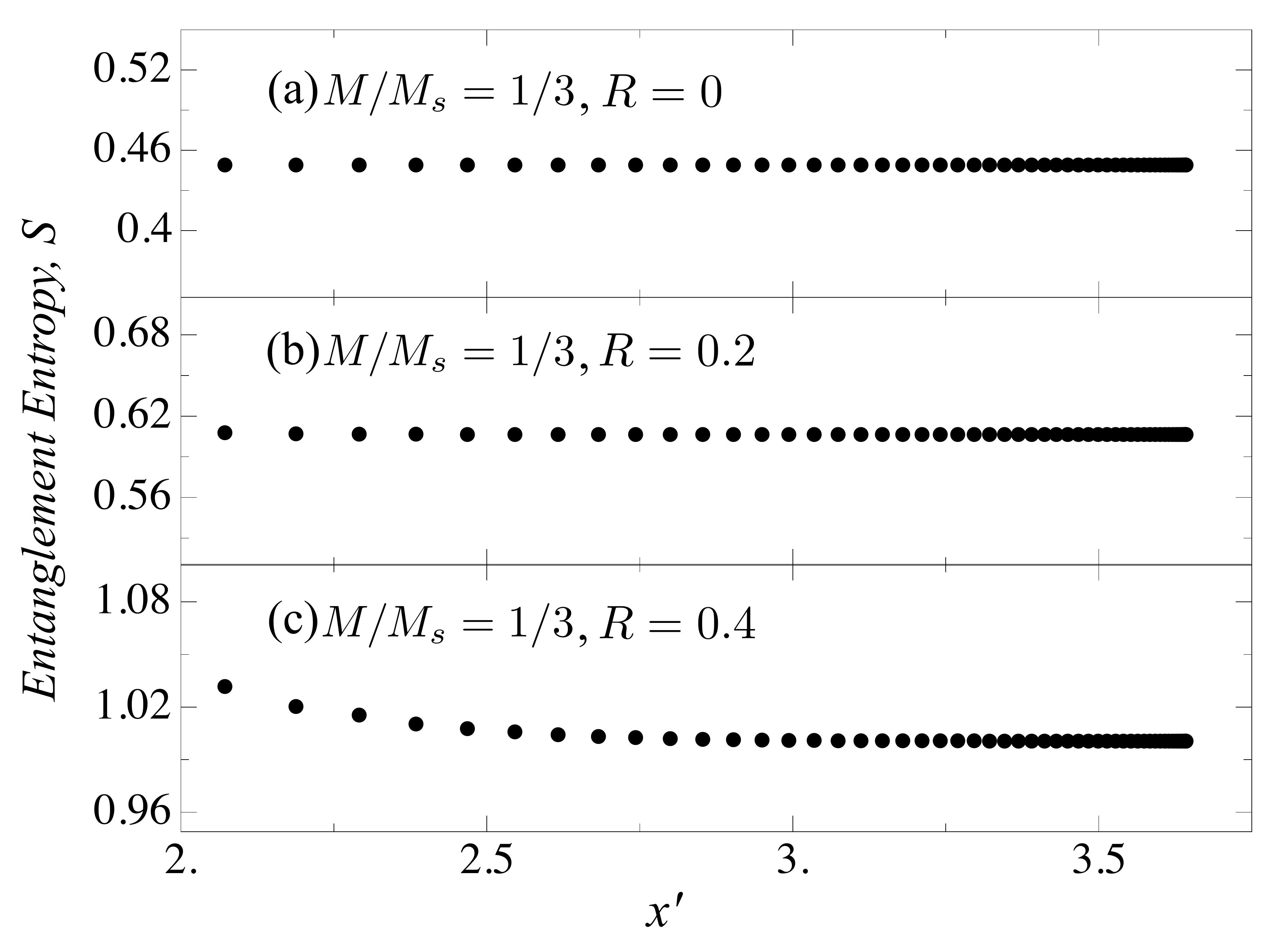}}
  \end{center}
  \caption{Entanglement entropy at (a)$R=0$, (b)$R=0.2$ and (c)$R=0.4$ in the 1/3 plateau state.
        We see that the entanglement entropy approaches a constant for large $x'$, which corresponds to a central charge of $c = 0$ in the ordered state.}
  \label{fig:ee-plateau}
\end{figure}

An actual plateau occurs for a given value of magnetization
characterized by integers $k,p$ only if the associated term, $t_k$, is
{\sl relevant} \cite{hikihara2010}, when considered as a perturbation to the low energy
Hamiltonian of the SDW state, which is just the free massless field
theory for $\varphi$.  The scaling dimension of the
operator in Eq.~\eqref{eq:54} is easily obtained as $\Delta_{3k} = 9 k^2 \eta_{\rm sdw} = 3\pi
k^2 \kappa_0/\beta^2$, c.f. Eq.~\eqref{eq:48}, and therefore, under RG, we find
\begin{equation}
  \label{eq:56}
  t_k(b) = t_k b^{2-\Delta_{3k}} = t_k b^{2-9k^2\eta_{\rm sdw}}.
\end{equation}
Here, $t_k$ is relevant, and a magnetization plateau appears when
$\Delta_{3k}<2$.  Consider the case $k=1$, which corresponds to the case $M=M_{\rm
  sat}/3$, and small $J'/J$.  There (recall
Sec.~\ref{sec:dmrg-results-sdw}) $\eta_{\rm sdw} \approx 0.23$ so
$\Delta_3 \approx 2.07>2$, and thus $t_1$ is {\sl irrelevant}.
Because $\Delta_{3k}$ increases quadratically with $k$, clearly all other
potential plateau with larger $k$ are absent in the quasi-1d limit.
Thus we expect that for $J'/J \ll 1$, the SDW state remains stable, and
there are no magnetization plateaux.

\begin{figure}[t]
  \begin{center}
  \scalebox{1}{\includegraphics[width=\columnwidth]{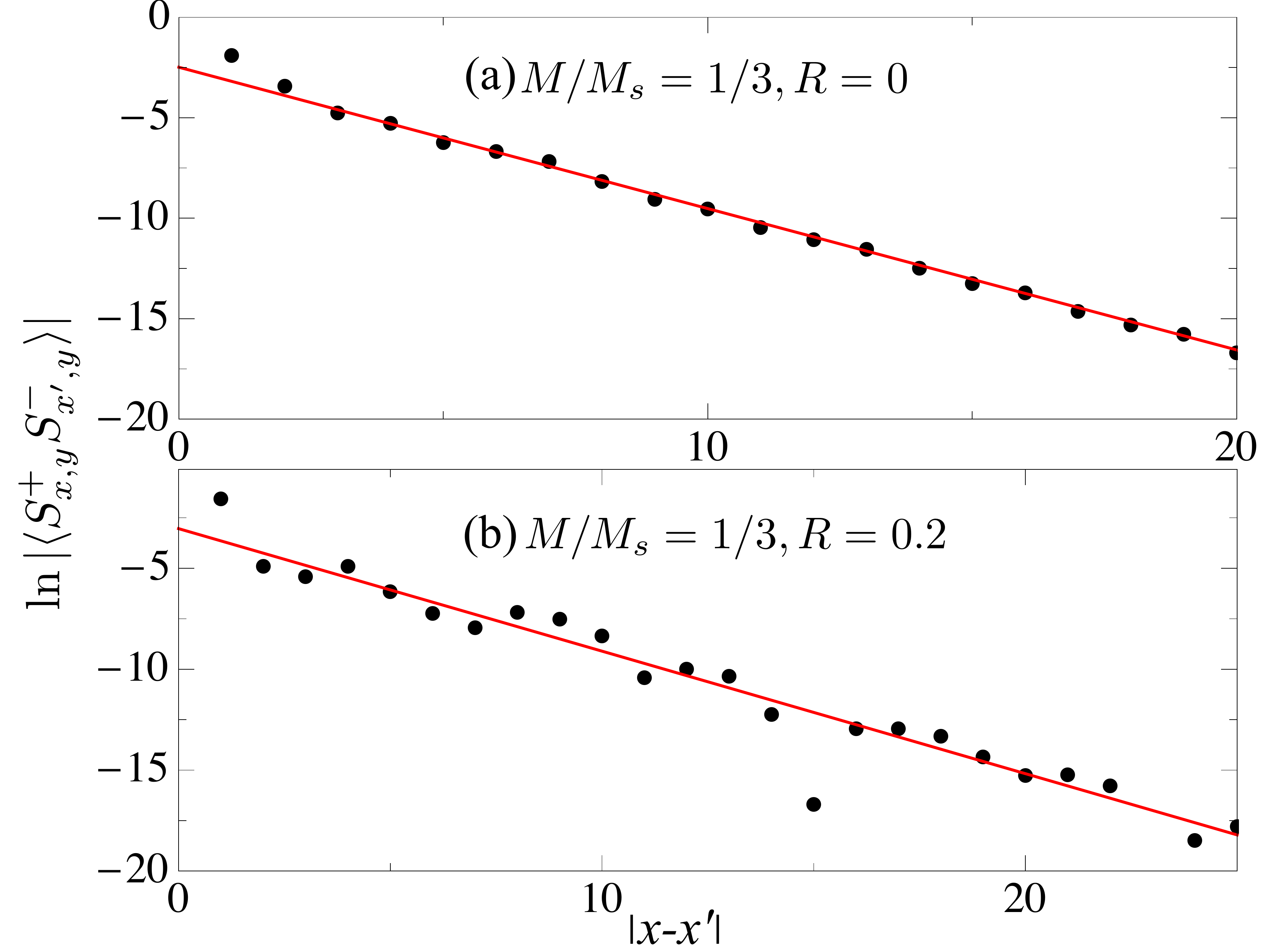}}
  \end{center}
  \caption{(Color online) Log plot for transverse spin-spin correlation function at (a)$R=0$ and (b)$R=0.2$ in the 1/3 plateau state for symtem size $N_x=120$ and $x'=N_x/2$. Data points are shown as (black) circles while the (red) line is a fit to the exponential function. }
  \label{fig:cc-plateau}
\end{figure}

With increasing $J'$, however, $\eta_{\rm sdw}$ decreases, owing to
its dependence on $\kappa_0$ in Eq.~\eqref{eq:48}.  Including this
dependence, and using the quasi-1d formula for $\kappa_0$ (in the text
following Eq.~\eqref{eq:sdw12}), we obtain the condition that
$t_1$ becomes relevant, i.e. $\Delta_3 <2$, when $J'/J > 0.17$.
We believe that this is still in the domain where the quasi-1d
approach is valid.  The result predicts that the 1/3
plateau appears only for $R<0.83$ in the TST.  At fixed $M=M_{\rm
  sat}/3$, the transition from the gapless SDW to gapped plateau state
at this value of $R$ or $J'/J$ is in the Kosterlitz-Thouless
universality class, as is well-known for the quantum sine-Gordon
model.  Consequently, the gap vanishes exponentially on approaching
the transition from the more isotropic side, and the ground state
energy itself shows only an unobservably weak essential singularity at
the transition.  We note that other potential plateaux with $n = 3k \geq 6$ are so
strongly suppressed by fluctuations that we do not expect any to
occur, at least in the quasi-1d regime.

\begin{figure}[t]
  \begin{center}
  \scalebox{0.95}{\includegraphics[width=\columnwidth]{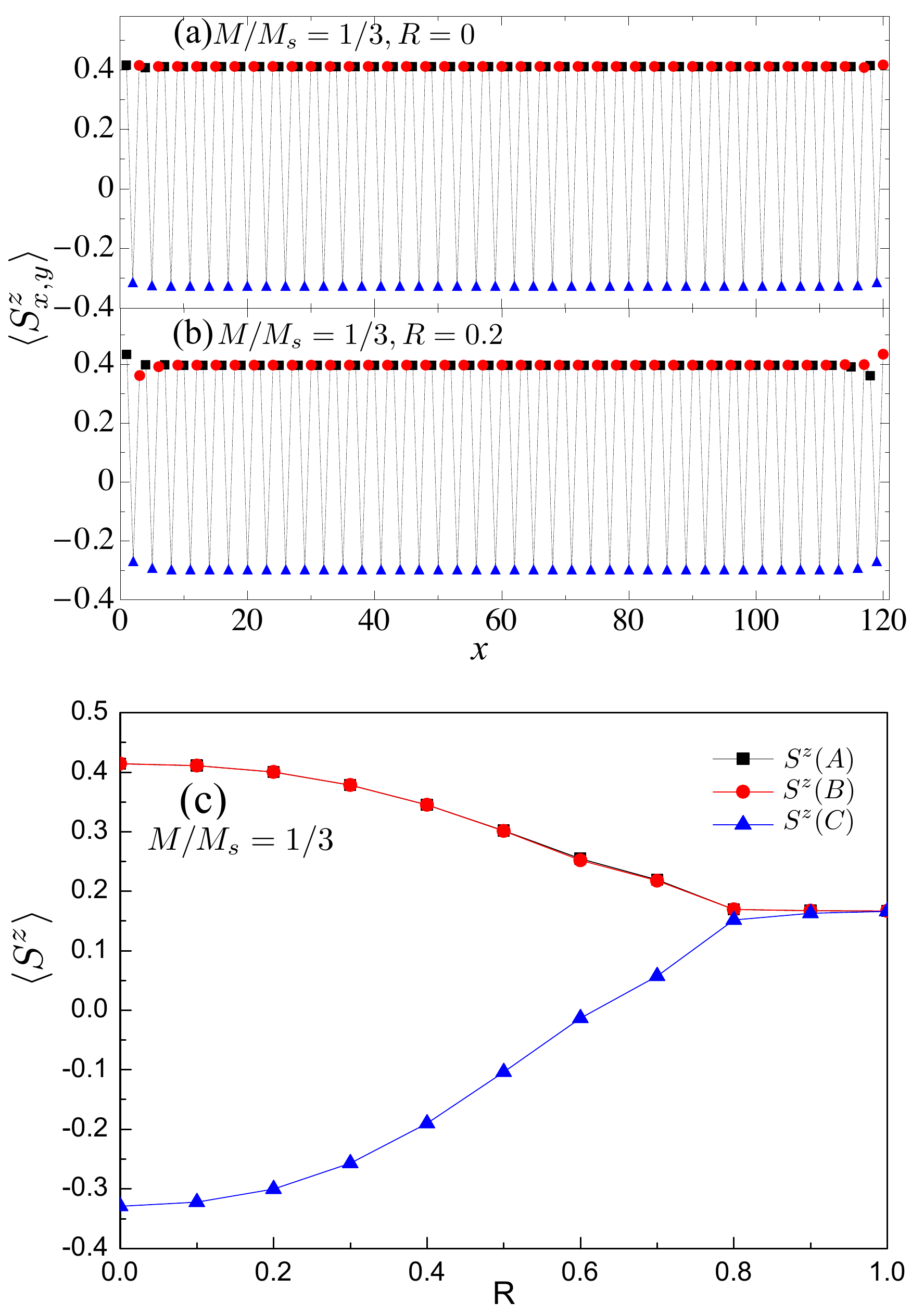}}
  \end{center}
  \caption{$S^z$ profile for the 1/3 plateau state at
    (a)$R=0.0$ and (b)$R=0.2$.
    We show data for the three sublattices, showing the up-up-down structure, as square (black), circle (red) and triangles (blue).
    There are slight boundary effects near the end of the chains, near $x = 0, 120$; however, the chains are well-ordered toward the center.
    (c) shows the $S^z$ profile at $M/M_s=1/3$ as a function
    of $R$. We observe that the width of the plateau decreases and eventually vanishes near $R \approx 0.8$.}
  \label{fig:Szprofile-plateau}
\end{figure}

It is interesting to consider the spin structure on the plateau.  This
depends on the sign of $t\equiv t_1$.  For $t>0$, the $\sin 3\varphi$
pinning term in Eq.~\eqref{eq:54} is minimized by three values with
equal energy, $\varphi = -\pi/6 + 2\pi n/3$, with $n=0,1,2$.  For
these values, using Eq.~\eqref{eq:52}, the spin density profile takes
the form
\begin{equation}
  \label{eq:57}
  \langle S_y^z(x)\rangle_{t>0} = \tilde{M} + A_1 \sin \big[ \tfrac{\pi}{6}+
  \tfrac{2\pi}{3}(x-y-n)\big].
\end{equation}
This equation describes a three sublattice structure with two spins
``up'', i.e. with $\langle S_y^z(x)\rangle > \tilde{M}$, when
$x-y-n=0,1\, (\textrm{mod 3})$ and one spin ``down'', when $x-y-n=2\,
(\textrm{mod 3})$. This is the semi-classical up-up-down state,
and has precisely the same qualitative structure as predicted
semiclassically in the isotropic limit $J'=J$.

For the other case, $t<0$, the minima occur for $\varphi = +\pi/6 +
2\pi n/3$, and the spin density profile becomes
\begin{equation}
  \label{eq:58}
   \langle S_y^z(x)\rangle_{t<0} = \tilde{M} - A_1 \sin \big[ \tfrac{\pi}{6}-
  \tfrac{2\pi}{3}(x-y-n)\big].
\end{equation}
This describes instead a three sublattice structure with two spins
nominally ``down'', with $x-y-n=0,2\, (\textrm{mod 3})$, and the
remaining one up.  This state does not have a natural semiclassical
picture, and instead corresponds to the `quantum' version of the
plateau, discussed for the two-dimensional lattice in
Ref.~\onlinecite{starykh2010extreme}.  A caricature of this state is a three site
unit cell with two sites forming a spin singlet entangled pair, and
the third (the ``up'' site) polarized along the field.  Our DMRG
results are consistent with the up-up-down configuration,
Eq.~\eqref{eq:57}, suggesting that $t > 0$ case is realized.

We should stress that, apart from the quantitative estimate of
$\kappa_0$, nothing in this subsection depends upon the quasi-1d
approach.  The conditions for the existence and stability of the
plateaux arising out of the SDW state are otherwise completely general
results based only on symmetries of the TST and general arguments.

\begin{figure}[t]
  \begin{center}
  \scalebox{0.99}{\includegraphics[width=\columnwidth]{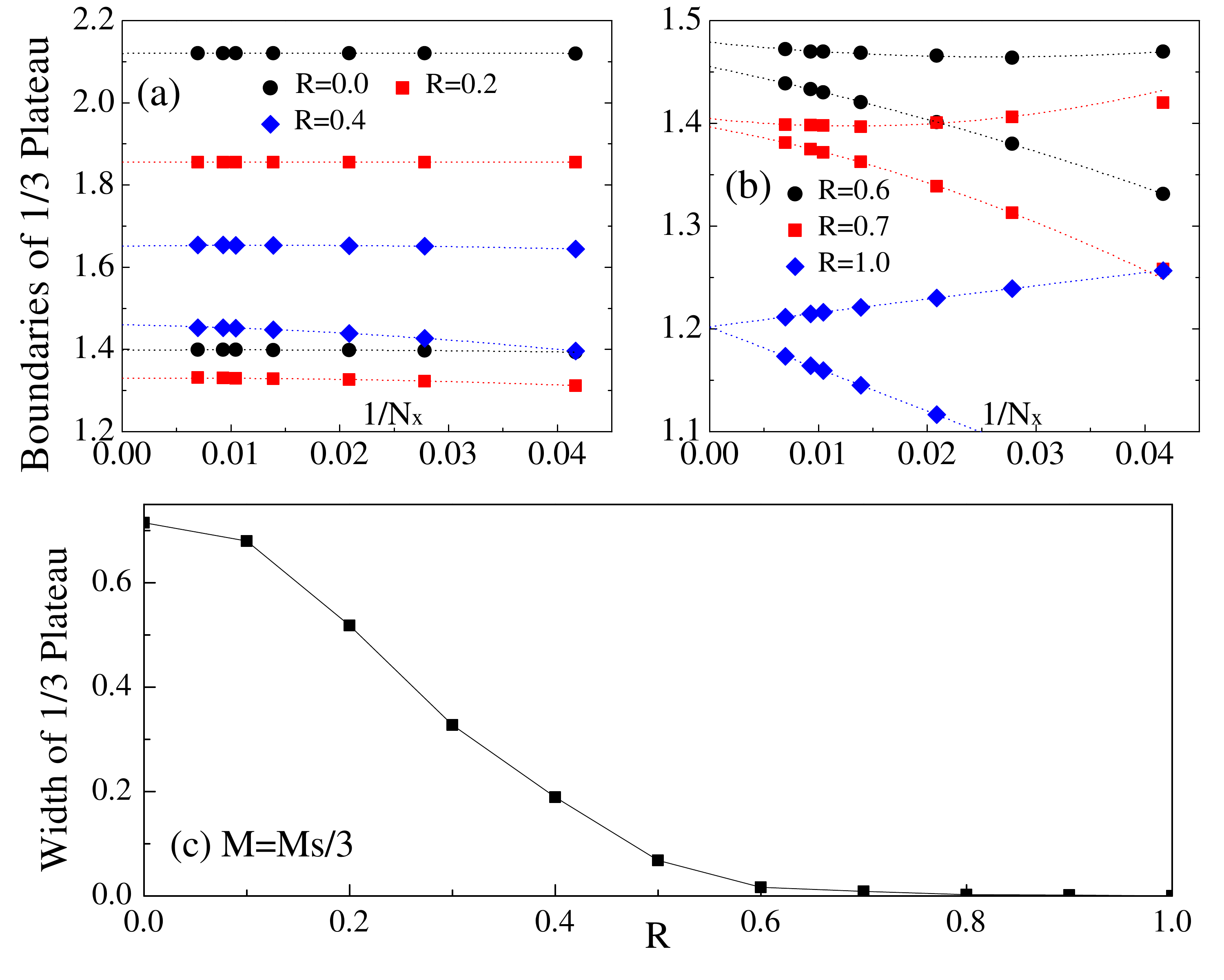}}
  \end{center}
  \caption{Finite-size scaling of the boundaries of the 1/3 plateau for $L_y=3$ TST for
  different anisotropies, (a) $R=0.0, 0.2, 0.4$ and (b) $R=0.6, 0.7, 1.0$.
  (c) Width of the 1/3 plateau as a function of $R$.}
  \label{fig:DMRG-plateau}
\end{figure}

\subsection{DMRG results for plateau}%
\label{subsec:1/3fromDMRG}%

In this section, we discuss how we use DMRG to probe into the 1/3 plateau.
The first observation of its existence is the constant entanglement entropy for the ranges of $R$ on the 1/3 plateau, as shown in Fig.~\ref{fig:ee-plateau}.
This shows an ordered state which corresponds to central charge $c = 0$, in Eq.~\eqref{eq:ent}.
Furthermore, we can measure the transverse spin-spin correlations, which should decay exponentially in the 1/3 plateau.
We show this measurement in Fig.~\ref{fig:cc-plateau}, for $R = 0.2$ as well as the isotropic case $R = 0$.

In Fig.~\ref{fig:Szprofile-plateau}(a,b), we plot the $S^z$ profile of the spins forming the three sublattices on the 1/3 plateau.
Near $x = L/2$, we see a perfect up-up-down structure, with some boundary effects on the edges of the chain.
This gives definitive evidence of the robustness of the 1/3 plateau in these ranges of anisotropies.
Moreover, in Fig.~\ref{fig:Szprofile-plateau}(c), we see that the plateau persists up until $R \approx 0.8$, at which point, the system undergoes a Kosterlitz-Thouless transition that destroys the 1/3 plateau.
As described in the previous subsections, this is a signature of the 1d TST only: in 2d, the plateau is even more robust, extending down to $R = 1$. This is further discussed in Sec.~\ref{sec:discussion2D}.

To characterize the properties of the plateau as well as its
width, we will adopt the following method, which takes advantage of the
total spin conservation due to the presence of the $U(1)$ symmetry
with a magnetic field along $z$-axis. In this case, we can work in a
given total spin sector $S^z=\sum_i S^z_i$, and get the
corresponding ground state energy $E(S^z)$
\begin{eqnarray}
    E(S^z,h)=E(S^z)-h\cdot S^z.%
\end{eqnarray}%
Then the energy difference between two adjacent spin $S^z$
sectors is given by
\begin{eqnarray}
    \delta E(S^z,h)=E(S^z+1,h)-E(S^z,h).%
\end{eqnarray}
Generally, at small magnetic field $h$,
$E(S^z+1,h)>E(S^z,h)$, so $\delta E(S^z,h)>0$. However,
$E(S^z+1,h)\leq E(S^z,h)$ when $h$ is large enough, so $\delta
E(S^z,h)\leq0$. Therefore, the boundaries of the plateau can be
determined when $E(S^z+1,h)=E(S^z,h)$, with the upper boundary
$h^2_c(S^z)$ and lower boundary $h^1_c(S^z)$ of the plateau given
by
\begin{eqnarray}
h^2_c(S^z) &=& E(S^z+1)-E(S^z), \nonumber\\%
h^1_c(S^z) &=& E(S^z)-E(S^z-1).\label{eq:plateauboundary}
\end{eqnarray}
Finally, the corresponding width of the plateau can also be
obtained as
\begin{eqnarray}
    W(S^z)=h^2_c(S^z)-h^1_c(S^z).\label{eq:plateauwidth}
\end{eqnarray}

In DMRG, the boundaries of the 1/3
plateau can be computed using Eq.~\eqref{eq:plateauboundary} by
fixing the total spin to $S^z=\frac{NM_s}{3}$. Here, $M_s=\frac{1}{2}$
is the saturation magnetization, and $N$ is the total number of
sites. As shown in Figs.~\ref{fig:DMRG-plateau}(a,b), both the
upper and lower boundaries of the 1/3 plateau are
determined using different system sizes and anisotropies. The
corresponding width of the plateau is also given in Fig.~\ref{fig:DMRG-plateau}(c) using
Eq.~\eqref{eq:plateauwidth}. From this, we can see that the
plateau is very robust and remains finite when the anisotropy, $R$, is
small, and decreases with increasing $R$. Interestingly, the
plateau still remains finite even $R$ is very large, i.e., $R=0.7$,
although the width $W$ is very small. In the region $0.7<R\leq 1$,
finite-size scaling of the data shows that the width of the plateau is zero
within the numerical error, for example, at the decoupled chain
limit $R=1$.

\section{Low field regime}
\label{sec:lowfield}

At zero field, there is already considerable work on the spatially
anisotropic Heisenberg model in two dimensions
\cite{weng2006spin,yunoki2006two,pardini2008magnetic,bishop2009,heidarian2009spin,tay2010variational,SRWhite2011}.
Away from the quasi-1d
region, i.e. for $0 < R \lesssim 0.8 $, the ground state of the 2d model
is unambiguously magnetically ordered, in a coplanar spiral with an
incommensurate wavevector that varies continuously with $R$.  With
increasing anisotropy, the ground state is less clear, and is quite
difficult to resolve numerically, owing to the fact that correlations
between chains set in only at extremely long length scales for small
$J'/J$.   A controlled renormalization group approach predicts, however,
that in the limit $0< J'/J \ll 1$,  the system develops a {\sl
  collinear} magnetic state instead of the spiral one \cite{starykh2007ordering}.  Such a collinear
state is qualitatively distinguished from the spiral one by its pattern
of symmetry breaking, which leaves a residual U(1) spin rotation
symmetry about the ordering axis, in contrast to the spiral state which
fully breaks SU(2) symmetry with no residual continuous invariance
remaining.

Here we turn to the situation in the one-dimensional TST.  We argue
that in this case the spiral order is converted by 1d quantum
fluctuations into a fully gapped state with spontaneous {\sl staggered
  dimerization}.  The argument is quite general and is expected to
hold for any 1d system with local non-collinear order and a
half-integer spin per unit cell.  Furthermore, specifically for the
TST, we show that the tendency to
short-range spiral order is {\em more} robust than in 2d, and unlike
in 2d, it prevails over collinear order even in the limit of
arbitrarily small $J'/J$.  Thus staggered dimerization is predicted at
zero field for all $0 \leq R < 1$ for the TST. See Appendix~\ref{sec:zero-field-analysis}
for an alternative calculation that leads to the same conclusion as the one presented below.

Given the presence of dimerization in zero field, we can discuss the
behavior in low fields, or more properly for small magnetization, in
terms of the elementary excitations of the symmetry broken dimerized
state, which are domain wall solitons.  We obtain in this way
different gapless phases at low field, including the SDW state
discussed previously from the quasi-1d point of view.

\subsection{Zero field dimerization from spiral order}
\label{sec:dimer-from-spir}

In the following, we assume that on short space and
time scales, the spins establish a similar spiral order to that of the
2d system.  This notion can be made more systematic by considering
spin tubes made by wrapping the triangular lattice into cylinders with
larger circumference.  Once the circumference is large enough compared
to the correlation length of the spiral order, the latter should
become well-established.  It seems reasonable to regard this as being
the case already for the circumference three TST studied here.  This
is corroborated also by the close correspondence of the phase diagram
in the weakly anisotropic limit, $R \ll 1$, and the expected
semi-classical one, as discussed already in Sec.~\ref{sec:iso}.

With this assumption, the description of the TST should be that of a
Non-Linear $\sigma$-Model (\nlsm) for the spiral order, confined to the finite
width cylinder.  This starting point is similar to the one of Haldane \cite{haldane1988}
applied to unfrustrated spin chains of spin $S$, which locally establish
collinear N\'eel order.  From this formulation, Haldane established the
existence of a featureless gapped state for integer $S$, while it is
known that chains with half-integral $S$ harbor a gapless Bethe
chain-like phase instead.  The case of the TST is distinct from
Haldane's analysis, however, owing to the different symmetry of the
order.  While the collinear N\'eel case is described by a vector O(3)
\nlsm, the spiral case is instead described by a \nlsm with a {\sl
  matrix} SO(3) order parameter \cite{dombre1989}.  Here the matrix may be constructed
from the local spin order,
\begin{equation}
  \label{eq:59}
  {\bf S}_i \sim m ({\bf\hat n}_1 \cos {\bf q}\cdot {\bf r}_i + {\bf\hat n}_2 \sin {\bf q}\cdot {\bf r}_i),
\end{equation}
where ${\bf\hat{n}}_1$ and ${\bf\hat{n}}_2$ specify the plane of the
spiral, with ${\bf\hat{n}}_1\cdot{\bf\hat{n}}_2=0$, ${\bf q}$ the spiral
wavevector, and $m$ the amplitude of the quasi-static moment.  One can
construct from this the SO(3) matrix
\begin{equation}
  \label{eq:60}
  {\mathcal O} = \left( {\bf\hat{n}}_1 | {\bf\hat{n}}_2 | {\bf\hat{n}}_3 \right),
\end{equation}
with ${\bf\hat{n}}_3 = {\bf\hat{n}}_1 \times {\bf\hat{n}}_2$.

If on short space and time scales, spiral order is present, we expect
that an appropriate effective \nlsm\ action is given by
\begin{eqnarray}
\label{eq:71}
&&   S_{NL\sigma M}  =  \\
&& \frac{1}{2g} \int \! dx \, d\tau\, \Big\{ \frac{1}{v}
    {\rm Tr} \left[\partial_\tau {\mathcal O}^T \partial_\tau
      {\mathcal O}\right] + v  {\rm Tr} \left[\partial_x {\mathcal O}^T \partial_x
      {\mathcal O}\right]\Big\}.\nonumber
\end{eqnarray}
Note that, for a quasi-1d system with circumference $L_y$, the effective
coupling constant $g\sim c/L_y \ll 1$ for large $L_y$, with some
constant two-dimensional coupling constant $c$.

Famously, in Haldane's analysis of spin chains with a vector O(3)
order parameter, the na\"ive \nlsm\ action must be supplemented by a
topological term \cite{haldane1988}.   Topology of the order parameter is also important
here, but its nature is rather distinct from Haldane's case.  For
clarity, we compare and contrast the two situations here.   The vector
O(3) order parameter comprises a manifold isomorphic to the sphere
$S^2$.  Its topology is summarized by the homotopy groups $\Pi_1(S^2) =
0$ and $\Pi_2(S^2) = \mathbb{Z}$.  The former implies that there are no
non-trivial loops on the sphere, and correspondingly no {\sl singular}
point defects in two dimensions.  The latter, second homotopy group
implies that there are classes of non-trivial {\sl smooth}
configurations of the order parameter in two dimensions, parametrized by
an integer.  These configurations are skyrmions, lacking any
singularity.  Because of the lack of any singularity, the skyrmions
appear in a continuum limit of the O(3) vector \nlsm\ , and modify the
physics of the \nlsm\ through a topological $\theta$-term, which gives a
phase factor to configurations with non-zero skyrmion number.  Based on
this \nlsm\ with $\theta$-term, Haldane postulated distinctly different
behavior for integer and half-integer spin chains.

In the matrix SO(3) case, the order parameter manifold is
$S^3/\mathbb{Z}_2$, and the corresponding homotopy groups are
$\Pi_1(S^3/\mathbb{Z}_2) = \mathbb{Z}_2$ and $\Pi_2(S^3/\mathbb{Z}_2)
=0$.  The trivial second fundamental group means that {\sl non-singular}
configurations of the order parameter have no topological distinctions.
This implies that a continuum limit exists in which there are no
topological defects and there is no topological term.  Instead, the
non-vanishing first homotopy group implies that there are {\sl singular}
point defects in two dimensions, with an Ising character.  Note that in
our theory, these are point defects in space-time, or {\sl instantons}.
Such defects are well-known in classical two-dimensional non-collinear
magnets, and are known as $\mathbb{Z}_2$ vortices \cite{kawamura1984}.  They do not appear
in the continuum \nlsm\ , but are allowed in a lattice theory.  Instead,
the proper way to treat them is to {\sl embed} the continuum theory in a
larger one in which the defects appear as {\sl operator insertions},
with some fugacity and selection rules.  This situation is familiar from
the Kosterlitz-Thouless analysis of the classical XY model, in which the
na\"ive continuum theory is just the Gaussian spin-wave line, and the
defects are point vortices which are treated as a kind of Coulomb gas \cite{chaikin2000principles}.
It occurs also in the quantum analysis of 2+1 dimensional collinear
antiferromagnets, where the singular defects are hedgehogs or monopoles.
The separation of these defects and the continuum theory is the basis of
the theory of deconfined quantum criticality \cite{senthil2004}.

With this understanding, we may first consider the SO(3) matrix \nlsm\
{\sl without} any $\mathbb{Z}_2$ vortices, which is simply described by
Eq.~\eqref{eq:59}.  There is no topological term.  This SO(3) \nlsm\ is,
like all \nlsm's in two dimensions for non-abelian groups,
asymptotically free.  Lacking any quantum phase factors, we expect
simply that it develops a gap at a length scale $\xi \sim e^{g_0/g} \sim
e^{\frac{g_0}{c} L_y}$, and that order parameter (hence spin)
correlations decay exponentially beyond this scale.  The gap itself
behaves as $\Delta \sim v/\xi$.  Note the difference from Haldane's
case, where the $\theta$ term, which is non-trivial for half-integer
spin, fundamentally alters the behavior of the continuum \nlsm, leading
to gapless behavior in the half-integer spin case.  Here, there is no
topological term, and the system is always gapped with exponential spin
correlations.

Now we can consider the role of the $\mathbb{Z}_2$ vortex instantons.
Such a vortex is described in the field theory by an operator, $\psi$,
which inserts the vortex at a particular space-time point.  It is
crucial to consider the quantum numbers of a $\mathbb{Z}_2$ vortex, i.e.
how the operator $\psi$ transforms under physical symmetries.  The
relevant operations are time-reversal, translation, and inversion.  It
can be argued (we discuss this in Appendix~\ref{sec:transf-prop-mathbbz}) that the vortex operator
is invariant under time-reversal and translations along $y$, and
transforms under the other two operations, translation along $x$,
$T_x$, and inversion, $P$, according to
\begin{eqnarray}
  \label{eq:63}
& &  T_x: x \rightarrow x+1,  \;  \psi \rightarrow
  (-1)^{L_y} \psi, \\
  & & P: x\rightarrow -x, y\rightarrow -y,  \; \psi
  \rightarrow (-1)^{L_y} \psi.\label{eq:64}
\end{eqnarray}
From the above properties, we see that {\sl for odd $L_y$, $\psi$ has
  the transformation properties of a staggered dimerization operator}.
In general, two operators with the same symmetry are expected to have
non-zero overlap in the operator sense, and their correlations will be
proportional.  Thus, for odd $L_y$, the $\mathbb{Z}_2$ vortex operator
$\psi$ can be viewed as a staggered dimerization order parameter.

Let us consider the correlations of $\psi$.  Its two-point correlation
function is obtained by inserting two $\mathbb{Z}_2$ vortices in the
system at separated space-time points.  When they are widely separated,
the result should be just the product of two independent $\mathbb{Z}_2$
vortices.  Naively, using Eq.~\eqref{eq:59}, such a vortex has an action
which diverges logarithmically with the system size.  However, its {\sl
  effective} action is expected to be finite, due to the vanishing order
and stiffness beyond the scale $\xi$.  Roughly, the effective action for
a single vortex is thus obtained by replacing the system size by $\xi$,
so $S_v \sim \frac{1}{g} \ln \xi \sim g_0/g^2$.  Then we expect that
\begin{equation}
  \label{eq:72}
  \lim_{x\rightarrow \infty} \langle \psi(x) \psi(0)\rangle \sim
  e^{-2S_v} \sim e^{-g_0/g^2} \sim e^{-c L_y^2},
\end{equation}
with some constant $c$.  The saturation to a finite value as
$x\rightarrow \infty$ implies $\langle \psi\rangle \neq 0$, and hence,
for odd $L_y$, the existence of staggered dimer order.  For even $L_y$,
there is no connection of $\mathbb{Z}_2$ vortices to dimerization, so
although the former are present, the system forms simply a featureless
gapped state.

\begin{figure}[t]
  \begin{center}
  \scalebox{0.95}{\includegraphics[width=\columnwidth]{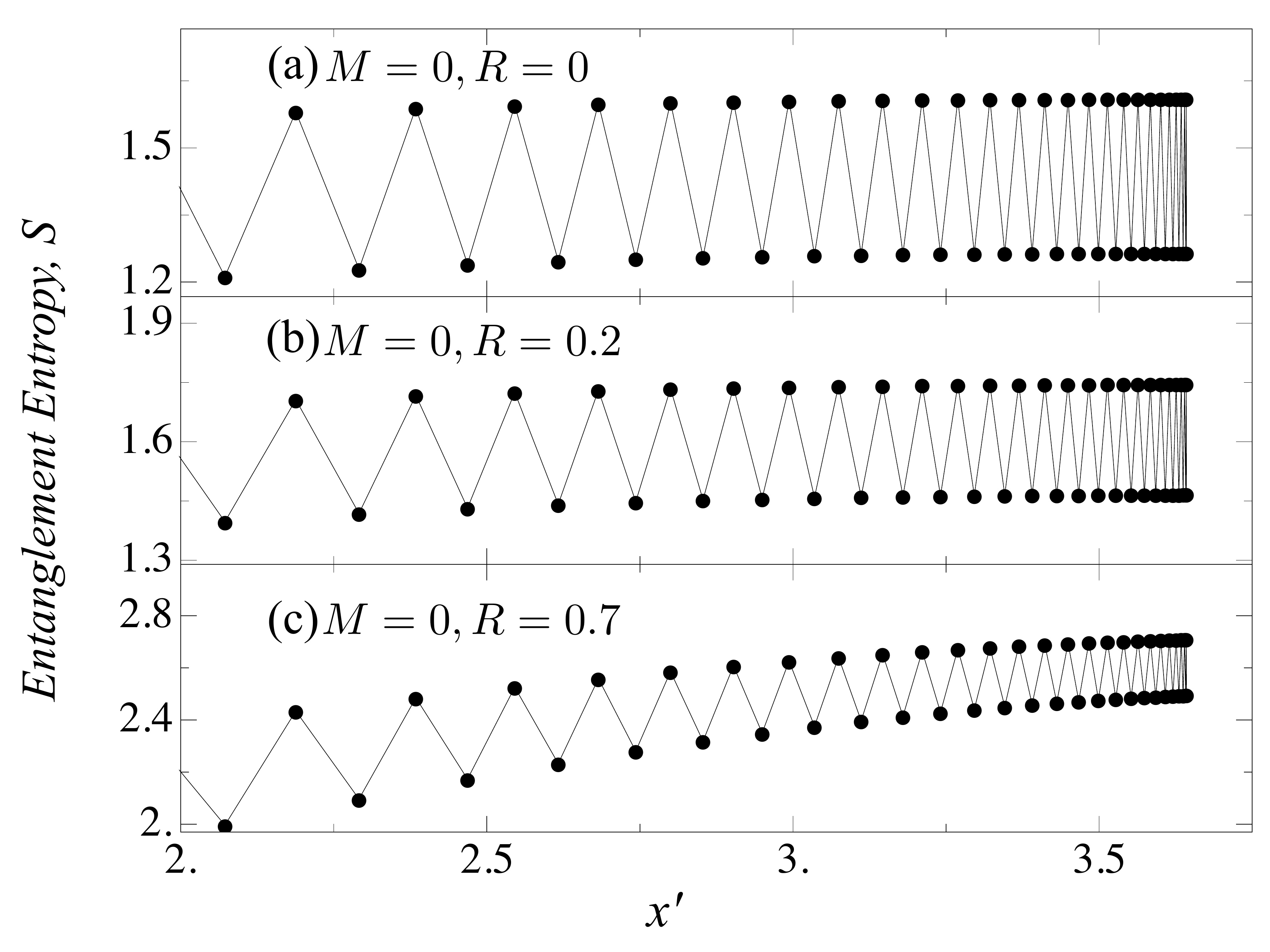}}
  \end{center}
  \caption{Entanglement entropy at (a) $R=0$, (b) $R=0.2$ and (c) $R=0.7$ at zero field. The oscillatory behavior with periodicity 2 shows the dimerized ground state. }
  \label{fig:EEm0_0}
\end{figure}

We can probe into this state by measuring the entanglement entropy in DMRG for a range of anisotropies at zero field.
We show this in Fig.~\ref{fig:EEm0_0}, where an oscillatory behavior of period 2 gives clear evidence of the dimerized phase described above.

\subsection{Gapless states in low but non-zero field}
\label{sec:gapless-states-low}

As argued in the previous subsection, the ground state in zero field is
a non-magnetic dimerized state with a gap to all excitations.  As a
consequence of the gap, the ground state is unchanged by application
of a sufficiently small field.  The ground state changes when the
field is large enough that a state with non-zero spin crosses the
energy of the spin zero ground state.  Generally, if the transition to
a state of non-zero magnetization occurs continuously, we can think
that the state with non-zero magnetization consists of a dilute set of
elementary excitations above the zero field ground state.

\begin{figure}[t]
  \begin{center}
  \scalebox{1}{\includegraphics[width=\columnwidth]{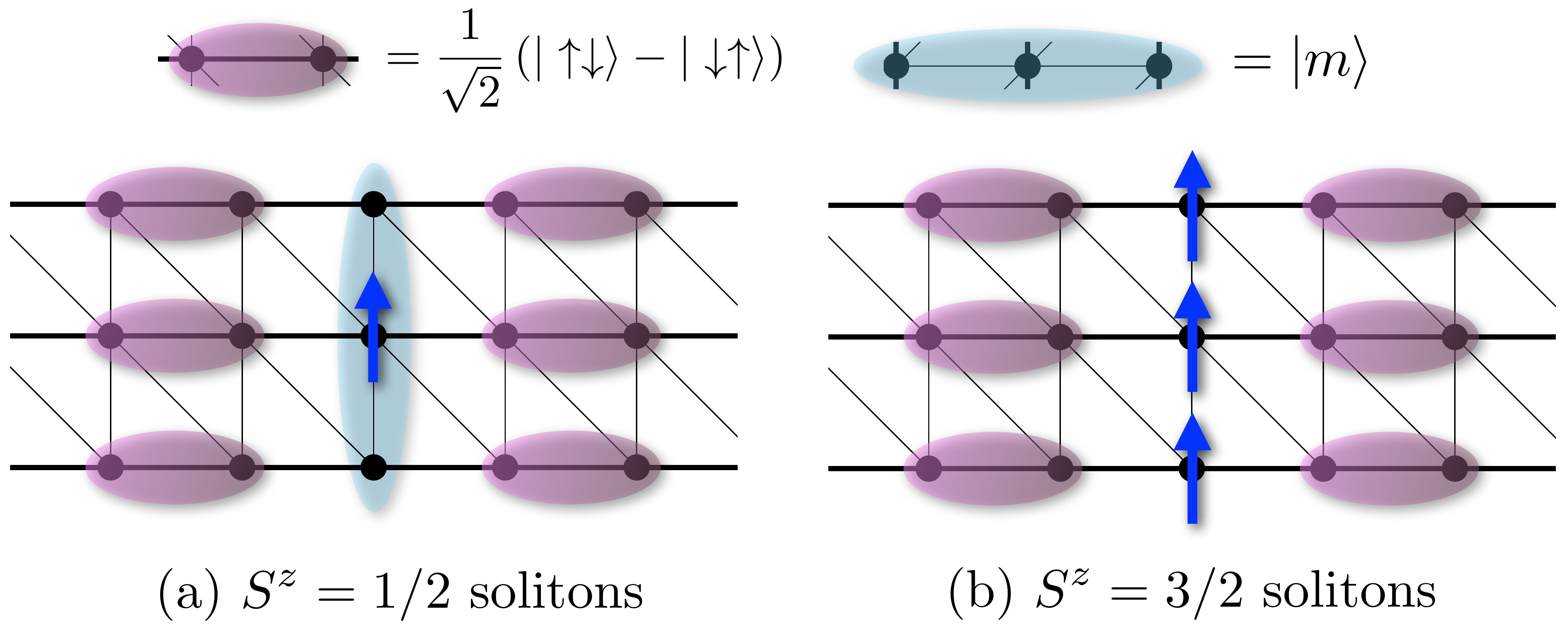}}
  \end{center}
  \caption{(Color online) Toy picture of solitons. The blue shade, covering three sites, in 
  (a) corresponds to the $S_z=1/2$ soliton in Eq.~\eqref{eq:66} while the three blue arrows in (b) corresponds to a single $S_z=3/2$ soliton.}
  \label{fig:solitons}
\end{figure}

We must consider therefore the elementary excitations of the dimerized
state, and in particular those which carry non-zero spin (as these
couple to the field).  The most important such excitations are the
topological {\sl soliton} excitations which are characteristic of the
broken Ising symmetry of the dimerized state.  Such solitons are
domain walls, connecting the two distinct dimerized ground states.  As
is well-known from the study of the Majumdar-Gosh chain \cite{sutherland1981}, solitons of
this type carry spin, and in particular for the TST, one can readily
argue that the solitons carry half-integer spin, namely $S^z =$ 1/2,
3/2, as shown in Fig.~\ref{fig:solitons}.  Both values of the spin are
possible, and generally differ in energy.  The solitons are
topological excitations insofar as they are non-local: they cannot be
created by the action of any local operator on a dimerized ground
state.  In addition to the topological soliton excitations,
non-topological excitations carrying spin $S^z=1$ also exist.  They
can be visualized either by replacing a singlet dimer by a triplet of
aligned spins, or as a bound pair of $S^z=1/2$ solitons.

Generally, if the magnetized state is realized as a dilute system of
{\em non}-topological $S^z=1$ triplons, then the dimerization is not
disrupted and must persist for $M>0$.  Numerically, however, the
dimerization appears to be disrupted at all non-zero $M$.  We
will assume henceforth that the magnetized state (at small $M>0$)
should be regarded as a collection of topological soliton excitations,
and neglect the $S^z=1$ triplons.

In general, the excitations can be characterized by spatial quantum
numbers in addition to spin.  For an excitation localized in $x$ in
the TST, we may consider the transformations under translations along
$y$, $T_y$, and under inversion, $P$.  From
Fig.~\ref{fig:solitons}, it is clear that the $S^z=3/2$ soliton is
invariant under both.  However, this is not the case for the $S^z=1/2$
soliton, which has additional structure.  In general, out of the three
non-dimerized spins in the ``core'' of the domain wall, we can form
three linearly independent states with $S^z=1/2$,
\begin{equation}
  \label{eq:66}
  |m\rangle = \frac{1}{\sqrt{3}}\left[
   \zeta^m \begin{pmatrix}
      \downarrow \\ \uparrow \\ \uparrow
    \end{pmatrix}
+ \begin{pmatrix}
      \uparrow \\ \downarrow \\ \uparrow
    \end{pmatrix}
+ \frac{1}{\zeta^{m}} \begin{pmatrix}
      \uparrow \\ \uparrow \\ \downarrow
    \end{pmatrix} \right],
\end{equation}
where $\zeta= e^{2\pi/3}$ and $m=0,\pm 1$.  These are simply momentum
eigenstates along the 3 site chain.  The state $|0\rangle$ is
invariant under the $T_y$ and $P$
operations, while the {\sl chirality} eigenstates $|\pm\rangle$ form a
two-dimensional irreducible representation.  In general, the chirality
states would differ in energy from the scalar one.  If we crudely
model the soliton core as a three-site antiferromagnetic Heisenberg
chain, then we see that the chirality states have lower energy, so we
expect that the elementary solitons take this form.  Consequently,
there are two chirality ``flavors'' to the $S^z=1/2$ solitons.

To understand the impact of the solitons, we will need the relation
between the microscopic lattice operators and those which describe the
solitons.  The simplest to consider is the dimerization operator, or
the bond kinetic energy, $B_{x,y} = \vec{S}_{x,y}\cdot
\vec{S}_{x+1,y}$.  This is negative on singlet bonds and has zero
average on bonds with uncorrelated spins.  In a ground state, it
oscillates with period $2$ in the $x$ direction. However, the singlets
are shifted over by one sublattice on crossing a soliton, so
\begin{equation}
  \label{eq:79}
  B_{x,y} \sim \overline{B} + (-1)^{x+N(x)} \epsilon_0,
\end{equation}
where $\overline{B}$ is the non-zero average, and $\epsilon_0$ is the
amplitude of the bond modulation.  We have defined $N(x) = \sum_{x'<x}
a_{+,x'}^\dagger a_{+,x'}^{\vphantom\dagger} + a_{-,x'}^\dagger
a_{-,x'}^{\vphantom\dagger} + a_{3,x'}^\dagger
a_{3,x'}^{\vphantom\dagger}$, which is the number of solitons to the
left of the position $x$.  The $N(x)$ factor accounts for the shift in
the singlet position on crossing each domain wall.

Next we turn to the spin density operator $S^z_{x,y}$.  We
are interested in its action on states which consist of a low density
of solitons.  It is helpful to consider a caricature of these states
in which solitons are described by a wavefunction which is a product
of columns of singlets, spaced by occasional non-singlet columns with
either the chiral $S^z=1/2$ form, or fully aligned $S^z=3/2$ spins, as
shown in Fig.~\ref{fig:solitons}.  If the operator $S_{x,y}^z$ acts on
a column $x$ which is part of a singlet, it converts that singlet to
an $S^z=0$ triplet state.  This triplet costs a non-zero energy equal
to the zero field spin gap, and having $S^z=0$ gains no energy back
from the magnetic field.  Thus if we restrict our description to a low
energy one, below the zero field spin gap, we can simply take
$S^z_{x,y}$ to annihilate the state in this case.  If, however, $x$ is
located at the position of a soliton, then $S_{x,y}^z$ gives back a
low energy state, which consists either of the original soliton or one
with reversed chirality.  Notably, in moving down the 1d system,
solitons alternate between odd and even columns of the lattice.  Thus
a non-zero spin is only measured when $S^z_{x,y}$ acts on an even or
odd site, if the number of solitons to the left of the position $x$ is
fixed.  This lets us write the following expression for the spin
operator,
\begin{eqnarray}
  \label{eq:75}
&&   S_{x,y}^z \sim  \left[1 + (-1)^{x+N(x)}\right] \big[ a_{+,x}^\dagger
    a_{+,x}^{\vphantom\dagger} + a_{-,x}^\dagger
    a_{-,x}^{\vphantom\dagger}  \nonumber \\
&&\; + \zeta^y a_{+,x}^\dagger
    a_{-,x}^{\vphantom\dagger} + \zeta^{-y} a_{-,x}^\dagger
    a_{+,x}^{\vphantom\dagger} + a_{3,x}^\dagger a_{3,x}^{\vphantom\dagger}\big],
\end{eqnarray}
where $a_{+,x}, a_{-,x}$ are annihilation operators for
chiral $S^z=1/2$ solitons, and $a_{3,x}$ is an annihilation operator
for an $S^z=3/2$ soliton.

Finally we consider the spin raising operator, $S_{x,y}^+$, containing the XY
components of the spin.  Acting on a site which is part of a singlet
bond, the raising operator converts the singlet to an $S^z=1$ triplet,
with amplitude $\mp 1/\sqrt{2}$ depending upon whether the site is the
left or right member of that singlet.  The triplet with $S^z=1$ has
overlap with the state of two adjacent $S^z=1/2$ solitons (as well as
other states not in the low energy sector).    Simple algebra shows
that, for instance
\begin{eqnarray}
  \label{eq:76}
  && \begin{pmatrix}
     | s \rangle \\ |\uparrow\uparrow\rangle \\ |s\rangle
  \end{pmatrix}
  = \\
  && \nonumber
  \frac{1}{3}\left[ |+\rangle|+\rangle + |-\rangle|-\rangle -
    \frac{1}{2}\left( |+\rangle|-\rangle + |-\rangle|+\rangle\right)\right]+\cdots,
\end{eqnarray}
where on the left hand side, $|s\rangle$ represents the singlet state,
and the columns represent the three columns in the TST.  One the right
hand side, the state has been decomposed into soliton states, and the
ellipses represent higher energy states.  Here we took the triplet to
reside in the middle row.  The other triplets can be obtained by
translation, as the chirality states are translational eigenstates.
From this construction, we obtain the analogous relation to
Eq.~\eqref{eq:75},
\begin{eqnarray}
  \label{eq:77}
  S^+_{x,y} &\sim & (-1)^{x+N(x)} \sum_{m=\pm}  \Big[ \zeta^{m y} a_{m,x}^\dagger
  a_{m,x+(-1)^{x+N(x)}}^\dagger \nonumber \\
  & & + a_{m,x}^\dagger a_{-m,x+(-1)^{x+N(x)}}^\dagger \Big].
\end{eqnarray}

The low energy excited eigenstates will not consist of localized
quasiparticles but delocalized ones, as solitons may hop
between columns of the same sublattice,  i.e. even or odd $x$.  As a
consequence, the states are eigenstates of the $x$-momentum $k_x$,
which is defined {\sl modulo $\pi$} rather than the usual $2\pi$, due
to the doubled background unit cell of the dimerization.  In the dilute limit we should consider only the states near the
minimum energy of the corresponding energy bands.  For the $S^z=3/2$
solitons, which are inversion symmetric, if this minimum is
non-degenerate it must occur at $k_x=0$ or $k_x=\pi/2$.  We expect it
to occur at the latter, $k_x=\pi/2$ value, owing to the dominant
antiferromagnetic spin correlations.  For the
$S^z=1/2$ solitons, inversion symmetry implies instead that if the
positive chirality ($q=+1$) soliton has minimum energy at $k_x=q_0$,
then the negative chirality soliton has its minimum energy at
$k_x=-q_0$.  We are not aware of a general argument to fix the momentum
$q_0$, however, and expect it is generically non-zero.  We have checked
this by a crude and uncontrolled variational calculation of the
soliton dispersion, which indeed gives minimum energy states with
opposite non-zero momenta for opposite chirality  (this calculation
gives $q_0=\pi/6$, but we do not expect this to be accurate).

With this in mind, we focus only on the minimum energy states and take
a continuum limit, writing
\begin{eqnarray}
  \label{eq:78}
  a_{\pm,x} & \sim & \psi_\pm(x) e^{\pm i  q_0 x}, \\
  a_{3,x} & \sim  & \Psi(x) e^{i\frac{\pi}{2} x},
\end{eqnarray}
where $\psi_m(x)$ and $\Psi(x)$ are taken as slowly varying continuum
boson fields.  Then Eqs.~\eqref{eq:75}\eqref{eq:77} become
\begin{eqnarray}
  \label{eq:80}
  S^z_{x,y} & \sim & \left[1 + (-1)^{x+N(x)}\right] \big[\sum_{m=\pm}
  \psi_m^\dagger \psi_m^{\vphantom\dagger} \nonumber \\
  && + \sum_m e^{i m (2q_0 x + \frac{2\pi}{3} y)} \psi_m^\dagger
  \psi_{-m}^{\vphantom\dagger} + \Psi^\dagger
  \Psi^{\vphantom\dagger}\big], \\
  S^+_{x,y} & \sim & 2i \sin q_0 \sum_m e^{i m
    (2q_0 x + \frac{4\pi}{3} y)} m (\psi_m^\dagger)^2 \nonumber \\
  && + 2\cos q_0 (-1)^{x+N(x)} \sum_m e^{i m
    (2q_0 x + \frac{4\pi}{3} y)} (\psi_m^\dagger)^2 \nonumber \\
  && + 2 \cos q_0 (-1)^{x+N(x)} \psi_+^\dagger \psi_-^\dagger
\end{eqnarray}

\begin{widetext}
  We are now in a position to write down an effective continuum theory
  to describe the low magnetization state in terms of bosonic field
  operators $\psi_m$ for $S^z=1/2$ solitons with chirality $m$ and
  $\Psi_m$ for the $S^z=3/2$ solitons, all taken near their band
  minima.  By symmetry, it takes the form
\begin{eqnarray}
  \label{eq:65}
  H_{\rm low} & = & \int \! dx\, \Big\{ \sum_{m=\pm } \psi_{m}^\dagger
  \big( -\frac{1}{2m_1}\partial_x^2+
  \epsilon_{1/2 }-h/2\big)
  \psi_{m}^{\vphantom\dagger} + \Psi^\dagger
  \big( -\frac{1}{2m_2}\partial_x^2+
  \epsilon_{3/2 }-3h/2\big)
  \Psi^{\vphantom\dagger} + V[\psi_+^\dagger \psi_+^{\vphantom\dagger}, \psi_-^\dagger
  \psi_-^{\vphantom\dagger}, \Psi^\dagger \Psi^{\vphantom\dagger} ]
  \Big\}. \nonumber \\
&&
\end{eqnarray}
\end{widetext}
Here $V$ is a general potential of quartic order and higher in the
fields, representing interactions of the solitons.  We have dropped
terms above which mix the different soliton species, e.g. ones which
might annihilate one $S^z=3/2$ soliton while creating 3 $S^z=1/2$
solitons.  Most such terms, at least at low order, are prohibited by
various symmetries, such as translation and inversion symmetry, at
least for a generic incommensurate wavevector $q_0$ for the $S^z=1/2$
solitons.

Consider increasing the magnetic field $h$ from zero.  The ground
state remains the soliton vacuum, i.e. the dimerized state, until the
energy of a state with non-zero solitons crosses the energy of the
vacuum.  Assuming repulsive interactions between solitons, this occurs
when the energy of a single soliton vanishes, and this type of soliton
will enter the system.  We must compare the energies $\epsilon_{1/2} -
h/2$ and $\epsilon_{3/2}- 3h/2$, and see which vanishes first on
increasing $h$.  If the $S^z=3/2$ soliton energy is large,
$\epsilon_{3/2}>3 \epsilon_{1/2}$, then the $S^z=1/2$ solitons will
appear, at $h=2\epsilon_{1/2}$.  Conversely, if
$\epsilon_{3/2}<3 \epsilon_{1/2}$, then the $S^z=3/2$ solitons will
appear, at $h=2\epsilon_{3/2}/3$.  The critical ratio
$\epsilon_{3/2}/\epsilon_{1/2} = 3$ is valid at infinitesimal soliton
density, i.e. $M\rightarrow 0^+$.  At larger magnetization,
interactions amongst solitons may become important, and will probably
tend to disfavor the $S^z=1/2$ solitons further, since these must
occur at a higher density and hence interact more strongly.
Since in any case we do not know the energies $\epsilon_{3/2},
\epsilon_{1/2}$, we cannot actually use this criteria quantitatively.
Instead, we simply consider both types of soliton liquids as
possibilities, and determine their properties at a phenomenological
level.

Let us consider first the $S^z=1/2$ case.  Then we can neglect the
$\Psi$ particle, which has an energy gap even when the $\psi_q$
solitons enter the system.   The structure of the solitonic state is
determined to a degree by the potential $V$ in Eq.~\eqref{eq:65}.  By
symmetry, it has the form
\begin{equation}
  \label{eq:74}
  V[n_+,n_-,0] =  \frac{a}{2} (n_+^2 + n_-^2) + b n_+ n_-.
\end{equation}
With $a>0$ for stability, the state depends upon the coefficient $b$.
If we assume $b<a$, then it is favorable for both solitons to enter
the system in equal amounts, and the system forms a one dimensional
Bose liquid of particles with two flavors.  Owing to the strong
quantum fluctuations in one dimension, this is a Luttinger liquid
phase with two independent massless bosonic modes, associated to the two
conserved densities.  In the CFT terminology, this is a state with
central charge $c=2$.   If instead $b>a$, it is preferable for the
system to choose one state of soliton only.  In this
case there is a spontaneously broken discrete symmetry (inversion
$P$), and only a single massless bosonic mode, or $c=1$.  We focus on
the former case, $b<a$, which we argue {\sl describes the same phase}
as the semi-classical incommensurate planar state.

To see this, we show that the spin correlations in the two-flavor
$S^z=1/2$ soliton liquid have the same form as those in the 1d
incommensurate planar phase, described in Sec.~\ref{sec:CI}.  In the
soliton liquid, we can use the usual bosonization of bosons for each
of the two species, $\psi_m \sim \sqrt{\bar{n}_s/2}e^{-i\theta_m}$,
$\psi_m^\dagger \psi_m^{\vphantom\dagger} \sim \bar{n}_s/2
+ \partial_x \phi_m/\pi$ (and $\Psi^\dagger \Psi=0$), where $\phi_m$
is the dual field to the boson phase $\theta_m$.  With this, we may
conveniently represent the non-local operator $N(x) = \bar{n}_s x +
\sum_m \phi_m/\pi$, where $\bar{n}_s$ is the mean soliton density.
Note since each soliton carries $S^z=1/2$ spread over the TST of width
$3$, the average magnetization {\sl per site} is $M=
\frac{1}{3}\bar{n}_s/2 = \bar{n}_s/6$.  Then
\begin{eqnarray}
  \label{eq:81}
  B_{x,y} & \sim & \overline{B} + \epsilon_0 \cos [(\pi + 2\delta)x +
  \varphi], \\
  S^z_{x,y} & \sim & \left(1+ \cos[(\pi + 2\delta)x +
    \varphi]\right) \Big( M + \frac{\partial_x \varphi}{6\pi} \nonumber \\
  && + n_s\cos [\theta_+-\theta_- +
  2q_0 x + \frac{2\pi}{3}y]\Big), \\
  S^+_{x,y} & \sim & 2i\sin q_0 \sum_m e^{im(2q_0 x +
    \frac{2\pi}{3}y)} m e^{2i\theta_m} \nonumber \\
  && + 2\cos q_0 \cos[(\pi + 2\delta)x +
    \varphi] \Big( e^{i(\theta_+ + \theta_-)} \nonumber \\
  && + \sum_m e^{im(2q_0 x +
    \frac{2\pi}{3}y)} e^{2i\theta_m} \Big).
\end{eqnarray}
Here $2\delta = \pi \bar{n}_s = 2\pi M/3$ and $\varphi = \phi_+ +
\phi_-$.   We can compare the above to the semi-classical result.  In
the semi-classical limit, the bosonic phases $\theta_\pm$ are weakly
fluctuating, while $\phi_\pm$ and hence $\varphi$ are strongly
fluctuating.  Then the dominant terms in the spin operators, with
smallest scaling dimension, are those which do not contain any of the
strongly fluctuating phases,
\begin{eqnarray}
  \label{eq:82}
  S^z_{x,y} & \sim &  M + n_s\cos [\tilde\theta +
  2q_0 x + \frac{2\pi}{3}y], \\
  S^+_{x,y} & \sim & -4\sin q_0\, e^{i\theta} \sin [\tilde\theta +
  2q_0 x + \frac{2\pi}{3}y],
\end{eqnarray}
where we defined $\theta=\theta_++\theta_-$ and $\tilde\theta =
\theta_+ - \theta_-$.  This can be directly compared to
Eqs.~\eqref{eq:73} of
Sec.~\ref{sec:CI}.   We see that the {\sl form} of the spin operators
is identical to that in the incommensurate coplanar state.   Thus we
can regard the $S^z=1/2$ chiral soliton liquid as another limit of the
same phase.

Let us turn to the case of the $S^z=3/2$ soliton liquid.  As there is
no chirality quantum number in this case, the state can be simply
viewed as a Luttinger liquid without spin, and is expected to be
described by a $c=1$ theory of a single massless boson.  We argue that
this $S^z=3/2$ soliton liquid is in fact another SDW phase very
similar to the one obtained by the quasi-one-dimensional approach of
Sec.~\ref{subsec:sdw}.  While one might have expected to find the {\sl
  identical} SDW phase in this way, we instead find that the $S^z=3/2$
soliton liquid is an SDW state with a different SDW wavevector, in
particular with $Q_y=0$, contrasting with the value $Q_y=2\pi/3$
obtain from the quasi-1d approach.   If the $S^z=3/2$ liquid indeed
occurs, therefore, we presumably require a phase transition to the
other SDW state upon increasing magnetization.

To observe the SDW structure of the $S^z=3/2$ soliton liquid, we again
consider the spin correlations.  Now we have no chiral solitons,
$\psi_m^\dagger \psi_m^{\vphantom\dagger}=0$.  This immediately
implies that {\em there are no low energy excitations with spin
  $S^z=1$ and hence no low energy content to the $S^\pm$ operators}.
Thus XY correlations decay exponentially in this phase, exactly as in
the the SDW phase.  To examine the $S^z$ correlations, we can bosonize
the non-chiral bosons.  This gives $\Psi \sim
\sqrt{\overline{n}_s}e^{i\vartheta}$,
$\Psi^\dagger\Psi^{\vphantom\dagger} \sim \overline{n}_s + \partial_x
\varphi/\pi$, with dual phases $\varphi,\vartheta$.  Now $N(x) =
\overline{n}_s x + \varphi/\pi$, and we note the relation between the
magnetization and soliton density is changed to $M= \overline{n}_s/2$,
since the solitons have spin $S^z=3/2$.  We see then that
\begin{equation}
  \label{eq:83}
  S^z_{x,y} \sim (1 + \cos[(\pi + 2\delta)x + \varphi]) (M +
  \frac{\partial_x\varphi}{2\pi}).
\end{equation}
Higher harmonics of the above cosine also appear in a more careful
treatment.  Note that the incommensurability is different in this
case: $2\delta = \pi \overline{n}_s = 2\pi M$.  Eq.~\eqref{eq:83} can
be compared to the corresponding formula, Eq.~\eqref{eq:52}, for the
quasi-1d SDW state in the TST.  We see that it is identical, save for
the presence of a factor $2\pi y/3$ inside the cosine in the quasi-1d
case.  This shows that the two states have the same structure, save
for a difference in the SDW wavevector, as mentioned above.

\section{Discussion}
\label{sec:discussion}

In this paper, we have presented a comprehensive analysis of the
field-anisotropy phase diagram of the three-leg spin-1/2 triangular spin tube,
of interest primarily as an approximation to the corresponding two
dimensional Heisenberg model on the anisotropic triangular lattice.
Pronounced quantum effects, strongly deviating from the expectations
based on classical analysis, occur throughout the phase diagram.  In
this section, we will discuss the implications of our results for two
dimensions, and how robust these quantum effects are to other
modifications to the model.

\subsection{Implications for two dimensions}
\label{sec:discussion2D}

Throughout the paper we have commented on how results obtained in the
one-dimensional TST geometry apply to the two-dimensional spin-1/2
system. Here we summarize these connections, with particular attention
to the phase diagram in 2d.  With a few exceptions, the phases
we obtained for the TST have straightforward analogs in 2d, and
consequently we expect the 2d diagram to be only slightly modified.

\begin{figure}[t]
  \begin{center}
  \scalebox{0.95}{\includegraphics[width=\columnwidth]{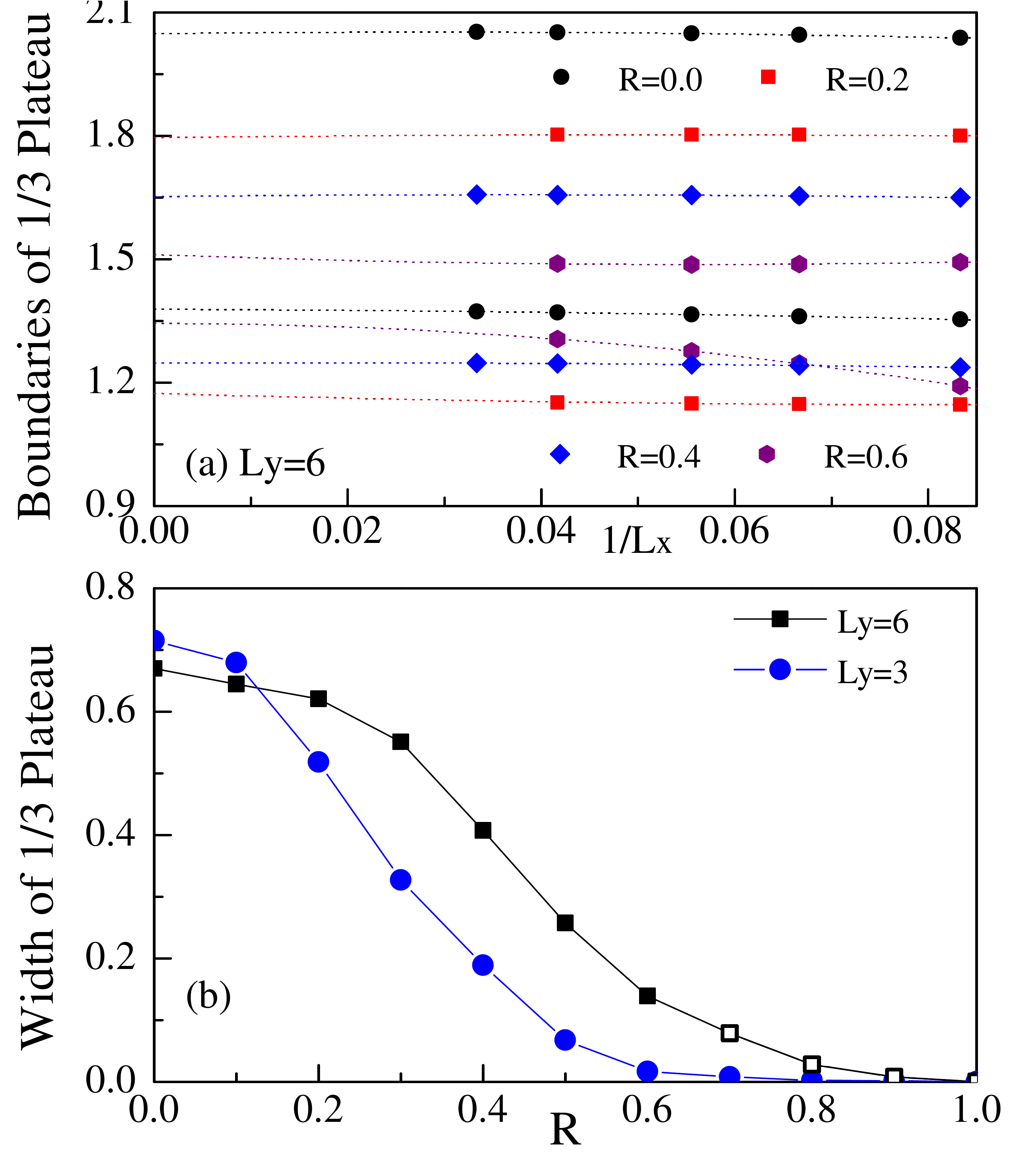}}
  \end{center}
  \caption{(Color online) Finite-size scaling of the boundaries of the
    1/3 plateau for cylinders of width $L_y =6$, and different
    anistropies: (a)R=0.0, 0.2, 0.4, 0.6. (b) Width of the 1/3 plateau
    as a function of R for cylinder of width $L_y=6$, shown as (black)
    squares.  The data points at $R>0.6$ (hollow square) are based on
    a preliminary finite-size scaling for quasi-2D system with
    $L_x\sim L_y$. The plateau width for $L_y=3$ (blue circles), from
    Fig.~\ref{fig:DMRG-plateau}, is shown for comparison.}
  \label{fig:plateau6leg}
\end{figure}
\begin{figure}[t]
  \begin{center}
  \scalebox{0.95}{\includegraphics[width=\columnwidth]{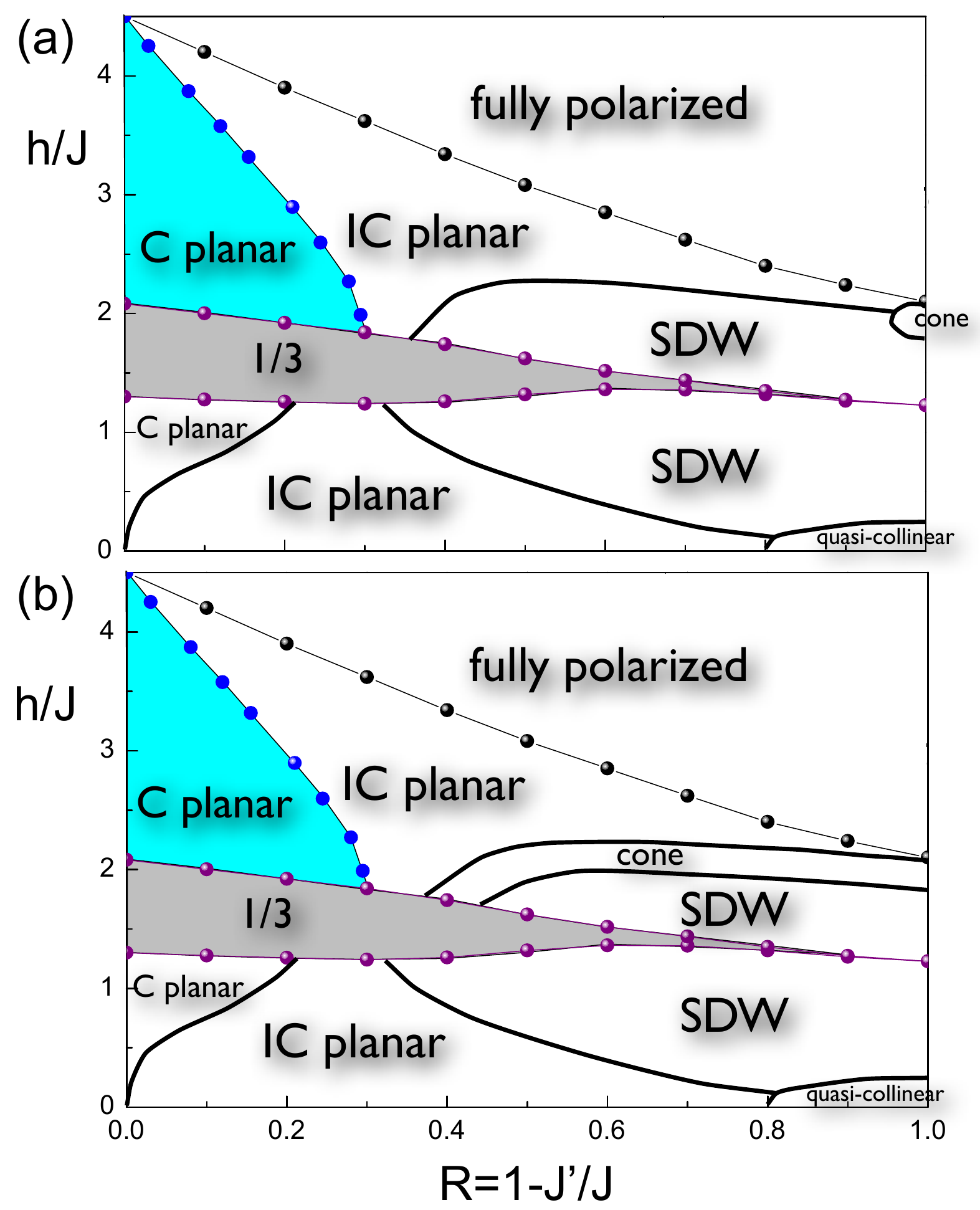}}
  \end{center}
  \caption{Schematic phase diagrams for the two dimensional, $S=1/2$
    system.  The shaded regions and boundaries containing full circles
    are based on preliminary DMRG results for circumference $L_y=6,9$
    systems in addition to the TST with $L_y=3$.  Other boundaries are
    drawn by hand using the considerations described in the text.  Two
    possible schematics are drawn, differing in the extent of the cone
    phase.  In (a) it is limited to the quasi-1d regime, while in (b) it
    extends to enclose the SDW state.  The latter possibility is more
    classical.  Intermediate or more complex cases are also possible.
    See text for further explanations. }
  \label{fig:schematicpds}
\end{figure}

For example, on the isotropic line, $R=0$, away from very small field,
all the phases we found are precisely those expected from the
semiclassical analysis of
Refs.~\onlinecite{alicea2009quantum,chubukov1991quantum}.  We expect
the semiclassical analysis to only work better in 2d, so the same
coplanar and plateau states, and their incommensurate analogs for
small anisotropy, $0<R\ll 1$, should occur there as well.  We note
that all coplanar states are intrinsically stabilized by quantum effects.

Perhaps the most striking feature amongst these states is the 1/3
magnetization plateau, which extends well beyond the semiclassical
regime in our phase diagram for the TST,
Fig.~\ref{fig:DMRG-plateau}. In addition to the TST, we have also
studied the 1/3 magnetization plateau for $L_y=6$ cylinders, see
Figure~\ref{fig:plateau6leg}. Close to the isotropic limit, $R\ll 1$,
the plateau width is only slightly changed by the increase in width
from $L_y=3$ to $L_y=6$, and its value $\Delta h \approx 0.7 J$ agrees
well with previous numerical studies
\cite{richter2009,tay2010variational,sakai2011,miyashita2004}.  This
trend in width is consistent with our picture that for small $R$ the
phases proximate to the plateau are commensurate planar {\sl ordered}
ones in the 2d limit.  The broken $U(1)$ symmetry of these phases
makes them sensitive to infrared quantum fluctuations in the 1d
geometry, since of course continuous symmetries are unbroken in 1d.
Hence in the thinner cylinders, the commensurate plateau state
competes slightly more effectively against the planar phases than in
two dimensions, leading to a wider plateau for smaller circumference.

In the intermediate region,
$0.2\lesssim R \lesssim 0.7$, the trend is much more striking and {\em
  opposite} to that for small $R$: the plateau width is seen to increase significantly
compared with that for $L_y=3$. The same is true in the larger
anisotropy limit, $0.7\lesssim R <1$, for which our preliminary
results, based on the finite-size scaling for quasi-2D systems with
$L_x \approx L_y$ (for such highly anisotropic systems, we were unable
to converge the $L_x \rightarrow \infty$ limit), still suggests a
finite 1/3 plateau, consistent with analytical arguments put forward
in \cite{starykh2010extreme} and in Sec.~\ref{sec:plateau}.  The
increase of the plateau width is understood as being due to a greater
stability of crystal phases in two dimensions.  Our DMRG results
strongly support existence of the 2d magnetization plateau state for
{\em all} values of spatial anisotropy $0 < R < 1$.


Several experimental spin-1/2 materials with the triangular lattice
structure have indeed been observed to support a 1/3 magnetization
plateau, including the well documented material Cs$_2$CuBr$_4$
\cite{ono2003,fortune2009} as well as Ba$_3$CoSb$_2$O$_9$, studied
more recently \cite{shirata2012}.  A notable exception is
Cs$_2$CuCl$_4$, which is isostructural to Cs$_2$CuBr$_4$, but does
{\sl not} exhibit a magnetization plateau\cite{tokiwa2006}. In our
opinion, as explained in detail in Ref.~\onlinecite{starykh2010extreme}, the
plateau is destabilized in this case by three-dimensional coupling,
which is stronger (relative to the appropriate $J$) in the Cs-based
magnet in comparison with the Br-based one \cite{ono2005}, with
perhaps strong Dzyaloshinskii-Moriya (DM) interactions in
Cs$_2$CuCl$_4$ playing an additional role \cite{povarov2011}.

The SDW phase dominates a large fraction of the phase diagram for the
TST.  This is an entirely quantum phase (since it requires modulation
of the {\sl length} of the static moments), which in 2d exhibits
incommensurate collinear long-range order along the field direction.
Being of quantum origin, one may wonder whether the SDW persists into
2d.   Based on renormalization group arguments, discussed extensively in
Ref.~\onlinecite{starykh2010extreme}, we know that the SDW indeed must
exist in the quasi-1d regime, $J'\ll J$, when inter-chain correlations
are relatively weak.  We expect that the region
occupied by the SDW may be somewhat curtailed in 2d relative to that in the
TST, but that it still is quite large.  This is based on intuition and numerical
evidence that inter-chain correlations remain suppressed for
relatively large $J'$ due to frustration.

Experimental verification of this novel magnetic state is clearly
called for. In this regard we would like to point out a recent series
of experiments on quasi-1d spin-1/2 material LiCuVO$_4$. While much of
the interest in this material stems from the high-field nematic phase
predicted \cite{mzh2010} and observed \cite{svistov2011} to occur near
the saturation field, several experimental studies
\cite{buttgen2007,masuda2011,svistov2012,takigawa2012} have found
strong evidence in favor of an incommensurate longitudinal SDW phase
in the intermediate range of magnetic fields. To understand this
finding better, it is important to realize that the inter-chain
exchange in this material is of zig-zag (triangular) type albeit of
predominantly ferromagnetic sign \cite{enderle2005}.  The
considerations of Section \ref{sec:weak-coupled} make it clear that
the SDW phase is not sensitive to the sign of inter-chain $J'$ and
should appear in the model with ferromagnetic $J'$ as well, see for
example Ref.~\onlinecite{sato2012} for explicit calculations.  We thus would like
to posit that a recent neutron scattering study \cite{mourigal2012},
which observed longitudinal spin fluctuations but no
transverse ones, is very much consistent with SDW phase
scenario. Like the spin nematic phase, which is expected to occur at
much higher magnetic fields, the SDW phase does not support
low-energy transverse spin excitations.
It would also be interesting
to seek evidence of an SDW state in Cs$_2$CuBr$_4$.

The above aspects of the TST and 2d phase diagrams are qualitatively
similar.  Qualitative differences are expected at low and high
fields.  At zero field, the TST exhibits a dimerized phase, which we
attribute (Sec.~\ref{sec:lowfield}) to quantum fluctuation effects
specific to one dimension.    In 2d, most of the zero field line should
exhibit incommensurate spiral order, with a small region of collinear
antiferromagnet at small $J'/J$, as argued in
Ref.~\onlinecite{starykh2007ordering}.  At high field, near saturation, where
the TST shows both coplanar and cone phases, we saw in
Sec.~\ref{sec:BS-2d-lattice} that in 2d only the coplanar state
occurs.  This is a rather surprising result, since the coplanar state
might be considered more quantum than the cone.  This observation
poses a tricky problem of connecting the limit of field approaching
saturation at fixed small
$J'$, where the coplanar state is expected, to the limit of vanishing $J'$
at fixed field slightly below saturation, where we instead expect a
cone state.  In 2d, therefore, a phase boundary must emanate from the
saturation point at $J'=0$, and we do not presently understand where
this boundary extends to.

Putting together all these considerations, we can construct schematic
phase diagrams for two dimensions.  The two simplest possibilities we
could construct are shown in Fig.~\ref{fig:schematicpds}.  The quasi-1d
analysis, which was carried out directly in 2d in
Ref.~\onlinecite{starykh2010extreme}, demands the cone, SDW, and plateau
phases at non-zero field and small $J'/J$.  It also requires a collinear
anti-ferromagnetic state at zero field and small $J'/J$.  This collinear
state is expected to be rapidly destroyed in favor of the SDW as the
field is imposed.  It is likely to become canted as it does so, but in
the absence of a detailed description of this narrow region descending
from the collinear antiferromagnet at zero field, we label it
``quasi-collinear'' in the figures.  Near the isotropic line, the
semi-classical description requires commensurate (C) planar and
incommensurate (IC) planar states, as well as the 1/3 plateau.  Finally,
near saturation, the dilute spin-flip approach becomes exact, and the
solution of the BS equation required the IC planar phase.  The shaded phases and the
boundaries containing circles are taken from preliminary
DMRG results for more two-dimensional system with $L_y=6,9$ lattice
spacings around the circumference.   The remaining phase
boundaries are drawn arbitrarily to connect the known regions demanded
by the above reasoning in the simplest possible manner consistent
with scaling.  The principal uncertainty in the diagrams is the extent
of the cone phase.  We expect it to occupy a relatively small portion of
the phase space, despite the fact that it is the classical ground state
everywhere below saturation except on the $R=0$ line!  In the first
schematic, Fig.~\ref{fig:schematicpds}a, the cone state occupies the
minimum possible area, while a more semi-classical situation might be as
shown in Fig.~\ref{fig:schematicpds}b.

\subsubsection{Comparison to other work}
\label{sec:comp-other-work}

It is interesting to compare our results to those of Tay and
Motrunich\cite{tay2010variational}, which is the only other
comprehensive study of the full anisotropy-field phase diagram of which
we are aware.  We caution that a strict comparison is not possible
because both their and our predictions for 2d are somewhat schematic,
being based on conjectural extrapolation of results for the 1d TST (us)
and finite clusters (them).  Nevertheless, one notices immediately
similarities between their schematic 2d phase diagram, Fig. 10 of their
paper, and our Fig.~\ref{fig:phase}.  First, the region near the
isotropic line is in both cases quite close to semiclassical
predictions.  Small differences appear at low fields, where indeed
quantum effects of the finite systems studied in both works are probably
maximal. Second, near the saturation field, they also find a wide range
of incommensurate planar phase (called incommensurate V in their study).
Our analytical BS analysis indicates that this phase in fact extends
over the full range of anisotropy, a fact which was not resolved in
their diagram.  Third, both studies indicate the robustness of the 1/3
plateau. As already mentioned above, our results for the width of the plateau
$\Delta h \approx 0.8 J$ at the isotropic point $R=0$ agree well with those
of Refs.~\onlinecite{richter2009,tay2010variational}. The more recent
exact diagonalization study \cite{sakai2011} predicts smaller width,
about $0.5 J$, but this is based on extrapolating $\Delta h$ from
small-size clusters. For $R > 0$, Ref.~\onlinecite{tay2010variational}
is the only one we can compare with, and qualitative agreement is quite good.
Our DMRG work completes the phase diagram, demonstrating the $1/3$ plateau
existence for all $J'>0$.

The major distinction between the two works
is in our finding of the SDW state in a wide field anisotropy range,
where Tay and Motrunich postulate separate spin liquid, spiral
(corresponding to our cone state), and quasi-1d regimes.  In our work,
renormalization group arguments rather clearly establish the SDW phase
in the small $J'/J$ regime in 2d, which is the quasi-1d region of Tay and
Motrunich.  We think it likely that even in 2d, the SDW phase extends to
$R \approx 0.5$.

\subsection{Suppressing the quantum effects}
\label{sec:suppr-quant-effects}

As remarked above, we predict two types of quantum states -- coplanar
phases and collinear SDWs -- in the 2d S=1/2 model.  While remarkably
robust in this case, these quantum phases can be suppressed by other
changes to the model: larger spins $S>1/2$, three-dimensional
coupling, and Dzyaloshinskii-Moriya (DM) interactions.

\subsubsection{Higher spin}
\label{sec:highS}

We first consider $S>1/2$, and find that the quantum phases are
strongly suppressed.  We begin with the vicinity of the saturation
field.  In Sec.~\ref{sec:BS-2d-lattice}, we showed that for $S=1/2$
the system forms a coplanar state in this limit for all $0<J'/J \leq
1$.  This is surprising since except for the isotropic case, the
coplanar phase is not a classical ground state.  Using the
calculations sketched below, we find that with increasing $S$, the
classical results are recovered, with the coplanar phase restricted to
increasingly narrow region near the isotropic limit, where it occurs
due to classical degeneracy.

To do so we use the
representation below\cite{batyev1986} , which is more convenient than
the Holstein-Primakoff one:
\begin{eqnarray}
\label{eq:laS1}
S_{\bf\sf r}^\dagger &=& \sqrt{2 S} [1 + (K_s - 1) b_{\bf\sf r}^\dagger b_{\bf\sf r}] b_{\bf\sf r} , \\
S_{\bf\sf r}^z &=& S - b_{\bf\sf r}^\dagger b_{\bf\sf r} ,\nonumber
\end{eqnarray}
where $K_s = \sqrt{1 - 1/(2S)}$. This expression reproduces the matrix
elements of spin raising and lowering operators between states with
different magnetization {\em exactly} within the two-magnon (two spin
flip) subspace.  The advantage of this form is that it requires no
$1/S$ expansion.   Note that for $S=1/2$ \eqref{eq:laS1}
reduces to \eqref{eq:bec1}, thanks to the hard-core condition
$(b_{\bf\sf r})^2 =0$, while for large $S\gg 1$ we recover Holstein-Primakov
asymptote $K_s \sim -1/(4 S)$. Note that for $S \geq 1$ the hard-core
constraint is not required and as a result the $U$-term is absent from
the two-magnon Hamiltonian \cite{kolezhuk2012}.  The Hamiltonian
within the two-magnon subspace retains the form in \eqref{eq:bec2} but now
the interaction term is a bit more complicated,
\begin{eqnarray}
&&V({\bf\sf k},{\bf\sf k}',{\bf\sf q}) = \frac{1}{2}\Big(J({\bf\sf q}) + J({\bf\sf k}+{\bf\sf q}-{\bf\sf k}')\Big) \\
&& - S K_s \Big( J({\bf\sf k}+{\bf\sf q}) + J({\bf\sf k}'-{\bf\sf q}) + J({\bf\sf k}) + J({\bf\sf k}')\Big), \nonumber\\
&&J({\bf\sf k}) = 2 J \cos[{\sf k}_x] + 4 J' \cos[\frac{{\sf k}_x}{2}] \cos[\frac{\sqrt{3}{\sf k}_y}{2}]. \nonumber
\end{eqnarray}
Numerical solution of the BS equation \eqref{eq:bec9} for the
two-dimensional triangular lattice, which proceeds along the same
lines as in Sec.~\ref{subsec:planar to cone}, finds that for higher
spins $S\geq 1$, near the saturation field the coplanar phase near the
isotropic limit is limited to a region $J'>J'_{\rm cr}>0$, with a cone
phase obtaining instead for $J'<J'_{\rm cr}$.  The critical value
monotonically increases with $S$, taking the values $J'_{\rm cr}/J
\approx 0.1, 0.5, 0.61$ for $S=1$, $3/2$ and $2$, respectively.  These
findings show that the absence of the cone state for $S=1/2$ found
here is a very unusual feature of the most quantum case. Larger, more
classical spins, do recover the classically expected state, although
still in a limited range of $J'/J$.

We next turn to the SDW phase.  Since this state is rooted in the
one-dimensional limit, we consider just the limit of weakly coupled
chains, for $S>1/2$, and in particular $S=1$.  We find that the SDW is
completely absent in this case.

To see this, we consider a magnetic field above the lower critical
field $h_\Delta$ needed to overcome the non-zero Haldane gap
($\Delta_{s=1} \approx 0.41 J$ for $J' \ll J$).  This turns the gapped
(and, essentially, decoupled -- see
Refs. \onlinecite{sato07,pardini08} spin-1 chains into critical
Luttinger liquids\cite{konik02,fath03,kolezhuk05}. It turns out that
these critical chains are characterized by a Luttinger parameter $K =
1/(4 \pi {\cal R}^2)\geq 1$ for all values of the magnetic field above
the gap-closing $h_\Delta$.\cite{konik02,fath03} This immediately
implies that the scaling dimension of the longitudinal spin density
operator $\mathcal{S}_{\pi-2\delta}^z(x)$ in \eqref{eq:bosontoS} is $K
> 1$ as well, which makes inter-chain SDW coupling in
\eqref{eq:hamisdw} (which has twice this scaling dimension) strictly
irrelevant.  As a consequence the SDW phase does not occur in the
quasi-1d limit.  Since this was its most stable regime in the $S=1/2$
case, it may well be that the SDW phase is totally absent for $S=1$!
It would be interesting to check this in future simulations.

What replaces the SDW? The large value of $K$ implies an increased tendency to
spin ordering transverse to the field direction, and indeed the twist
term (4th term in \eqref{eq:perturbH1}) is instead always relevant,
leading to stabilization of the cone state.  This result
is supported by analytical \cite{kolezhuk05} and numerical
\cite{mcculloch08} studies of the spin 1 zigzag ladder. For example,
Ref.~\onlinecite{mcculloch08} finds a finite vector chirality (that is,
a cone state) for all values of the magnetization in the case of $J_1 -
J_2$ spin-1 chain, along the $J_1 = J_2$ line.

Note that above, we found that the cone state was also stabilized for
small $J'/J$ in the vicinity of saturation.  It is likely then that
the cone phase evolves smoothly between the 1d limit $J'/J = 0^+$ and
the approach to saturation at finite $J'/J$.  Moreover, the presence
of the cone state at small $J'$ implies the absence of any
magnetization plateau in that regime.  The predictions appear quite
similat to those of the semiclassical analysis of
Ref. \onlinecite{alicea2009quantum}, which suggests that the full
phase diagram for $S=1$ might be well described semiclassically.  It
is clear that in particular the 1/3 plateau must terminate at some
finite (and perhaps not particularly small) value of the $J'/J$ ratio
in this case.

\subsubsection{Three dimensional coupling}
\label{sec:three-dimens-coupl}

Another experimentally-relevant modification of the spin-$1/2$
Hamiltonian is three dimensional coupling.  We
consider the simplest case of unfrustrated antiferromagnetic inter-plane
exchange interaction $J''$ between identical triangular layers.
Provided the three dimensional coupling is unfrustrated, we expect
that the particular form is not too important.  Such an interaction
is expected to make the spin system more classical and thus to
promote the classical cone state over the coplanar one.

Considering again the regime near saturation, one may readily solve
the BS equation, appropriately modified to the three-dimensional
situation.  We indeed find that high-field co-planar configuration
changes to the cone one for sufficiently large $J''/J$ ratio.  When
the triangular lattice is isotropic, $J'=J$, this occurs for
$(J''/J)_{\rm cr} \approx 0.2$, in agreement with the calculation in
Ref.~\onlinecite{nikuni1995hexagonal}.  Not unexpectedly, the critical
$J''$ becomes smaller for weaker inter-chain exchange $J'$. For
example, for $J' /J = 0.75$, as perhaps appropriate for
Cs$_2$CuBr$_4$, we find $(J''/J)_{\rm cr} \approx 0.15$ while for $J'
/J = 0.34$ (the Cs$_2$CuCl$_4$ case), $(J''/J)_{\rm cr} \approx
0.034$.  One-dimensional scaling arguments, described in
Appendix~\ref{sec:BS-1d}, suggest that $(J''/J)_{\rm cr} \sim
(J'/J)^2$ when $J'/J \ll 1$, in agreement with the numerical values
listed above.

In the 1d limit, $J'/J \ll 1$, introduction of unfrustrated $J''/J \ll
1$ disfavors SDW order in favor of a cone phase.  This is discussed
in detail in Sec.V of Ref.\onlinecite{starykh2010extreme}.  Thus
three-dimensional coupling, if unfrustrated, tends to remove all
quantum features of the phase diagram.

\subsubsection{Dzyaloshinskii-Moriya interactions}
\label{sec:dzyal-moriya-inter}

A variety of DM interactions can be present in anistotropic triangular
lattice systems, depending upon the crystal symmetry and microscopic
details.  This can lead to diverse effects which are difficult to
discuss without being more specific.  For the materials Cs$_2$CuCl$_4$
and Cs$_2$CuBr$_4$, the symmetry allowed DM interactions were obtained
and discussed in detail in Ref.\onlinecite{starykh2010extreme}.  Here
we describe only the effects of the dominant DM term in those
materials, which can be written as
\begin{equation}
  \label{eq:92}
  H_{\rm DM} = \sum_{x,y} {\bf D} \cdot {\bf S}_{x,y}\times \left(
    {\bf S}_{x-1,y+1} - {\bf S}_{x,y+1}\right),
\end{equation}
in the notation of this paper, with the DM-vector ${\bf D}= D {\bf\hat
  a}$ oriented
along the crystallographic $a$ axis, normal to the triangular planes.

Though small, a non-zero $D$ has significant
effects in both zero field and when a magnetic field is applied normal
to the triangular plane, i.e. parallel to the DM-vector.  In these
situations, unlike the $J'$ interchain coupling, it is not frustrated
either by the dominant chain interactions $J$ or by the applied
magnetic field.  It tends to favor the cone state (or a spiral in zero
field), and can obliterate the more quantum coplanar and SDW phases
completely if sufficiently strong in this field orientation.  Indeed,
with this field orientation, an
arbitrarily weak DM coupling inevitably forces the state in immediate
proximity to the saturated state to be a cone phase, for all values of
$J'/J$.  This occurs because the DM coupling splits the degeneracy of
the two minimum energy spin wave modes, already at the single spin wave
level, making a two-component condensate impossible when the spin flip
magnons are sufficiently dilute.

We note, however, that when the magnetic field is applied normal to the
$a$ axis, i.e. in the triangular plane, it itself frustrates the DM
interaction.  In this situation, the DM interaction is largely
ineffective and has only minimal perturbative effects on the spin
correlations.  These field orientations are therefore optimal for
observing quantum effects.

\subsection{Experimental implications and future directions}
\label{sec:future-directions}

Our study indicates that a number of ``quantum'' ordered states may be
found in $S=1/2$ anisotropic triangular lattice systems.  These states
are not so exotic as quantum spin liquids, and are well characterized by
their symmetries and associated order parameters.  They are instead
quantum in the weaker sense that they cannot be obtained in the
classical limit.  Most notably, we obtained a SDW
state whose order involves (quasi-)periodic modulation of the {\sl
  length} of the spin expectation value, along the field direction.  We
suggest this state occupies a wide swath of the field-anisotropy phase
diagram, provided perturbations to our model are not too strong.

The particular material Cs$_2$CuBr$_4$ appears a good candidate for the
observation of the SDW state, since three-dimensional coupling is known
to be relatively weak there, and experiments have already identified the
1/3 magnetization plateau.  Direct observation of the SDW would consist
of observing the incommensurate ordering wavevector evolving
monotonically with field, for fields above and below the plateau, and
correlating this wavevector with the average magnetization.  We expect
it to approximately follow the 1d relation, $q= \pi (1- M/M_s)$, away
from the plateau.  Given its 1d origin, one might well also expect that
the inelastic spectra retain 1d features, such as spinon continua, in
the SDW state and even in the plateau state above the gap.  Of course,
at low energy, in the vicinity of the SDW wavevector, we expect the
collective phason mode to dominate.  There must therefore be significant
rearrangement of the spectra on passing from low to high energy.  A more
detailed understanding of the spectral evolution with energy, field, and
anisotropy may make an interesting subject for future study.

In Cs$_2$CuBr$_4$, many additional features suggestive of phase
transitions were identified above the 1/3 plateau in the magnetization
process with an in-plane field.\cite{fortune2009}\  Our study indicates
that few such transitions should be expected in the pure $J-J'$ model.
Likely additional DM interactions (beyond the one given in
Eq.~\eqref{eq:92}) and perhaps further-neighbor couplings are at play.
Study of their effects is a possible avenue for more research.

More generally, the richness and surprisingly quantum nature of
field-anisotropy phase diagram of the relatively weakly frustrated
triangular lattice suggests that the behavior on more frustrated
lattices such as the kagom\'e and pyrochlore may be even more
interesting.   The methods used here should be helpful in attacking
these problems.

\acknowledgements We would like to thank A. Chubukov, R. Coldea,
A. Kolezhuk, M. Mourigal, F. Mila, M. Takigawa, and M. Zhitomirsky for
discussions and communications. We acknowledge support from the Center
for Scientific Computing at the CNSI and MRL: an NSF MRSEC
(DMR-1121053) and NSF CNS-0960316. This research was supported in part
by the National Science Foundation under Grants NSF DMR-1206809 (LB,
RC, and HJ), NSF PHY11-25915 (HCJ), and NSF DMR-1206774 (OAS).
\appendix

\section{Sine-Gordon model and commensurate-incommensurate transitions}
\label{sec:sine-gordon-model}

In this appendix, we summarize the Commensurate-Incommensurate
Transition (CIT) within the sine-Gordon model, which appears in
multiple places throughout the manuscript.  We consider the sine-Gordon
action in $d+1$ dimensions, with the form
\begin{eqnarray}
\label{eq:84}
  \mathcal{S}_{\rm sg} & = & \int d^d{\bf x}\, d\tau\, \Bigg\{ \frac{\kappa}{2}
    (\partial_\tau \vartheta)^2+ \sum_{\mu} \frac{\rho_{\mu}
    }{2} (\partial_\mu\vartheta)^2
    \nonumber \\
    && - \lambda \cos \left[ n (\vartheta-q x)\right] \Bigg\} ,
\end{eqnarray}
where $\vartheta$ is the sine-Gordon field.  We can write an alternative
expression in terms of the shifted field, $\hat\vartheta=\vartheta -
qx$, so that
\begin{eqnarray}
  \label{eq:85}
   \mathcal{S}_{\rm sg} & = & \int d^d{\bf x}\, d\tau\, \Bigg\{ \frac{\kappa}{2}
    (\partial_\tau \hat\vartheta)^2+ \sum_{\mu} \frac{\rho_{\mu}
    }{2} (\partial_\mu\hat\vartheta)^2
    \nonumber \\
    && + \delta \partial_x \hat\vartheta - \lambda \cos \left[ n \hat\vartheta\right] \Bigg\} ,
\end{eqnarray}
with $\delta = \rho_x q$.  In general, large $\delta$ prefers an
incommensurate state, where the field $\hat\vartheta$ is non-uniform and
unpinned, while for small $\delta$, a commensurate phase occurs, where
$\hat\vartheta$ is pinned to a fixed value by the cosine term.  The
detailed nature of the sine-Gordon model depends upon dimensionality, so
we treat the $d=1$ and $d \geq 2$ cases separately.

\subsection{$d\geq 2$: mean-field transition}
\label{sec:dgeq-2:-mean}

For $d \geq 2$, the fluctuations of the phase field $\hat\vartheta$ are
small even in the absence of the sine-Gordon term, i.e. for
$\lambda=0$.  This can be seen from the fact that, already at the
Gaussian level, the free boson propagator is non-divergent at small
momentum for $d\geq 2$.  This implies that the fluctuations of
$\vartheta$ are bounded, and one can therefore treat the entire problem
by a saddle point approximation.  Moreover, one can show that
fluctuation effects are negligible in the (quantum) CIT for $d\geq 2$.
More formally, $D=d+1=2+1$ is the upper critical dimension for the CIT.

Therefore in this case we may proceed by simply minimizing the action in
Eq.~\eqref{eq:85}.   The minimum action configuration is
independent of the $d-1$ coordinates normal to $x$ and $\tau$.  This gives
\begin{equation}
  \label{eq:15}
  {\mathcal S}_{sg} = L_\perp^{d-1}\beta E_{1d},
\end{equation}
5
where $L_\perp$ is the system width in the directions normal to $x$, and $\beta$ is
the length of the imaginary time integration.  The one-dimensional
energy is then
\begin{equation}
  \label{eq:86}
  E_{1d} = \int dx\, \left\{ \frac{\rho}{2}
    (\partial_x\hat\vartheta)^2 + \delta \partial_x\hat\vartheta -
    \lambda\cos ( n\hat\vartheta)\right\},
\end{equation}
where $\rho = \rho_x$.  Notice that $\delta$ only appears as a
boundary term, which means that the energy depends on $\delta$ only through the
winding number, $N = \left(\hat\vartheta( x = L ) - \hat\vartheta( x =
  0)\right)\frac{n}{2\pi}$.  Consider the case $N = 0$. Then, the
solution is uniform, i.e. $\hat\vartheta = 2\pi k/n$, with $k = 0,1,2 ...$.
With $N=1$, one obtains a well-known soliton solution of the
sine-Gordon model\cite{chaikin2000principles}, which reads
\begin{equation}
  \label{eq:ic7}
  \hat\vartheta(x) = \frac{4}{n} \arctan \left\{ e^{\pm n
      \sqrt{\frac{\lambda}{\rho}}(x - x_0)} \right\},
\end{equation}
where $x_0$ is the location of the center of the soliton. Note that
the soliton has a width $w \sim \sqrt{\rho/\lambda} $, and energy $E
\sim \sqrt{\rho\lambda}$. This gives a critical value,
\begin{equation}
  \label{eq:87}
  \delta_c = 4 \sqrt{\rho
  \lambda}/\pi,
\end{equation}
such that, for $\delta < \delta_c$, domain wall solitons cost positive
energy and so, are unfavorable, resulting in a commensurate wavevector.
For $\delta > \delta_c$, it is favorable for solitons to
be present, and the minimum energy configuration will be an array of
solitons which characterizes an incommensurate phase.

Eq.~\eqref{eq:87} defines the {\em location} of the CIT phase
boundary.  We may also discuss its critical properties.  On the
commensurate side, no solitons are present, which implies the winding
number $N=0$ precisely, and the ground state energy and field configuration are
independent of $\delta$.  Thus, there is no visible critical behavior
in the ground state (hence in equal time correlations) in the
commensurate phase.   On the incommensurate side, however, the minimum
energy configuration of $\hat\varphi(x)$ depends upon $\delta$.  It
can be considered as an array of solitons, whose main characteristic
is the spacing $\ell$ between solitons.  This spacing is determined by
the balance of the negative energy to introduce a soliton (which
favors many solitons with a short spacing) and the repulsive energy of
interaction between solitons (which favors large spacing).  The
repulsive interaction is exponentially small in the separation $\ell$
in units of the width $w$. Hence the energy of the array is
\begin{equation}
  \label{eq:88}
  E_{1d} = E_{1d}^{C} - (\delta-\delta_c) \frac{2\pi L}{n \ell} + c
 \sqrt{\rho\lambda} \frac{L}{\ell} e^{-\ell/w},
\end{equation}
where $c$ is an unimportant constant, and $L/\ell$ is the total number
of solitons.  Minimizing this over $\ell$, one finds the critical
behavior, to leading logarithmic accuracy,
\begin{equation}
  \label{eq:89}
  \ell \sim w \ln
  \left[\frac{\delta_c}{\delta-\delta_c}\right],
\end{equation}
for $0<\delta-\delta_c \ll \delta_c$.
The presence of the soliton array implies that the average gradient of
the phase $\hat\vartheta$ is non-zero, which defines the {\sl
  incommensurability} wavevector $\overline{q}$:
\begin{equation}
  \label{eq:90}
  \overline{q} = \overline{\partial_x \hat\vartheta} = \frac{2\pi}{n\ell} \sim
  \frac{1}{w|\ln[\delta-\delta_c|/\delta_c]|} \Theta(\delta-\delta_c).
\end{equation}
The incommensurability $\overline{q}$ in the incommensurate phase gives the shift of the ordering
wavevector from its commensurate value.  Other critical properties at
the CIT in $d\geq 2$ are readily obtained from the results above.  For
example, the ground state energy density is simply the saddle point
value of $E_{1d}$, which scales as
\begin{equation}
  \label{eq:91a}
  \frac{E}{L} \sim -\frac{\delta-\delta_c}{|\ln(\delta-\delta_c)|} \Theta(\delta-\delta_c).
\end{equation}

\subsection{$d=1$: quantum fluctuations}
\label{sec:d=1:-quant-fluct}

In the case $d=1$, fluctuations of the phase field cannot be
neglected.  This can be anticipated from the Gaussian level result
that, in the absence of a sine-Gordon term, the free boson Green's
function is logarithmically divergent at small momentum, signalling
large fluctuations of $\vartheta$.  Hence we must deal directly with
the 1+1-dimensional action,
\begin{eqnarray}
  \label{eq:91}
&&  \mathcal{S}_{\rm sg}  = \int \! dx\, d\tau\! \Bigg\{ \frac{\kappa}{2}
    (\partial_\tau \hat\vartheta)^2+ \frac{\rho}{2} (\partial_x\hat\vartheta)^2
    + \delta \partial_x \hat\vartheta - \lambda \cos n\hat\vartheta \Bigg\} .\nonumber \\
\end{eqnarray}
Once again, $\delta$ is the
coefficient of a pure boundary term, which simply counts the number of
solitons in the system.  A finite density of solitons will be
generated, provided the energy of a soliton for $\delta=0$ is
compensated by this boundary energy, which equals $2\pi \delta/n$.
Thus we need the energy of a soliton at $\delta=0$, i.e. in the
pure quantum sine-Gordon model.

We estimate this as
follows.  The scaling dimension of the cosine term, $\Delta_n$, is
easily calculated, and is equal to
\begin{equation}
  \label{eq:24}
  \Delta_n = \frac{n^2}{4\pi \sqrt{\kappa\rho}}.
\end{equation}

The cosine is relevant when $\Delta_n<2$, and irrelevant if
$\Delta_n>2$.  When it is irrelevant, there is no pinning of the phase
field at low energies.  A state of this type is known as a ``floating
phase'', and because of the lack of pinning, the state becomes
immediately incommensurate for any non-zero $\delta$,
i.e. $\delta_c=0$, and there is no CIT.

When the cosine is relevant, then when $\delta=0$, the phase is pinned
at low energies, and the energy of a soliton is non-zero.  We need to
estimate this energy to locate the value $\delta_c$ which defines the
CIT.    We do this by renormalization
group (RG) arguments.  Renormalizing out to a length $\xi$, the cosine is
reduced by fluctuations by an amount proportional to
$\xi^{-\Delta_n}$, so $\lambda_{\rm eff} \sim \lambda
\xi^{-\Delta_n}$.  For a possible soliton of width $\xi$, the energy
cost is of order
\begin{equation}
  \label{eq:25}
  \epsilon_s \sim \frac{\rho}{\xi} \left(\frac{2\pi}{n}\right)^2 -
  \lambda_{\rm eff} \xi.
\end{equation}
The actual soliton size is determined by optimizing this over $\xi$,
which gives
\begin{equation}
  \label{eq:26}
  \xi \sim \left( \frac{\rho}{\lambda n^2}\right)^{\frac{1}{2-\Delta_n}},
\end{equation}
and thus an energy cost for the soliton of order
\begin{equation}
  \label{eq:27}
  \epsilon_s \sim \lambda^{\frac{1}{2-\Delta_n}} \left( \frac{\rho}{n^2}\right)^{\frac{1-\Delta_n}{2-\Delta_n}}.
\end{equation}
This energy should equal the energy gain $2\pi\delta_c/n$  from the boundary term at the
CIT, which gives
\begin{equation}
  \label{eq:28}
  \delta_c \sim \sqrt{\lambda\rho}
  \left(\frac{\lambda}{\rho}\right)^{\frac{\Delta_n}{4-2\Delta_n}}.
\end{equation}
Note that this approaches the mean-field result of the previous
subsection when $\Delta_n \rightarrow 0$, and becomes very suppressed
when $\Delta_n \rightarrow 2^-$ (since we must assume $\lambda<\rho$
for consistency of the treatment).

We now turn to the critical behavior, which in 1+1 dimensions is a
storied problem in critical phenomena.  It is sometimes
referred to as a Pokrovsky-Talapov transition, due to the solution by
those authors.\cite{PhysRevLett.42.65}\  We recapitulate the essence of
the argument.  As in the mean-field case, for $\delta<\delta_c$, there
are no solitons in the system, and the ground state energy is independent of
$\delta$, i.e. there is no sign of criticality in any static quantity.
However, the excitation gap for creating a soliton vanishes linearly
with $\delta_c-\delta$.  For $0<\delta-\delta_c \ll \delta_c$, we expect
a low density of solitons to be present in the system, again determined
by the balance of the (negative) single soliton energy and the repulsive
soliton-soliton interactions.

We must, however, in this case treat the problem quantum mechanically.
In particular, we must consider the effects of interactions properly in
the low density limit.  In this limit, the kinetic energy and momentum
of individual solitons is vanishingly small, and well-known results for
low energy scattering apply.  In particular, for short-range repulsively
interacting particles in one dimension, the probability of transmission
{\em vanishes} in the low energy limit.  Thus effectively, regardless of
the microscopic strength of the interaction, or of its short distance
structure, the solitons behave at low densities as though they were {\em
  hard core} particles, which cannot pass one another.  To model this
behavior, we can treat the solitons as {\em fermions}.   Interactions at
longer distances beyond the local hard core are weak and unimportant, so
the fermions are effectively {\em free}.

The free fermion problem is trivially soluble, so we can easily
obtain the critical behavior.   When $\delta>\delta_c$, we simply fill
the negative energy fermion states to form a Fermi sea.  The sine-Gordon
model has Lorentz invariance, so the dispersion of the solitons must be
relativistic, hence the energy for a single soliton is
\begin{equation}
  \label{eq:86a}
  E_{\rm sol} = \sqrt{\epsilon_s^2 + v^2 k^2} - \frac{2\pi}{n} \delta,
\end{equation}
where the velocity $v=\sqrt{\rho/\kappa}$, and $\epsilon_s = 2\pi
\delta_c/n$.  The Fermi momentum $k_F$ is determined by the condition
$E_{\rm sol}=0$.  It will be small near the CIT, so we may expand the
relativistic dispersion into its non-relativistic limit
\begin{equation}
  \label{eq:87a}
    E_{\rm sol}(k_F) = -\frac{2\pi}{n} (\delta-\delta_c) +
    \frac{k_F^2}{2m} = 0,
\end{equation}
with $m= \epsilon_s/v^2$.  This determines the Fermi momentum
\begin{equation}
  \label{eq:88a}
  k_F = \left[ \frac{4\pi m}{n} (\delta-\delta_c) \right]^{1/2} \sim \sqrt{\delta-\delta_c}.
\end{equation}
The density of solitons is just $k_F/\pi$, as usual for spin-less
fermions, so the incommensurability is thus
\begin{equation}
  \label{eq:89a}
  \overline{q} = \frac{2\pi}{n} \frac{k_F}{\pi} = \frac{2k_F}{n} \sim \sqrt{\delta-\delta_c}.
\end{equation}
The square-root behavior is quite distinct from the logarithmic one in
$d\geq 2$.  We may also easily obtain the behavior of the ground state
energy density, as the total energy of the Fermi sea,
\begin{eqnarray}
  \label{eq:90a}
  \frac{E}{L} & = & \int_{-k_F}^{k_F} \frac{dk}{2\pi} \left[ \frac{k^2}{2m} -
    \frac{2\pi}{n} (\delta-\delta_c)\right] \nonumber \\
  &\sim & -(\delta-\delta_c)^{3/2} \Theta(\delta-\delta_c).
\end{eqnarray}
Many more results, e.g. for correlations in the incommensurate phase,
can be readily obtained from the free fermion formulation, but we leave
this to the reader to discover for themselves in the literature.

\section{Detailed calculations of BS}
\label{app:ideal2d}

In this appendix, we present our solutions to the Bethe-Salpeter (BS) equation in Eq.~\eqref{eq:bec9}.
This equation applies only near saturation field, where the system can be modeled as dilute (hard core) bosons.
We substitute our ansatz, Eq.~\eqref{eq:ansatz}, into the BS equation.
With the constraint equation, Eq.~\eqref{eq:bec10}, which enforces $s=1/2$, we obtain a set of linear equations for the constants $A_i$, which can be written in a matrix form as
\begin{widetext}
\begin{center}
\begin{equation}
\label{eq:mat}
\begin{pmatrix}
 \tau_{11} & \tau_{12} &\tau_{13} &\tau_{14} & \tau_{15} &\tau_{16} & \tau_{17} \\
2J\tau_{21} & 2J\tau_{22}+1 &2J\tau_{23} &2J\tau_{24} & 2J\tau_{25} &2J\tau_{26} & 2J\tau_{27}\\
2J\tau_{31} & 2J\tau_{32} &2J\tau_{33}+1 &2J\tau_{34} & 2J\tau_{35} &2J\tau_{36} &2J\tau_{37}\\
2J'\tau_{41} & 2J'\tau_{42} &2J'\tau_{43} &2J'\tau_{44}+1 & 2J'\tau_{45} &2J'\tau_{46} & 2J'\tau_{47}\\
2J'\tau_{51} & 2J'\tau_{52} &2J'\tau_{53} &2J'\tau_{54} & 2J'\tau_{55}+1 &2J'\tau_{56} &2J'\tau_{57} \\
2J'\tau_{61} & 2J'\tau_{62} &2J'\tau_{63} &2J'\tau_{64} & 2J'\tau_{65} &2J'\tau_{66}+1 &2J'\tau_{67} \\
2J'\tau_{71} & 2J'\tau_{72} &2J'\tau_{73} &2J'\tau_{74} & 2J'\tau_{75} &2J'\tau_{76} &2J'\tau_{77}+1 \\
\end{pmatrix}
\left( \begin{array}{c} A_0 \\ A_1 \\A_2 \\A_3 \\A_4 \\A_5 \\A_6 \end{array} \right)=
\left( \begin{array}{c} 1 \\ 2J \\ 0 \\ 2J' \\ 0 \\2J' \\0\end{array} \right)
\end{equation}
\end{center}
\end{widetext}
where we have defined
\begin{eqnarray}
\label {eq:ru1}
\tau_{lm}(k,k';\Omega) &\equiv& \int_{q}\frac{{\bf T}_l(q){\bf T}_m(q)}{\epsilon(k+q)+\epsilon(k'-q)+\Omega} \\
{\bf T}(q)&=&(1,\cos q_x , \sin q_x , \cos q_y, \sin q_y,   \nonumber \\
&&\cos (q_x-q_y),\sin(q_x-q_y)),
\end{eqnarray}
and $\Omega \propto | h - h_{\rm sat}|$.
Although the $\tau_{lm}'s$ are integrals over simple trigonometric functions and other known quantities, e.g. the dispersion, these integrals are divergent in both one and two dimensions and must be treated with care.
It is possible, however, to analyze them asymptotically.
Once these integrals are evaluated, we can solve for the constants $A_i$ to obtain $\Gamma(q)$ from our ansatz, Eq.~\eqref{eq:ansatz}.
In the next two subsections, we take the reader through our asymptotic analysis.

\subsection{Asymptotic behavior of $\tau_{lm}$ for the 2d case}
\label{app:2d}

In this section, we calculate the $\tau_{lm}$'s for the 2d case.
As aforementioned, we are interested in performing asymptotic analysis in the limit $\Omega \to 0$, as the full integrals are too complicated to evaluate fully.
We can partition the integrals into the the first two subleading terms, $B_{lm} \ln(\Omega)+C_{lm}$, where the constants $B, C$ are independent of $\Omega$.
We can consider two cases: one with the same incoming momenta, i.e. the cone phase with $\Gamma_1 = \Gamma({\bf Q},{\bf Q},0) = \Gamma(-{\bf Q},-{\bf Q},0)$,
and the other with different incoming momenta, i.e. the coplanar phase with $\Gamma_2 = \Gamma({\bf Q}, -{\bf Q}, 0) + \Gamma({\bf Q},-{\bf Q},-2{\bf Q})$.
Here, the wave vector ${\bf Q}$ minimizes the dispersion relation in Eq.~\eqref{eq:bec3}, which can now be substituted into Eq.~\eqref{eq:ru1}.
After some algebraic simplifications, we obtain
\begin{equation}
\label{eq:ru2}
\tau_{lm}=\frac{1}{4\pi^2}\int_0^{2\pi} dq_x \int_0^{2\pi} dq_y\frac{{\bf T}_l(p){\bf T}_m(p)}{a+b\cos q_y}.
\end{equation}
The exact forms of $a, b$ will depend on whether the incoming momenta are same or different.
In this appendix, we will only present our results for $l = m = 1$, in which case, we can integrate analytically over $q_y$ in Eq.~\eqref{eq:ru2}, and obtain the following
\begin{equation}
\label{eq:ru3}
\tau_{11}=\int_0^{2\pi} dq_x \frac{1}{\sqrt{a^2-b^2}}
\end{equation}
To proceed further, we need to specify the exact form of $a$ and $b$.

\begin{enumerate}
\item {\sl Same incoming momenta}:
For the same incoming momenta, $a,b$ take on the following form
\begin{eqnarray}
\label{eq:ru3}
&& a= \Omega+J(2+j^2-(2-j^2)\cos q_x) \nonumber \\
&& b= J (-2j^2 \cos(q_x/2)),
\end{eqnarray}
where we define $j \equiv J'/J$.
The integrand diverges near $q_x=0$ like $1/q_x$ in the limit $\Omega \rightarrow 0$, and thus, integral is logarithmically divergent.
After some analysis, the integral takes on the following form,
\begin{equation}
\label{eq: ru4}
\tau_{11} \sim -\frac{1}{2\pi j \sqrt{4-j^2}}\ln(\Omega)+\frac{\ln(2j(4-j^2))}{\pi j \sqrt{4-j^2}} .
\end{equation}

\item {\sl Different incoming momenta}:
For this case, $a,b$ are as follows
\begin{eqnarray}
\label{eq:ru5}
a&=& \Omega+J [2+j^2+(-2+j^2) \cos(q_x) \nonumber \\
&+&j\sqrt{4-j^2} \sin(q_x)] \nonumber \\
b&=& -2Jj \left[j\cos(q_x/2)+\sqrt{4-j^2}\sin(q_x/2)\right].
\end{eqnarray}
The integrand now has two divergent points at $q_{x}=0$ and $q_{x}=-2\arccos(1-j^2/2)$.
Therefore, in comparison with the previous case, the logarithmic term doubles, and the integral takes on the form
\begin{eqnarray}
\label{eq:ru6}
\tau_{11} \sim  -\frac{1}{\pi j \sqrt{4-j^2}}\ln(\Omega)+\frac{2\ln(j(4-j^2))}{\pi j \sqrt{4-j^2}} .
\end{eqnarray}
\end{enumerate}

\subsection{$\tau_{lm}$ for the TST case}
\label{app:tst}

In computing the $\tau_{lm}$'s for the TST, we turn the two-dimensional integral in the previous section into a single integral over $q_x$ and a sum over $q_y$.
As one can imagine, the asymptotic behaviors differs in the TST from the 2d, in that, in the limit $\Omega \to 0$, the integrals diverge as $1/\sqrt{\Omega}$.
Therefore, the two subleading terms of the integrals are $B_{lm}/\sqrt{\Omega}+C_{lm}$, where again, $B,C$ are independent of $\Omega$.
We present our results for $l=m=1$ for the two cases of the same and differing incoming momenta.

\begin{enumerate}
 \item For the {\sl same incoming momenta}, we obtain the following expression
\begin{widetext}
\begin{equation}
\label{eq:7ru}
\tau_{11} = \frac{1}{6 \sqrt[4]{j^2-j+1} \sqrt{\Omega }}+\frac{4}{3 \sqrt{9 j^2+24
   \sqrt{(j-1) j+1} j-24 j+\frac{36 (j-1)}{(j-1)
   j+1}+36}}+O(\sqrt{\Omega}),
\end{equation}
\end{widetext}
where again, $j\equiv J'/J$.

\item We now compute $\tau_{11}$ for the case of {\sl differing incoming momenta}, in which case, the integral evaluates to
\begin{widetext}
\begin{equation}
\label{eq:ru8}
\tau_{11}= \frac{1}{3 \sqrt[4]{(j-1) j+1} \sqrt{\Omega }}+\frac{1}{3 \sqrt{3} \sqrt{j\left(3 j+4 \sqrt{(j-1) j+1}-4\right)}}+O(\sqrt{\Omega}).
\end{equation}
\end{widetext}
\end{enumerate}

\subsection{Weakly coupled chains limit}
\label{sec:BS-1d}

In this appendix, we analytically check the results of Sec.~\ref{sec:BS-2d-lattice} in the limit of weakly coupled chains, $J' \ll J$.
Recall that the calculation was done for a full two-dimensional lattice.
Hereafter, we will use Cartesian coordinates, $({\sf x},{\sf y})$, for convenience.
In this limit, we can express the spin flip operator as a continuous function of ${\sl x}$, which is along the chain direction, while keeping the chain index ${\sf y} \in {\cal Z}$ discrete.
Then, from Eq.~\eqref{eq:spinops}, we write this operator as $\Psi_{\sf y}({\sf x}) \sim S^+_{{\sf y}, \pi}({\sf x})$, where its low energy theory is described by the following action
\begin{eqnarray}
\label{eq:bs1}
{\mathcal S}_{\rm 1d} & = &  \sum_{\sf y} \int d {\sf x} d\tau \Big\{ \Psi_{\sf y}^\dagger ( \partial_\tau -\frac{1}{2m}\partial_{\sf x}^2  - \mu)\Psi_{\sf y} \nonumber\\
&& - t (\Psi_{\sf y}^\dagger i \partial_{\sf x}  \Psi_{{\sf y}+1} + {\rm h.c.}) + u \Psi_{\sf y}^\dagger \Psi_{\sf y}^\dagger \Psi_{\sf y} \Psi_{\sf y} \nonumber\\
&& + v \Psi_{\sf y}^\dagger \Psi_{{\sf y}+1}^\dagger \Psi_{{\sf y}+1} \Psi_{\sf y} \Big\}.
\end{eqnarray}
The spin-flip (magnon) mass, $m = 1/J$, follows from the quadratic dispersion of the magnon mode near momentum $\pi$ in a fully polarized chain.
Additional interaction terms describe the hard-core constraint ($u$ term) as well as the transverse ($t = J'_{xy}/2$) and longitudinal ($v = 2 J'_z$)
parts of the interchain exchange interaction $J'$.
Note that $t$-term contains a spatial derivative with respect to ${\sf x}$, which reflects the frustration of the interchain exchange by the triangular geometry.
In addition, this term contains a factor of $i$ from the staggered factor $(-1)^x = e^{i \pi x}$ in Eq.~\eqref{eq:spinops}, and from the fact that ${\sf x}$ takes
half-integer values on odd chains (see Eq.~\eqref{eq:9}, Fig.~\ref{fig:lattice}(a), and Appendix D6 of Ref.~\onlinecite{starykh2010extreme}).

We can analyze each term of Eq.~\eqref{eq:bs1} through simple dimensional analysis, which will deem all these terms to be relevant under RG.
Denoting the spatial scale along ${\sf x}$ as $L$, we can conclude that $\tau \sim L^2$, $\Psi_{\sf y}$ scale as $1/\sqrt{L}$, while
thee three interaction terms, $t, u$ and $v$, scale as $L$.
Hence, these are {\sl relevant} interactions and must be included in our analysis of the low energy theory.

We can Fourier transform Eq.~\eqref{eq:bs1} and write the Hamiltonian that corresponds to this action,
\begin{eqnarray}
\label{eq:bs2}
H_{\rm 1d} &=& \sum_{\bf\sf k} \Psi_{\bf\sf k}^\dagger (\frac{{\sf k}_x^2}{2 m} + 2 t {\sf k}_x \cos[{\sf k}_y] - \mu)\Psi_{\bf\sf k} \nonumber\\
&&+ \frac{1}{2N}\sum_{{\bf\sf k},{\bf\sf k}',{\bf\sf q}} V({\bf\sf k},{\bf\sf k}',{\bf\sf q}) \Psi_{{\bf\sf k} + {\bf\sf q}}^\dagger
\Psi_{{\bf\sf k}' - {\bf\sf q}}^\dagger \Psi_{{\bf\sf k}'} \Psi_{{\sf\bf k}}.
\end{eqnarray}
Here $V({\bf\sf k},{\bf\sf k}',{\bf\sf q}) = V({\bf\sf q}) = 2 u + 2 v \cos[{\sf q}_y]$.
Note that while the range of ${\sf k}_x$ is not restricted, $-\infty < k_x < \infty$, that of ${\sf k}_y$ is limited by the lattice, $-\pi \leq {\sf k}_y \leq \pi$.
This single particle dispersion contain two degenerate moment at ${\bf\sf Q}_1 = (-2 t m, 0)$ and ${\bf\sf Q}_2 = (2 t m, \pi)$.

The single particle dispersion has two degenerate minima, at ${\bf\sf Q}_1 = (-2 t m, 0)$ and ${\bf\sf Q}_2 = (2 t m, \pi)$.
We can now compute the renormalized couplings $\Gamma_1, \Gamma_2$ in a similar manner as the previous subsections.
However, we alter our ansatz of the BS equation, Eq.~\eqref{eq:bec9}, to take the form $\Gamma({\bf\sf q}) = A_0 + A_1 \cos[{\sf q}_y]$, because the odd contribution, $\propto \sin[{\sf q}_y]$, vanishes under the integral as the denominator in Eq.~\eqref{eq:bec9} is even for all combinations of incoming and transferred momenta.

Computing $\Gamma_1, \Gamma_2$ requires one to solve two linear equations for $A_0, A_1$, which involve 2d integrals over functions with denominators like $[{\sf k}_x^2/m + 16 m t^2 \sin^2[{\sf k}_y/2] + \Omega]$ (for $\Gamma_1$) and $[({\sf k}_x - 4 m t \sin^2[{\sf k}_y/2])/m + 4 m t^2 \sin^2[{\sf k}_y] + \Omega]$ (for $\Gamma_2$).
We first evaluate these integrals analytically by separating out the leading terms in $\ln[\Omega/(mt^2)]$, then taking the limit $u \to \infty$ to, again, enforce the $s=1/2$ constraint.
The expressions are as follows
\begin{eqnarray}
\label{eq:bs3}
\frac{\Gamma_1}{8 \pi t} &=& \frac{1 + \frac{4}{3}\gamma}{(1 + \frac{4}{3}\gamma) \ln\Upsilon + 4 \ln 2 + 4 \gamma (\frac{4}{3}\ln 2 -1)},\\
\frac{\Gamma_2}{8 \pi t} &=& \frac{1}{\ln\Upsilon + 2 \ln 2},
\label{eq:bs4}
\end{eqnarray}
where $\gamma = v/(\pi t)$ and $\Upsilon = 16 m t^2/\Omega$.
Given these forms, we can conclude that $\Gamma_1 > \Gamma_2$ for $\gamma \geq \gamma_c = 3 \ln 2/(6 - 4 \ln 2) \approx 0.644$.
Since we are considering the isotropic Heisenberg model, where $\gamma = 4/\pi > \gamma_c$, we observe that the coplanar fan state prevails over the cone state in the $J' \ll J$ limit, in agreement with the full lattice approach in Eq.~\eqref{eq:gamma2d-decoup},
once the parameters $m$,$t$,$v$ are expressed in terms of exchange integrals.

With this approach, we can also estimate the width of the planar fan state near saturation field through simple dimensional analysis of Eq.~\eqref{eq:bs1}.
Since the chemical potential, $\mu = h_{\rm sat} - h$, scales as $L^2$ and the $t$ interaction scales as $L$,
the phase boundary between the planar and the lower-field phase must scale as $\Delta h \sim (J')^2/J$.
This boundary separates the planar fan phase from the cone phase, a region in which a standard bosonization description of
Sec.~\ref{sec:weak-coupled} becomes appropriate. Details of this analysis are presented in Appendix~\ref{sec:cone}.

Similar reasoning allows one to estimate the stability of the planar fan state with respect to inter-layer coupling $J''$,
which is always present in real materials. It is clear that (non-frustrated) inter-layer coupling corresponds to adding
a simple single particle hopping term between layers with a different ${\sf z}$-coordinated
$\int d\tau d{\sf x} \sum_{\sf z} J'' (\Psi^\dagger_{{\sf y},{\sf z}} \Psi_{{\sf y},{\sf z}+1} + {\rm h.c.})$ term to the action
in Eq.~\eqref{eq:bs1}. Such a term also scales as $L^2$, which implies that the phase boundary
between the planar and the cone phase in the $J' - J''$ plane takes on a quadratic shape, $J'' \sim (J')^2/J$.

\section{Additional one dimensional analysis}
\label{sec:addit-one-dimens}

The purpose of this appendix is to show that the TST geometry with 3-legs is unique in that the renormalized couplings generated through RG produce significantly different physics for $N=3$ compared to that of $N>3$, in the limit $J' \ll J$.
Moreover, we show that the arguments given below further support our claims in Sec.~\ref{sec:lowfield} for the existence of a dimerized state near low field.
Finally, we conclude this appendix with a more thorough analysis of the cone state near high fields.

\subsection{Zero field analysis by quasi-1d methods}
\label{sec:zero-field-analysis}

We start with the zero field case of Eq.~\eqref{eq:hami} in the limit of decoupled chains $J' \ll J$, where each Heisenberg chain can be bosonized using the Wess-Zumino-Witten SU(2)$_1$ theory, with central charge $c = 1$.
In this theory, the spin operator can be decomposed into its uniform $\bM_y(x) = \bJ_{R,y}(x)+\bJ_{L,y}(x)$ and staggered $\bN_y(x)$ magnetizations
\begin{equation}
    \label{low:1}
    \mathbf{S}_{x,y} \to a_0 \left[  \mathbf{M}_y(x) + (-1)^x \mathbf{N}_y(x) \right],
\end{equation}
and its scalar product can be written in the continuum limit
\begin{equation}
    \label{low:2}
    \mathbf{S}_{x,y} \cdot \mathbf{S}_{x+1,y} \to (-1)^x \epsilon_y(x),
\end{equation}
where $\epsilon_y(x)$ is the staggered dimerization.
With $J' = 0$, this theory describes the Luttinger liquid fixed point of the decoupled chains.
The scaling dimensions of these continuum operators, $\bM, \bN,$ and $\ep$, determine the relevance of each operator as it perturbs this fixed point.
The uniform magnetization has scaling dimension 1, whereas both the staggered spin magnetization and the staggered dimerization have scaling dimension 1/2.
These three continuum operators form a closed operator algebra with well-defined operator product expansions (OPEs) used widely in literature~\cite{gogolin2004bosonization,senechal2004introduction,starykh2004dimerized,starykh2005anisotropic,starykh2007ordering,schnyder2008spatially,starykh2010extreme}.
For instance, the product of $\bJ_R$ and $\bN$ can be expanded as
\begin{equation}
    \label{eq:jn}
    J^a(x,\tau) N^b (x', \tau') = \frac{i \epsilon^{abc} N^c(x,\tau) - i \delta^{ab} \ep (x,\tau)}{4\pi\left( v (\tau-\tau') - i (x-x') + a_0 \sigma_\tau \right)},
\end{equation}
where $\tau$ is the imaginary time, $v = \pi J a_0/2$ is the spin velocity, and $a_0$ is the short-distance cutoff.

Let us now consider interchain Hamiltonian perturbing the decoupled Heisenberg chains,
\begin{equation}
    \label{low:3}
    V =  J' \sum_{y = 1}^3 \sum_x \mathbf{S}_y(x) \left( \mathbf{S}_{y+1}(x) + \mathbf{S}_{y+1} (x-1) \right).
\end{equation}
Perturbation theory is formulated by expanding the partition function $Z = \int e^{-S_0-\int d\tau V}$ up to quadratic order, i.e.
\begin{equation}
    \label{eq:pert}
    Z \simeq \int e^{-S_0} \left[ 1 - \int_\tau V + \frac{1}{2} \text{T} \int_{\tau_1}\int_{\tau_2} V(\tau_1) V(\tau_2) \right],
\end{equation}
with an implied short time cutoff $\alpha = a_0/ v$. Here, T is the time-ordering operator.
To utilize this perturbation theory and the OPEs, we express Eq.~\eqref{low:3} in terms of continuum operators, Eqs.~\eqref{low:1} and~\eqref{low:2},
\begin{eqnarray}
    \label{eq:v1}
    V_1  & = & 2a_0^2 J' \sum_{y = 1}^3 \sum_x \bM_y(x) \cdot \bM_{y+1}(x),\\
    V_2 & = &  - a_0^2 J'   \sum_{y = 1}^3 \sum_x \bM_y(x) \cdot \px \bM_{y+1}(x),\\
    V_3 & = & a_0^2 J'  \sum_{y = 1}^3 \sum_x \bN_y(x) \cdot  \px \bN_{y+1}(x), \\
    \label{eq:v4}
    V_4 & = & -a_0^2 J' \sum_{y = 1}^3 \sum_x \bN_y(x) \cdot \frac{1}{2} \px^2 \bN_{y+1}(x),
\end{eqnarray}
where $V = V_1 + V_2 + V_3+V_4$.
It is {\sl crucial} to realize that the periodic boundary conditions enforced in the $y$-direction by the TST system, c.f. Fig.~\ref{fig:lattice2}, allows us to rewrite any operator $\mathcal{O}$ as
\begin{equation}
\label{eq:v1a}
    \sum_{y=1}^3 \sum_x  \mathcal{O}_y \mathcal{O}_{y+1} =  \sum_{y=1}^3 \sum_x  \mathcal{O}_y \mathcal{O}_{y+2}.
\end{equation}
Using OPEs, one can show that the nearest neighbor chain couplings of the staggered magnetization and dimerization enter in the third power of $J'$,
\begin{equation}
    \label{eq:third}
    V = J_3 \sum_{y = 1}^3 \sum_x \left( \bN_y (x) \cdot \bN_{y+1} (x)  - \frac{3}{2} \ep_y (x) \ep_{y+1} (x) \right),
\end{equation}
where $J_3 > 0$ and $J_3 \propto (J')^3$.
This is done by first generating $\px \bN_{y-1} \px \bN_{y+1}$ by quadratic in $V_3+V_4$ terms.
Next, this term is fused with $V_1$ to generate the $J_3 \propto (J')^3$ interaction.
The calculations are similar to those described in Refs.~\onlinecite{starykh2004dimerized,starykh2005anisotropic,starykh2007ordering,schnyder2008spatially,starykh2010extreme} and refer
the reader to these papers for more details.

In a 2d system\cite{starykh2007ordering,starykh2010extreme}, however, we find that the generated term is instead quartic in $J'$, with
interaction constant $J_4 \sim (J')^4/J^3$ and is of the opposite (negative) sign $J_4 < 0$ in comparison with $J_3$ above.
It turns out that $J_3 \sim (J')^3 > 0$ is a feature of the $N=3$ TST model {\sl only}:
wider tubes with $N>3$ are anologous to the 2d case, where the renormalized couplings $\sim (J')^4/J^3 < 0$.
Note that this difference is important as it implies that spin tubes with $N>3$ are not frustrated by the periodic BC along the $y$-direction.

Going back to the $N=3$ TST, both of the generated interactions in Eq.~\eqref{eq:third} are strongly relevant (scaling dimension 1) and scale to strong coupling
under RG transformations. It would appear that because of the greater numerical coefficient of $\ep_y  \ep_{y+1}$
in Eq.~\eqref{eq:third}, it is the dimerized ground state that emerges from the competition in the strong coupling.
However, this argument is not complete as it neglects the crucially important effect of marginally irrelevant in-chain
backscattering term, $\propto {\bf J}_R \cdot {\bf J}_L$, which in fact breaks the symmetry between the
$\bN_y  \cdot \bN_{y+1}$ and $\ep_y  \ep_{y+1}$ interactions in favor of the first one\cite{starykh2007ordering}.
This outcome is not unexpected as it is well-known that in-chain marginal current-current interaction
spoils the extended $SU(2)_R \times SU(2)_L$ symmetry of the Heisenberg chain by subleading logarithmic corrections
which modify chain spin correlations as follows \cite{voit1988,affleck1989}
\begin{eqnarray}
\langle \bN_y(x) \bN_y(0)\rangle &=& (\ln[x])^{1/2} x^{-1}, \nonumber\\
\langle \ep_y(x) \ep_y(0)\rangle &=& (\ln[x])^{-3/2} x^{-1}.
\end{eqnarray}
Essentially, the same mechanism promotes interchain $\bN_y  \cdot \bN_{y+1}$ interaction over that of staggered dimerizations.
In the infinite 2d lattice, this leads to the stabilization of the collinear antiferromagnetic phase \cite{starykh2007ordering},
which, however, is not possible in the TST geometry.

It is important to realize at this point that the relevant $J_3 \sum_y \bN_y  \cdot \bN_{y+1}$ interaction, which
describes non-frustrated coupling of staggered magnetizations on neighboring chains, changes the geometry
of the system into that of a {\em rectangular} spin tube. The renormalized, relevant coupling, $J_3$, become comparable to the intrachain exchange $J$ under RG and
forces N\'eel vectors $\bN_{1,2,3}$ to order into the familiar $120^\circ$ pattern on every rung. Our 1d reasoning
stops at this scale, but further progress can be made by assuming that the spin tube with $J_3 \sim J$
can be accessed from the opposite limit of the strong rung exchange $J_\perp \gg J$\cite{schulz1997}. In this limit,
the spins on each rung form 3-spin triangles that interact via $J_\perp = J_3$, and are coupled to neighboring triangles by a weak exchange $J$.
The ground state of each triangle is 4-fold degenerate and is characterized by
{\em two} quantum numbers, total spin $s_{\rm rung} = 1/2$ and chirality $\tau$, which is itself another pseudo-spin $1/2$ object.
The physical meaning of $\tau$ is just a sense of either a clockwise or a counterclockwise rotation of the `unpaired' spin-$1/2$ in the ground state of the
individual triangle.
In other words, in addition to spin 1/2, the ground state now carries finite momentum $\pm 2\pi/3$ due to chirality.
Focusing on this low-energy subset of triangle's states, one can derive spin-orbital Hamiltonian \cite{kawano1997}
\begin{eqnarray}
H_{\rm s-o} &=& \frac{J_\perp}{N} \sum_x {\bf s}_{\rm rung}(x) \cdot {\bf s}_{\rm rung}(x+1) \times\nonumber\\
&&\times [1 + \alpha_N (\tau^+_x \tau^-_{x+1} + \tau^-_x \tau^+_{x+1})]
\label{eq:so}
\end{eqnarray}
describing correlated dynamics of spins and chiralities. For the triangular ladder considered here, $N=3$ and $\alpha_N = 4$.
The presented arguments remain valid for any {\em odd} $N$, however. See Ref. \onlinecite{kawano1997} for $N=5$
and Ref. \onlinecite{subrahmanyam1994} for $N>5$. Analytical \cite{schulz1997,orignac2000} and numerical \cite{kawano1997,pati2000,fouet2006frustrated}
studies of the model \eqref{eq:so} find dimerized ground state, in agreement with our consideration in Section~\ref{sec:dimer-from-spir}.
Fig.~\ref{fig:EEm0_0}, which shows oscillatory behavior of the entanglement entropy for different values of $R$, represents
clear evidence of the dimerized ground state.

Finally, we conclude by discussing the way to generate an interaction of the uniform magnetizations from the next neighboring chains.
This is done by fusing $V_1$ in Eq.~\eqref{eq:v1} with itself, which yields, under Eq.~\eqref{eq:pert},
\begin{eqnarray}
\label{eq:apB2}
    \delta H_{MM} =  -\frac{(2J')^2}{2} \sum_y \int_x \int_{x'}&& \langle M^z_y(x,\tau) M^z_y(x',\tau')\rangle \nonumber \\
            &&\times M^z_{y-1} M^z_{y+1},
\end{eqnarray}
Because the result is converging, the integral of the $y$-th chain correlation function can be extended to the full $x - v\tau$ plane.
This, using important short-distance cut-off $\sim a_0 \rm{sign}(\tau)$ and $y = v\tau$ (see Ref. \onlinecite{starykh2005anisotropic} for detailed discussion), leads to
\begin{equation}
\int_{-\infty}^\infty dx \int_{-\infty}^\infty dy \Big(\frac{1}{(y + i x + a_0 \rm{sign}(y))^2} + {\rm h.c.}\Big) = 4\pi.
\end{equation}
As a result we obtain for the amplitude of $\delta \gamma_{\rm MM} =  (J')^2/(\pi v)$, where $v$ is magnetization dependent
spin velocity.

\subsection{Cone state}
\label{sec:cone}

Now, turn on the magnetic field.
When a large enough magnetic field is applied to the TST, the ``twist" order, the fourth term in Eq.~\eqref{eq:perturbH1}, becomes more relevant than the SDW.
This was discussed in previous papers for the two-chain ladder\cite{kolezhuk05} as well as the 2d triangular lattice\cite{starykh2007ordering,starykh2010extreme}.
As both the SDW and the cone interaction amplitudes in \eqref{eq:perturbH1} are of
the order $J'$, the relative importance of the two interactions can be
estimated \cite{starykh2007ordering} from a comparison of their scaling
dimensions, $\Delta_{\rm saw} = 1/(2 \pi \mathcal{R}^2)$ and $\Delta_{\rm cone}
= 1 + 2\pi \mathcal{R}^2$.  These two dimensions are equal when $2 \pi \mathcal{R}^2 =
(\sqrt{5}-1)/2$, which takes place at sufficiently high magnetization
$M\approx 0.6 M_{\rm sat}$.  Because of rather steep dependence $M(h)$
of the magnetization on the magnetic field near the saturation, this
value of magnetization corresponds to $h\approx 0.9 h_{\rm sat}$, see
Fig. 2 in Ref. \onlinecite{starykh2007ordering}. A similar conclusion is obtained by
comparing mean-field transition temperatures of these two ordered
states as functions of magnetization, see Ref. \onlinecite{starykh2010extreme}.

These arguments, however, are not complete because they do not take into account the fluctuation-generated interactions
between spin densities on {\sl next-nearest} chains. The most important of these in the presence of an external magnetic field is
given  by
\begin{equation}
\label{eq:cone1}
V'_{\rm cone} = \delta \gamma_{\rm cone} \sum_y \int dx ~\mathcal{S}^+_{\pi, y} \mathcal{S}^{-}_{\pi, y+2} + {\rm h.c.}.
\end{equation}
Even though the generated coupling constant is small, $\delta \gamma_{\rm cone} \ll J'/J \ll 1$, this interaction
does not involve spatial derivatives and has scaling dimension $2\pi \mathcal{R}^2$ which approaches $1/2$ as $h \to h_{\rm sat}$.
Thus, this is a strongly relevant term.

In a 2d system\cite{starykh2007ordering,starykh2010extreme}, $\delta \gamma_{\rm cone} \sim (J')^4/J^3 < 0$ as discussed in the previous subsection.
(Note that \eqref{eq:cone1} is written in the `sheared' system of coordinates.) When translated into Cartesian coordinates,
it implies antiferromagnetic (positive) exchange interactions between spins on next-nearest chains at the same position ${\sf x}$
along the chain \cite{starykh2007ordering}. Crucially, as emphasized in the previous section, the TST geometry allows for a stronger renormalized coupling,
of the order of $\delta \gamma_{\rm cone} \equiv J_3 \sim (J')^3/J^2 > 0$.

The difference is due to slightly different routes to \eqref{eq:cone1} in 2d and $N=3$ TST geometries.
One can first show that, when you start from the original cone interaction
\begin{equation}
\label{eq:cone2}
V_{\rm cone} = \gamma_{\rm cone} \sum_y \int dx ~\mathcal{S}^+_{\pi, y} \partial_x \mathcal{S}^{-}_{\pi, y+1} + {\rm h.c.},
\end{equation}
one can couple the derivatives $\partial_x  \mathcal{S}^{\pm}_{\pi}$ on the next-nearest chains $y$ and $y+2$,
\begin{equation}
\label{eq:cone4}
V''_{\rm cone} \sim \frac{\gamma_{\rm cone}^2}{v} \sum_y \int dx ~\partial_x \mathcal{S}^+_{\pi, y} \partial_x \mathcal{S}^{-}_{\pi, y+2} + {\rm h.c.}.
\end{equation}
This step parallels calculations leading to Eq.~\eqref{eq:third} with minor variation due to $U(1)$ symmetry of the system in the
presence of an external magnetic field. In this situation the scaling dimension of the $\mathcal{S}_\pi$ field is smaller than $1/2$
which leads to a slightly different numerical pre-factor in the renormalization. However
the functional dependence on $J'$ remains the same.
Secondly, for all $N > 3$ one also needs to generate
\begin{equation}
\label{eq:cone3}
V'_{\rm MM} = -\delta \gamma_{\rm MM} \sum_y \int dx ~M^z_y M^z_{y+2},
\end{equation}
which was described in the end of the previous subsection, Sec.~\ref{sec:zero-field-analysis} . Here, $\delta \gamma_{\rm MM}  \sim (J')^2/J$.
Fusing next \eqref{eq:cone4} and \eqref{eq:cone3} together leads to the result \eqref{eq:cone1}.
In the $N=3$ TST, however, the second step is not required due to \eqref{eq:v1a}, and we end up with a larger coupling of
the order $\delta\gamma_{\rm cone} \sim (J')^3/J^2 > 0$ in \eqref{eq:cone1}.

To compare the original $V_{\rm cone}$ with the generated $V'_{\rm cone}$ quantitatively, we can estimate the RG scale $\ell$ at which
the coupling constant of the interaction becomes of the order one (in units of spin velocity $v$). For \eqref{eq:cone2} this is, with
logarithmic accuracy,
$\ell_{\rm cone} \sim -\ln(J')/(2 - \Delta_{\rm cone}) = -\ln(J')/(1 - 2\pi \mathcal{R}^2)$, while for \eqref{eq:cone1} it is
$\ell_3 \sim -3 \ln(J')/(2 - 2\pi \mathcal{R}^2)$. We immediately conclude that $\ell_3 < \ell_{\rm cone}$ for all values of $2\pi \mathcal{R}^2 \in (1,1/2)$,
i.e. that the generated cone interaction term \eqref{eq:cone1} is more relevant than the bare one
for all values of magnetization in the case of $N=3$ TST. Similar consideration allows us to analyze the competition between
the generated cone $V'_{\rm cone}$ interaction and the SDW one, which is characterized by the RG scale
$\ell_{\rm sdw} \sim - 2\pi \mathcal{R}^2 \ln(J' \sin[\delta])/(4\pi \mathcal{R}^2 - 1)$.
We find that $\ell_{\rm sdw} < \ell_3$ for $1 \geq 2\pi \mathcal{R}^2 \geq \sqrt{7}-2 \approx 0.65$, which corresponds to low-to-intermediate range of magnetization
$M\gtrsim 0.25$. At higher $M$, however, the modified cone interaction takes over the SDW one.
(For the 2d case, the comparison is less conclusive as the result sensitively depends on numerical factors inside the argument of the logarithm \cite{starykh2010extreme}.)

We now investigate the consequences of the strong $J_3\equiv \delta\gamma_{\rm cone}$ interaction in Eq.~\eqref{eq:cone1} for the TST problem.
In the high-field region where SDW fluctuations are suppressed, the Hamiltonian of the system is given by the sum
of $H_0$ in Eq.~\eqref{eq:Hchain}, the generated direct coupling $V'_{\rm cone}$ in Eq.~\eqref{eq:cone1}, and the original cone
interaction $V_{\rm cone}$ in Eq.~\eqref{eq:cone2}, which now is a subleading one in comparison with \eqref{eq:cone1}.
With this, we perform abelian bosonization form of the interaction potential and arrive at the following expression,
\begin{widetext}
\begin{eqnarray}
\label{eq:tst-cone}
H^{\rm TST}_{\rm cone} &&= J_3 \int dx \{ \cos[\beta(\theta_1 - \theta_2)] +  \cos[\beta(\theta_2 - \theta_3)] +\cos[\beta(\theta_3 - \theta_1)]\}\\
&&+ \frac{\beta J'}{2} \int dx \{ \partial_x(\theta_1 + \theta_2) \sin[\beta(\theta_1 - \theta_2)]
+\partial_x(\theta_2 + \theta_3) \sin[\beta(\theta_2 - \theta_3)] + \partial_x(\theta_3 + \theta_1) \sin[\beta(\theta_3 - \theta_1)] \}.\nonumber
\end{eqnarray}
\end{widetext}
For $J_3 \gg J'$,  which is the appropriate regime according to our RG arguments above, this potential is minimized by configurations
with $\cos[\beta(\theta_y - \theta_{y+1})] = -1/2$ for all $y$. This allows for two different values of sine terms,
$\sin[\beta(\theta_y - \theta_{y+1})] = \pm \sqrt{3}/2$. In fact, different signs describe states with different vector chiralities
defined as
\begin{equation}
\label{eq:tst-kappa}
\kappa_y^z = \Big(\mathbf{S}_y \times \mathbf{S}_{y+1}\Big)_z \sim \sin[\beta(\theta_y - \theta_{y+1})] .
\end{equation}
Thus, different signs of $\kappa_y^z$ correspond to different senses of rotation (clockwise or counterclockwise) of
$e^{i\beta \theta_y}$ as we go from one chain to the next.
These chiralities also represent useful order parameter describing two degenerate cone states \cite{sato07A}.

To account for the subleading twist terms with spatial derivatives in \eqref{eq:tst-cone}, we shift $\theta_y \to \theta_y + \upsilon x$,
where $\upsilon$ is determined by the requirement that in the new ground state, the bosonic field $\theta$ is twist-less, i.e. $\langle \partial_x \theta_y\rangle =0$.
Minimizing $H_0 + H^{\rm TST}_{\rm cone}$ over $\upsilon$, we find
\begin{equation}
\label{eq:tst-shift}
\upsilon = -\beta J' \langle \sin[\beta(\theta_y - \theta_{y+1})] \rangle \sim - J' \kappa_y^z
\end{equation}
This shows that the doubly-degenerate cone state is characterized by incommensurate transverse spin correlations,
by virtue of the relation ${\cal S}_y^+ = (-1)^x e^{i\beta \theta_y} \to \exp[i (\pi + \upsilon) x + i \beta \theta_y]$.
Depending on the spontaneously chosen vector chirality, Eq.~\eqref{eq:tst-kappa}, transverse spin correlations are picked at either
$Q_{1,x} = \pi + \upsilon$ (for $\kappa_y^z > 0$) or $Q_{2,x} = -\pi + \upsilon$ (for $\kappa_y^z < 0$) along the chain.


\section{Transformation properties of $\mathbb{Z}_2$ vortices}
\label{sec:transf-prop-mathbbz}

In this appendix, we address the transformation properties of the
$\mathbb{Z}_2$ vortex instanton operator $\psi$.  We give several
arguments.  First, these properties have been implicitly obtained in the
case of a three leg spin tube, slightly different from the one studied
here, in Ref.~\onlinecite{oshikawa2010}.  There, the authors explicitly evaluate the
Berry phase contribution to the action for instantons on the lattice.
Microscopically, the instantons are associated with columns of spatial
{\sl links} along the $x$-direction of the cylinder (see below how this arises
in another formulation).  They showed that, due to the Berry phase, a
single pair of instantons (an odd number of instantons cannot occur)
is accompanied by a weight,
\begin{equation}
  \label{eq:62}
  e^{iS_{BP}}  =e^{2\pi i S (x-x')},
\end{equation}
where $x$ and $x'$ are the locations of the instantons.  For half-integer
spins, this gives an oscillating factor equal to +1 or -1 if
the separation between instantons is even or odd, respectively.  From
this we can extract the transformation properties.  If we translate
{\sl one} of the instantons, $x \rightarrow x+1$, we see that the
weight in Eq.~\eqref{eq:62} changes sign.  This requires $\psi
\rightarrow -\psi$, in agreement with Eq.~\eqref{eq:63}.  Under
inversion, $P$, about a lattice site, the instantons, which live on
the links, change from the even to odd sublattice of bonds and
vice-versa.  Inverting a single instanton, therefore, changes the parity
of $x$, and hence also the sign of the weight in Eq.~\eqref{eq:62}.
Thus, again, $\psi$ is odd under inversion, in agreement with
Eq.~\eqref{eq:64}.  Since the instantons do not move under
time-reversal or translation along $y$, the invariance of $\psi$ under
these operations is obvious.  Thus, for the case $L_y=3$, for the
model studied in Ref.~\onlinecite{oshikawa2010}, the symmetry of the
instanton operator is determined as shown in the text.

We turn now to an alternative derivation of the transformation laws,
which gives the general result, and clarifies its generality.  Here we
follow the general strategy of Ref.~\onlinecite{PhysRevB.84.104430}, in
which the $\mathbb{Z}_2$ vortices are explicitly separated from the
smooth configurations of the SO(3) order parameter using a slave
particle construction.   This is achieved by writing the unit vectors
defining the SO(3) matrix in terms of a ``slave spinon'' $z_\alpha$:
\begin{equation}
  \label{eq:61}
  {\bf\hat{n}}_1 + i {\bf\hat{n}}_2 = \epsilon_{\alpha\beta} z_\beta
  {\boldsymbol \sigma}_{\alpha\gamma} z_\gamma,
\end{equation}
where the complex, two component vector $z_\alpha$ is constrained to
have unit norm, $\sum_\alpha z_\alpha^* z_\alpha^{\vphantom*}=1$.  This
representation faithfully reproduces the orthonormality constraints on
the ${\bf\hat n}_i$, but is two to one: the physical order parameter
${\mathcal O}$ is unchanged by the transformation $z_\alpha
\rightarrow - z_\alpha$.  This is actually a gauge invariance, since the
transformation is made locally.  The $\mathbb{Z}_2$ vortex is a
configuration in which, on encircling the center of the defect,
$z_\alpha$ returns not to itself but to $-z_\alpha$.

As explained in Ref.~\onlinecite{PhysRevB.84.104430}, a low energy
effective theory, appropriate to describe the regime with a local spiral
order, as well as a quantum disordered phase, is a 2+1 dimensional
$\mathbb{Z}_2$ gauge theory coupled to the spinon variables
$z_\alpha$.  We refer the reader to Ref.~\onlinecite{PhysRevB.84.104430}
for details.
The $\mathbb{Z}_2$ vortex in this theory appears as a
configuration of a spinon field which has a discontinuity $z_\alpha
\rightarrow -z_\alpha$ across a semi-infinite ``cut'' emanating from
the vortex.
This $\mathbb{Z}_2$ vortex is accompanied by an Ising vortex, the so-called ``vison",
which is itself a defect with a non-zero Ising gauge field crossing the same semi-infinite ``cut".
In this way, the topological defects
of the spiral magnet become identified with the visons of the
$\mathbb{Z}_2$ gauge theory.

The discussion in the previous paragraph applies to $\mathbb{Z}_2$
vortices in two-dimensional space, which are particles in the 2+1-dimensional theory.
We need to go from this to the description of
instantons in the 1+1-dimensional theory obtained by applying periodic
boundary conditions in the $y$ direction.  A 1+1-dimensional instanton
can be viewed as an event in which a pair of $\mathbb{Z}_2$ vortices is
nucleated: one of them winds around the cylinder and finally arrives
back at the other $\mathbb{Z}_2$ vortex and annihilates it.  We can, by
the previous argument, consider the particles nucleated and
annihilated to be visons in the gauge theory.

Such a process was considered in Ref.~\onlinecite{jiang2011spin} (in
the Supplementary Material), where it was shown that the operator
representing this process in the Ising gauge theory has the
transformation properties in Eqs.~\eqref{eq:63}, \eqref{eq:64}, i.e. this operator can
be viewed as a staggered dimerization operator for odd $L_y$.  There, a
rectangular lattice gauge theory was studied, but the basic physics is
quite general.  Let us consider the translation.  We ask about the
amplitude to first wind one vison around the cylinder at position $x$,
and then wind another at position $x+1$.   The overall phase of the
amplitude for both processes taken together gives the transformation
property of the instanton operator under translation. The visons reside at the
plaquette centers of the original lattice, and the winding trajectories
form closed circles at fixed $x$, circumnavigating the cylinder.  Together, these two events
form two such circles that enclose one column of sites in the lattice.
The fundamental property of a vison is that it has a mutual statistical
interaction with  ``electric'' gauge charges, with the
wavefunction acquiring a phase of $\pi$ whenever one encircles the
other.  For a $S=1/2$ system, a unit gauge charge is present at every
lattice site -- this represents the physical spin at each site.  The net
effect of the two events together is that one vison is wound around each
site of the lattice between the two circles, leading to an overall
amplitude of $(-1)^{L_y}$ for the two processes together. Here, $L_y$ is
the number of sites contained between the two circles.  This gives
the result in Eq.~\eqref{eq:63}.    Note that we may also roughly understand
this phase factor by considering the smooth rotations of
microscopic spins between the two contours, all of which rotate by
$2\pi$, and, due to their $s=1/2$ spinor transformation properties, each
acquire a minus sign.  A similar argument shows that spatial
inversion gives the same phase factor.  Explicit calculations for these
factors in the Ising gauge theory can be found in
Ref.~\onlinecite{jiang2011spin}.  Note that these arguments do not
depend at all on the interactions in the model, just the presence of
these symmetries and fundamental statistics of the particles.

\end{document}